\documentclass[10pt,a4paper]{article}
\usepackage[dvips]{color}
\usepackage{epsfig}
\usepackage{amsmath}
\usepackage{graphicx}

\begin{document}
\title{\bf Mutually interacting Tachyon dark energy with variable $G$ and $\Lambda$}
\author{{J. Sadeghi$^{a}$ \thanks{Email: pouriya@ipm.ir},\hspace{1mm} M. Khurshudyan$^{b}$ \thanks{Email: martiros.khurshudyan@nano.cnr.it},
\hspace{1mm} M. Hakobyan$^{c,d}$
\thanks{Email: margarit@mail.yerphi.am}\hspace{1mm} and H. Farahani$^{a,e}$ \thanks{Email:
h.farahani@umz.ac.ir}}\\
$^{a}${\small {\em  Department of Physics, Islamic Azad University - Ayatollah Amoli Branch,}}\\
        {\small {\em P.O.Box 678, Amol, Iran}}\\
$^{b}${\small {\em Department of Theoretical Physics, Yerevan State
University, 1 Alex Manookian, 0025, Yerevan, Armenia}}\\
$^{c}${\small {\em A.I. Alikhanyan National Science Laboratory,
Alikhanian Brothers St., Yerevan, Armenia} } \\
$^{d}${\small {\em Department of Nuclear Physics, Yerevan State
University, Yerevan, Armenia}}\\
$^{e}${\small {\em Department of Physics, Mazandaran
University, Babolsar, Iran}}}  \maketitle
\begin{abstract}
In this paper, we consider Tachyonic scalar field as a model of dark energy with
interaction between components in the case of variable $G$ and
$\Lambda$. We assume a flat Universe with specific form of scale
factor and study cosmological parameters numerically and
graphically. Statefinder analysis also performed as well.
In the special choice of interaction parameters we succeed to obtain analytical expression of densities.
We find that our model will be stable in the late stage but there is an instability at the early Universe. So we propose this model as a realist model of our Universe.
\end{abstract}

\section*{\large{Introduction}}
In order to explain recent observational data, which reveals
accelerating expansion of the Universe, several models were
proposed. One of the possible scenarios is the existence of a dark
energy with negative pressure and positive energy density giving an
acceleration to the expansion.\\
Various kinds of dark energy models have been proposed such as
cosmological constant [1], quintessence [2-4], k-essence [5-7],
tachyon [8-11], phantom [12-14], ghost dark energy [15-18], Chaplygin
gas and its extensions [19-28], quintom [29], holographic dark
energy [30-34], and extra dimensions [35, 36].\\
Among above models concerning to the nature of the dark component of
the Universe, in this article, we assume that it could be described
by a scalar field and we choose a scalar field called Tachyonic
field with the following relativistic Lagrangian,
\begin{equation}\label{eq:tach lag}
L_{TF}=-V(\phi)\sqrt{1-\partial_{\mu}\phi\partial^{\nu}\phi},
\end{equation}
which captured a lot of attention (see, for instance, references in
[37]). The stress energy tensor is given by,
\begin{equation}\label{eq:energy tensor}
T^{ij}=\frac{\partial L_{TF}}{\partial
(\partial_{i}\phi)}\partial^{j}\phi-g^{ij}L_{TF},
\end{equation}
which gives the energy density and pressure as the following
expressions,
\begin{equation}\label{eq:tachyonic density}
\rho=\frac{V(\phi)}{\sqrt{1- \partial_{i}\phi \partial^{i}\phi}},
\end{equation}
and,
\begin{equation}\label{eq:tachyonic pressure}
P=-V(\phi)\sqrt{1- \partial_{i}\phi \partial^{i}\phi}.
\end{equation}
Our next step is to decompose Eqs. (3) and (4) as follow,
\begin{eqnarray}\label{eq:presdens}
\rho &=& \rho_{\small{m}}+\rho_{\small{\Lambda}},\nonumber\\
P&=&P_{\small{m}}+P_{\small{\Lambda}},
\end{eqnarray}
with the following components,
\begin{eqnarray}\label{matter}
\rho_{\small{m}}&=&\frac{V(\phi)\partial_{i}\phi
\partial^{i}\phi}{\sqrt{1- \partial_{i}\phi \partial^{i}\phi}},\nonumber\\
P_{\small{m}}&=&0,\nonumber\\
\omega_{\small{m}}&=&0,
\end{eqnarray}
and,
\begin{eqnarray}\label{eq:cosconst}
\rho_{\small{\Lambda}}&=& V(\phi)\sqrt{1- \partial_{i}\phi
\partial^{i}\phi}\nonumber\\
P_{\small{\Lambda}}&=&-V(\phi)\sqrt{1-
\partial_{i}\phi \partial^{i}\phi},\nonumber\\
\omega_{\small{\Lambda}}&=&-1.
\end{eqnarray}
It means that we can consider Tachyonic scalar field as a
combination of a cosmological constant and pressureless matter with
$\omega_{m}=0$. From mathematical point of view this is not only one
possibility and rather different splitting could be considered.\\
We should note that, flat FRW metric with the following line
element,
\begin{equation}\label{eq:FRW metric}
ds^{2}=dt^{2}-a(t)^{2}\left(dr^{2}+r^{2}d\theta^{2}+r^{2}\sin^{2}\theta d\phi^{2}\right)
\end{equation}
will be used for our purposes.\\
Recently, Ref. [37] considered a model, where the components of a
Tachyonic scalar field interact mutually. Motivated by the idea of
that work, we would like to consider interaction
$Q=3Hb\rho+\gamma\dot{\rho}$ of the general form between components
in the case of variable $G$ and $\Lambda$. As we know the Einstein equation has two important parameters which are the gravitational constant $G$ and the
cosmological constant $\Lambda$. It is known that $G$ plays the role of a
coupling constant between geometry and matter in the Einstein equations. In an
evolving Universe, it appears natural to look at this constant as a function of time [38, 39]. Also, time-dependent cosmological constant has been considered by several works in various variable $G$ theories [40, 41].
It is possible to point out boundary on $G$ for instance, observation of spinning-down
rate of pulsar $PSR J2019+2425$ provides the result,
\begin{equation}\label{eq:gvar1}
\left|\frac{\dot{G}}{G} \right|\leq (1.4-3.2) \times 10^{-11} yr^{-1}.
\end{equation}
Depending on the observations of pulsating white dwarf star $G 117-B
15A$, the astroseismological bound may be [42],
\begin{equation}\label{eq:gvar2}
\left|\frac{\dot{G}}{G} \right|\leq 4.1 \times 10^{-10} yr^{-1}.
\end{equation}
Today, $\Lambda$ has the incredibly small value, $\Lambda<10^{-46} GeV^4$, whereas generic inflation models
require that $\Lambda$ has a large value during the inflationary epoch. This is the source of the
cosmological constant problem. We hope that consideration of variable $\Lambda$ could solve such problem. Also in a pioneer work on varying cosmological constant and its interaction with matter suggested to resolve the fine-tuning problem [43]. So, we apply these idea to the recent work [37] and extend this model.\\
Also, interacting models may solve the cosmic coincidence problem [44, 45]. In the Ref. [46] it is found that the interaction between dark sectors cannot ensure the dark energy to fully cluster along with dark matter. There are several possibilities to choose interaction term for example those introduced in the ref. [47].
It is also possible to construct holographic cosmological model where dark matter and dark energy interact non-gravitationally with
each other [48]. In the interesting work [49] the effects of interaction between dark matter and dark energy on the evolution of the gravitational and
the peculiar velocity fields investigated. In the recent work [50] it is concluded that an interaction is compatible with the recent observations and can provide a strong argument towards consistency of different values of cosmological parameters. All of these give us motivation to use interaction in the model to have a comprehensive model.\\
This paper organized as follows, in next section we will
introduce the equations which governs our model. Then, we study
statefinder diagnostics. Also we will consider mathematics and
solving strategy of the problem for non interacting case and will
present analysis of the model for special type of scale factor.
Model including interaction between components analyzed numerically
and cosmological parameters discussed graphically. Last section
includes discussions and conclusion.
\section*{\large{The field equations}}
Field equations that govern our model with variable $G(t)$ and
$\Lambda(t)$ (see for instance [41]) are,
\begin{equation}\label{eq: Fridmman vlambda}
H^{2}=\frac{\dot{a}^{2}}{a^{2}}=\frac{8\pi G(t)\rho}{3}+\frac{\Lambda(t)}{3},
\end{equation}
and,
\begin{equation}\label{eq:fridman2}
\frac{\ddot{a}}{a}=-\frac{4\pi G(t)}{3}(\rho+3P)+\frac{\Lambda(t)}{3}.
\end{equation}
Energy density (3) and pressure (4) of a Tachyonic field reduced to
following expressions,
\begin{equation}\label{eq:tachyonic density FRW}
\rho=\frac{V(\phi)}{\sqrt{1- \dot{\phi}^{2}}},
\end{equation}
and,
\begin{equation}\label{eq:tachyonic pressure FRW}
P=-V(\phi)\sqrt{1- \dot{\phi}^{2}}.
\end{equation}
Also, energy conservation $T^{;j}_{ij}=0$ reads as,
\begin{equation}\label{eq:conservation}
\dot{\rho}+3H(\rho+P)=0.
\end{equation}
In the case of conservation of particle number in Universe, combination
of (11), (12) and (15) gives the following relationship between
$\dot{G}(t)$ and $\dot{\Lambda}(t)$,
\begin{equation}\label{eq:glambda}
\dot{G}=-\frac{\dot{\Lambda}}{8\pi\rho}.
\end{equation}
Hereafter we will assume special forms of scale factor $a(t)$ and
cosmological constant $\Lambda(t)$. This assumption allows us to
determine $G(t)$, $\rho$, $\phi$ and $V(\phi)$. Before investigation
of these quantities we study statefinder diagnostics.
\section*{Statefinder diagnostics}
In the framework of general relativity it is accepted that a dark energy can explain
the present cosmic acceleration. Except cosmological constant, there
are many others candidates of dark energy. The property of dark
energy is model dependent and to differentiate different models of
dark energy, a sensitive diagnostic tool is needed.\\
Hubble parameter $H$ and deceleration parameter $q$ are very
important quantities which can describe the geometric properties of
the Universe. Since $\dot{a}>0$, hence $H>0$ means the expansion of
the Universe. Also, $\ddot{a}>0$, which is $q<0$ indicates the
accelerated expansion of the Universe. Since, the various dark
energy models give $H>0$ and $q<0$, they can not provide enough
evidence to differentiate the more accurate cosmological
observational data and the more general models of dark energy. For
this aim we need higher order of time derivative of scale factor and
geometrical tool. Ref. [51] proposed geometrical statefinder
diagnostic tool, based on dimensionless parameters $(r, s)$ which
are function of scale factor and its time derivative. These
parameters are defined as,
\begin{equation}\label{eq:statefinder}
r=\frac{1}{H^{3}}\frac{\dddot{a}}{a},
\end{equation}
and,
\begin{equation}\label{eq:statefinder}
s=\frac{r-1}{3(q-\frac{1}{2})},
\end{equation}
where the deceleration parameter given by,
\begin{equation}\label{eq:accparam}
q=-\frac{1}{H^{2}}\frac{\ddot{a}}{a}.
\end{equation}
It can be rewritten as the following,
\begin{equation}\label{eq:accchange}
q=\frac{1}{2}\left(1+3\frac{8 \pi G(t) P_{\Lambda} - \Lambda(t)}{8
\pi G(t) \rho + \Lambda(t)} \right).
\end{equation}
We give numerical description of statefinder parameters in the last
section.
\section*{\large{Method}}
In this paper we use the following forms of scale factor,
cosmological constant and interaction term.\\
We assume that the Universe is in a quasi-exponential expansion
phase with the following scale factor [37],
\begin{equation}\label{eq:scale Murli}
a(t) = a_{0}t^{n}\exp{(\alpha t)}.
\end{equation}
Also we assume the following scale factor-dependent cosmological
constant,
\begin{equation}\label{eq:cos}
\Lambda(t) = H^{2}+Aa^{-k}.
\end{equation}
Finally we consider the following interaction term,
\begin{equation}\label{eq:interaction}
Q=3Hb\rho+\gamma \dot{\rho}.
\end{equation}
In the $\gamma=0$ limit, the interaction term reduced to those for example used in the refs. [52, 53]. There are undetermined constants $n, \alpha, A, k, b$, and $\gamma$
in relations (21)-(23) which will be fixed in our numerical study.
Concerning to the forms and types of an interaction $Q$, we have
already discussed a lot of in our previous works [15, 21, 23]. Just
to mention to our readers about the form considered in this article
is that it carries a phenomenological character $\gamma
\dot{\rho}$, were introduced from units correctness point of view.
This form is going to be one of the forms intensively considered in
literature from different corners and found to be suitable for
cosmological problems. Generally in literature many authors are
taking such terms, which will simplify a problem and will have
analytical solutions.\\
In that case we obtain the following equation which gives dynamics
of $G$,
\begin{equation}\label{eq:dG}
\dot{G}+\frac{ ( Akt^{2} + 2n ( a_{0} e^{\alpha t} t^{n} )^{k} ) (n+\alpha t) }{ At^{3} -2t ( a_{0} e^{\alpha t} t^{n} )^{k} ( n+ \alpha t )^{2}} G=0.
\end{equation}
Absence of an interaction between components means that components
evolves separately i.e. the equation (15) separates into the
following equations,
\begin{equation}\label{eq: encperf}
\dot{\rho}_{\small{m}}+3H(\rho_{\small{m}}+P_{\small{m}})= 0,
\end{equation}
and,
\begin{equation}\label{eq: enctf}
\dot{\rho}_{\small{\Lambda}}+3H(\rho_{\small{\Lambda}}+P_{\small{\Lambda}})=0.
\end{equation}
From the equation (22), for the energy density of a Tachyonic
matter, we obtain,
\begin{equation}\label{eq:end TFM}
\rho_{\small{m}}=\rho_{\small{0m}}e^{\left[ -3 (\alpha t + n \ln{t}
)\right]}.
\end{equation}
For the pressure of a cosmological constant $P_{\small{\Lambda}}$ we
will have,
\begin{equation}\label{eq:pressure cosmconst}
P_{\small{\Lambda}}=\rho_{\small{0m}}e^{\left[ -3 (\alpha t + n
\ln{t} )\right]}+\frac{A ( a_{0} t^{n} e^{\alpha t} )^{-k} - 2
t^{-2} (n + \alpha t)}{8 \pi G}.
\end{equation}
For the Tachyonic field and potential we obtain,
\begin{equation}\label{eq:potential TF time}
\phi(t)=\int  {\sqrt{ 1-\frac{ 8 \pi G \rho_{\Lambda} } { 2 t^{-2} (n+\alpha t)^{2} - A ( a_{0} t^{n} e^{ \alpha t } )^{-k} } } ~dt },
\end{equation}
and,
\begin{equation}\label{eq:potnint}
V(\phi)=\frac{ \rho_{\Lambda} }{ \sqrt{1-\dot{\phi}^{2} } },
\end{equation}
where $\rho_{\Lambda} = -P_{\Lambda}$ is used. It means that $\dot{\phi}^{2}\leq1$.
In the special case of $\dot{\phi}^{2}=1$ Tachyon potential diverges and takes infinite value.\\
On the other hand, accounting interaction between cosmic components
modifies (25) and (26) in a such way that the conservation of energy
stays true. In that case we have,
\begin{equation}\label{eq:indiven}
\dot{\rho}_{\small{m}}+3H(\rho_{\small{m}}+P_{\small{m}})=Q,
\end{equation}
and,
\begin{equation}\label{eq:indiven}
\dot{\rho}_{\small{\Lambda}}+3H(\rho_{\small{\Lambda}}+P_{\small{\Lambda}})=-Q.
\end{equation}
Therefor, corresponding to our case with $\omega_{\small{m}}=0$ and
$\omega_{\small{\Lambda}}=-1$, we can obtain,
\begin{equation}\label{eq:final}
(1-\gamma)\dot{\rho}_{\small{m}}+3H(1-b)\rho_{\small{m}}=3Hb\rho_{\small{\Lambda}
} +\gamma \dot{\rho}_{\small{\Lambda}},
\end{equation}
and,
\begin{equation}\label{eq:final}
(1+\gamma)\dot{\rho}_{\small{\Lambda}}+3Hb\rho_{\small{\Lambda}
}=-3Hb\rho_{\small{m} }-\gamma \dot{\rho}_{\small{m} }.
\end{equation}
We use above relations to give numerical analysis of our system
to obtain behavior of some important cosmological parameters.
\subsection*{\large{Analytical results}}
Before numerical analysis of some important cosmological parameters
in general case we try to obtain time-dependent densities and
pressures in the special case where we restrict interaction
parameters as $b=\gamma$. This assumption helps us to decouple
equations given by (33) and (34) to extract $\rho_{\Lambda}(t)$ and
$\rho_{m}(t)$. Then, by using relations (31) and (32) we can obtain
pressure $P_{\Lambda}(t)$. In that case we can also investigate
stability of theory (see discussion section).\\
Under above
assumption we can obtain the following densities,
\begin{equation}\label{35}
\rho_{\Lambda}=ct^{-3n\gamma^{3}}e^{-3\alpha\gamma^{3}t},
\end{equation}
and,
\begin{equation}\label{36}
\rho_{m}=ct^{-3n}e^{-3\alpha
t}\left[1-\frac{\gamma^{3}+\gamma^{2}-1}{\gamma^{3}-1}t^{3n(1-\gamma^{3})}e^{-3\alpha(\gamma-1)(\gamma^{2}+\gamma+1)t}\right],
\end{equation}
where $c$ is an integration constant. These lead to the following
pressure,
\begin{equation}\label{37}
P_{\Lambda}=\frac{c(\gamma^{6}-\gamma^{5}-2\gamma^{3}+\gamma^{2}+1)}{\gamma^{3}-1}t^{-3n\gamma^{3}}e^{-3\alpha\gamma^{3}t},
\end{equation}
Therefore, we can write the following expressions of total density,
\begin{equation}\label{38}
\rho=c\left[t^{-3n\gamma^{3}}e^{-3\alpha\gamma^{3}
t}+\left(1-\frac{\gamma^{3}+\gamma^{2}-1}{\gamma^{3}-1}t^{-3n\gamma(\gamma^{3}-1)}e^{-3\alpha(\gamma-1)(\gamma^{2}+\gamma+1)t}\right)t^{-3n}e^{-3\alpha
t}\right].
\end{equation}
As we expected, $P_{m}=0$. Therefore, from the equation (28) and (37) one can obtain,
\begin{equation}\label{39}
8\pi G={\frac {{t}^{n \left( 3\,{\lambda}^{3}+2 \right) }{{\rm e}^{
\alpha\,t \left( 3\,{\lambda}^{3}+2 \right) }}-2\,{t}^{3\,n{\lambda}^{
3}+3\,n-2}{{\rm e}^{3\,\alpha\,t \left( {\lambda}^{3}+1 \right) }}n-2
\,{t}^{3\,n{\lambda}^{3}+3\,n-1}{{\rm e}^{3\,\alpha\,t \left( {\lambda
}^{3}+1 \right) }}\alpha}{ c{\lambda}^{3}{{\rm e}^{3\,
\alpha\,t}}{t}^{3\,n}-c{\lambda}^{2}{{\rm e}^{3\,\alpha\,t}}{t}^{3\,n}
-{t}^{3\,n{\lambda}^{3}}{{\rm e}^{3\,\alpha\,{\lambda}^{3}t}}-c{
{\rm e}^{3\,\alpha\,t}}{t}^{3\,n} }}.
\end{equation}

\subsection*{\large{Numerical results}}
In this section we, numerically, solve equations of above sections and
obtain potential and field, behavior of $G(t)$, deceleration
parameter $q$, and total equation of state which is given by,
\begin{equation}\label{41}
\omega_{tot}=\frac{P_{m}+P_{\Lambda} }{\rho_{m}+\rho_{\Lambda}},
\end{equation}
which, for the case of $P_{m}=0$, reduced to the following form
\begin{equation}\label{42}
\omega_{tot}=\frac{P_{\Lambda} }{\rho_{m}+\rho_{\Lambda}}.
\end{equation}
Reader should remember that $\omega_{tot} \neq
\omega_{1}+\omega_{2}.$ By using the scale factor given by the
equation (21), the Hubble parameter $H$ reduced to the following
relation,
\begin{equation}\label{eq:nHubble1}
H=\frac{n}{t}+\alpha,
\end{equation}
and the cosmological constant (22) takes the following form,
\begin{equation}\label{eq:nlambda1}
\Lambda(t)=A(a_{0}\exp{[\alpha t]}t^{n})^{-k}+\frac{(n+\alpha t)^{2}}{t^{2}}.
\end{equation}
Bellow, graphically we present behavior of $G$, $q$ and
$\omega_{tot}$. All parameters are fixed in order to obtain $V
\rightarrow 0$, when $t \rightarrow  \infty$.\\
First of all we consider non interacting case and draw $G(t)$,
$\omega_{tot}(t)$ and $q(t)$ for some fixed parameters in the
figures 1, 2 and 3 respectively. Then, in the figures 4, 5 and 6 we
obtain behavior of these quantities in presence of interaction term
given by the equation (23).\\
In the next step we draw $\phi$, $V$ and $P_{\Lambda}$ in terms of
time for the case of non interacting in the figures 7, 8 and 9
respectively. Then, extension to the case of interacting illustrated
in the figures 10, 11, 12, 13 and 14, where $\rho_{m}$ and
$\rho_{\Lambda}$ are also analyzed.\\
All figures contain four plots with different fixed parameters.
First, in the case of non interacting component, the first plot
(top, left), drawn for $n=3$, $\alpha=0.5$, $k=1.5$ and different
values of $A$. The second plot (top, right) drawn for $n=3$,
$k=1.5$, $A=2.5$ and different values of $\alpha$. The third plot
(bottom, left), drawn for $n=3$, $\alpha=0.5$, $A=2.5$ and different
values of $k$. Finally, the forth plot (bottom, right) drawn for
$k=1.5$, $\alpha=0.5$, $A=0.5$ and different values of $n$.\\
On the other hand, in the case of interacting components, the first
plot (top, left), drawn for $A=1$, $\alpha=1.1$, $\gamma=0.05$,
$k=0.5$, $b=0.04$ and different values of $n$. The second plot (top,
right) drawn for $n=2.5$, $k=0.5$, $A=1$, $\gamma=0.05$, $b=0.04$
and different values of $\alpha$. The third plot (bottom, left),
drawn for $n=1.5$, $\alpha=2.5$, $A=1.5$, $k=1.5$, $\gamma=0.05$ and
different values of $b$. Finally, the forth plot (bottom, right)
drawn for $k=1.5$, $\alpha=2.5$, $A=1.5$, $n=1.5$, $b=0.04$ and
different values of $\gamma$.\\
In the Fig. 15 we study statefinder parameters graphically. Also, in
the Fig. 16 stability of theory investigated.\\
In the next section
we give discussion about these figures and effects of parameters on the
cosmological quantities.

\begin{figure}[h]
 \begin{center}$
 \begin{array}{cccc}
\includegraphics[width=48 mm]{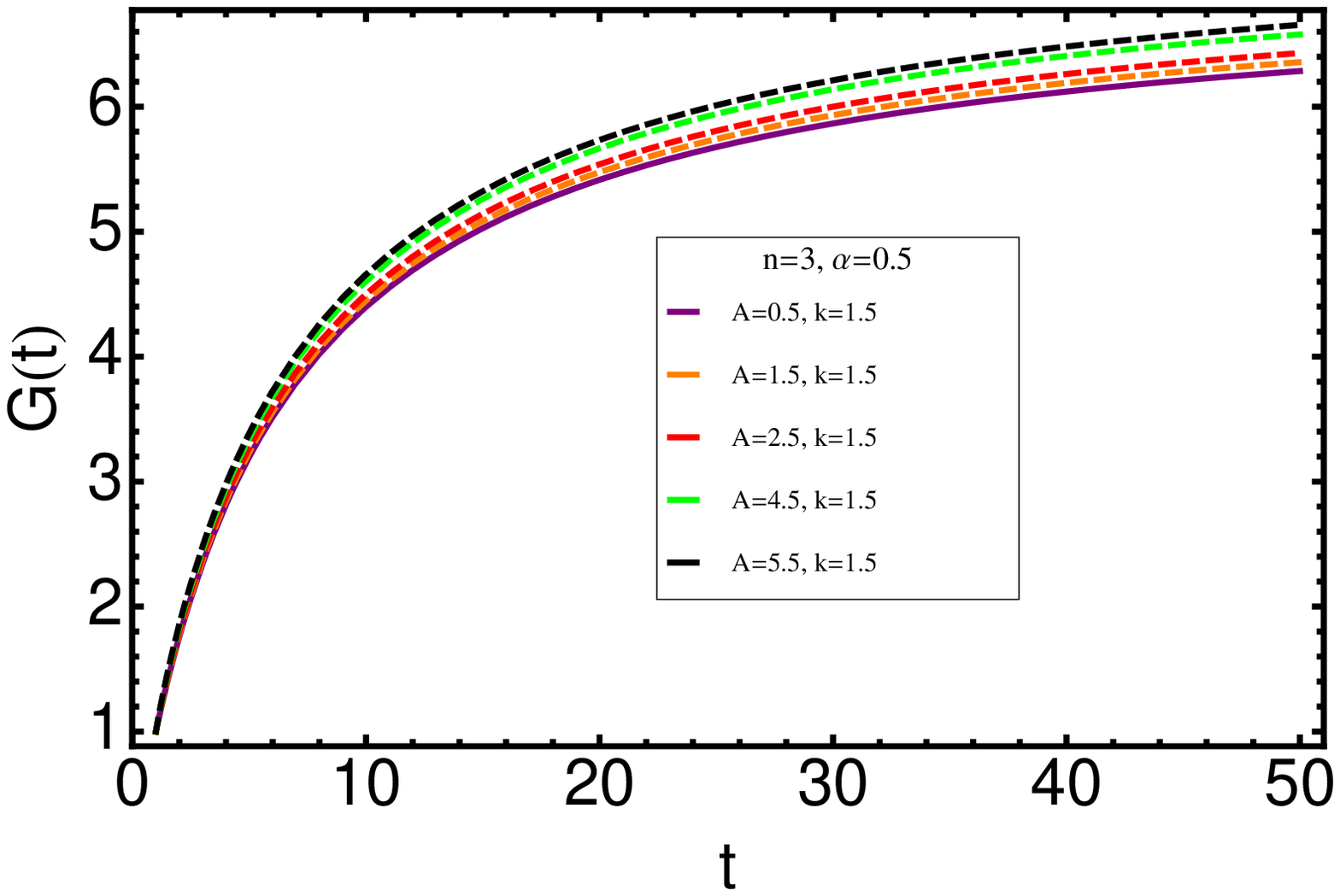} &
\includegraphics[width=48 mm]{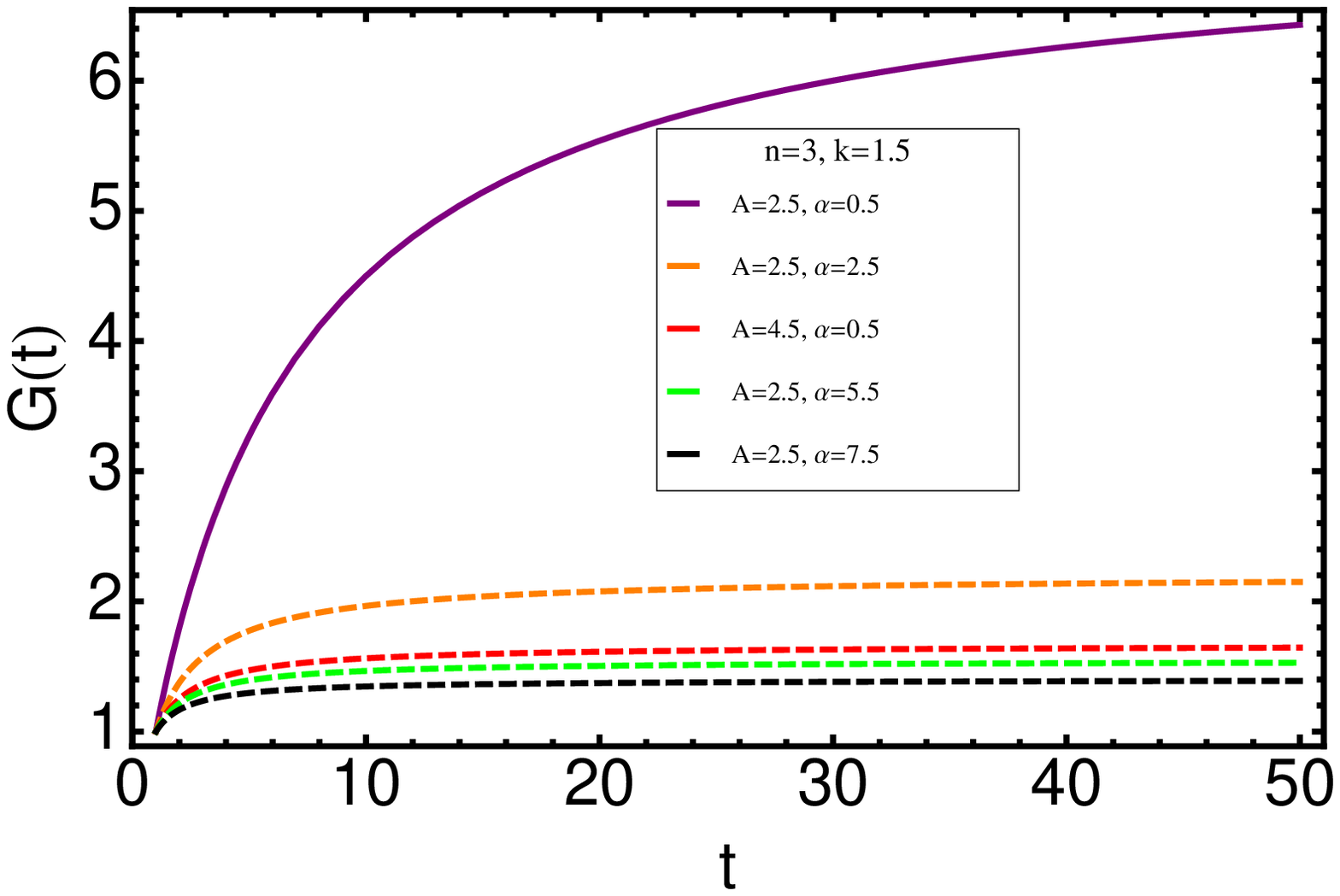}\\
\includegraphics[width=48 mm]{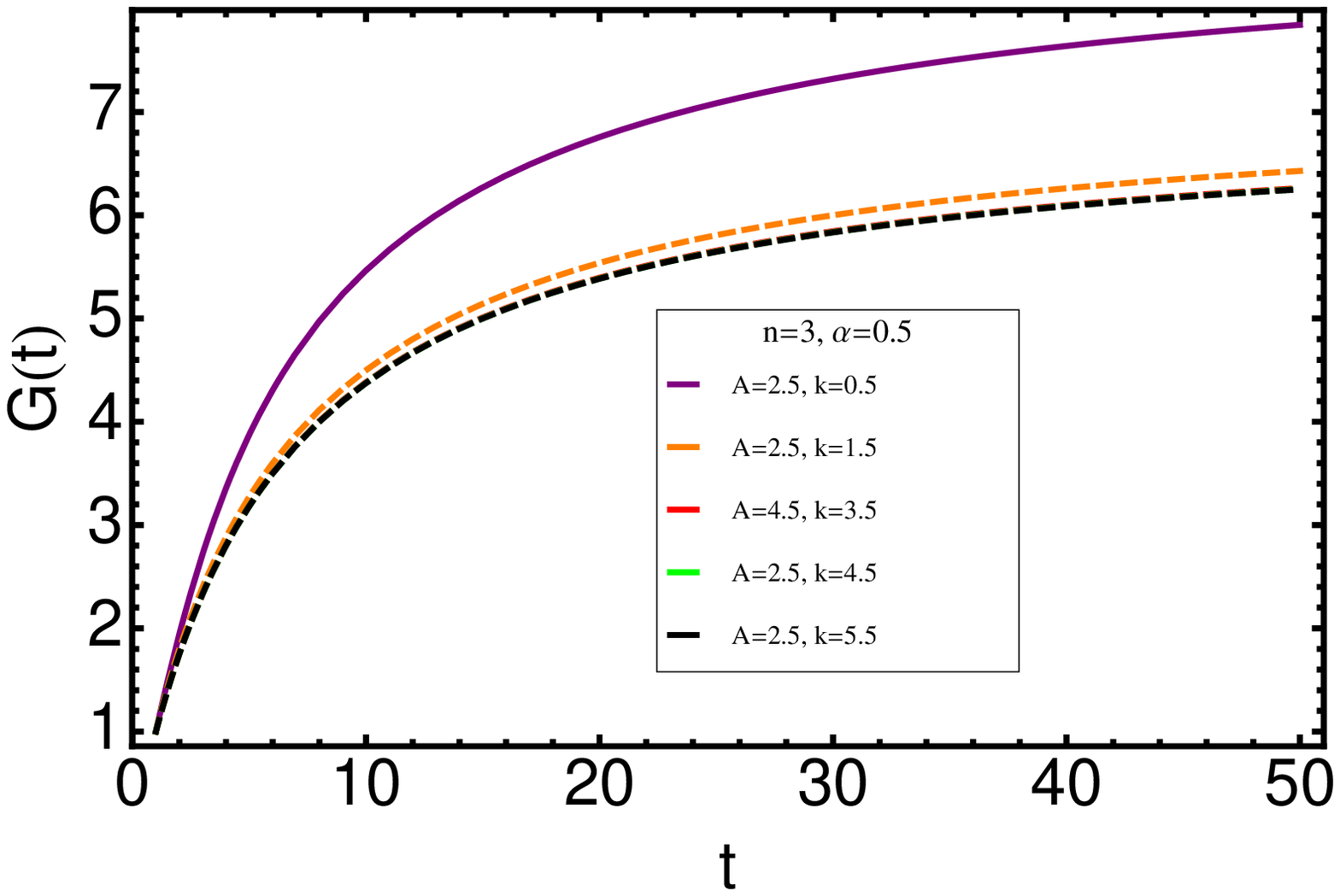} &
\includegraphics[width=48 mm]{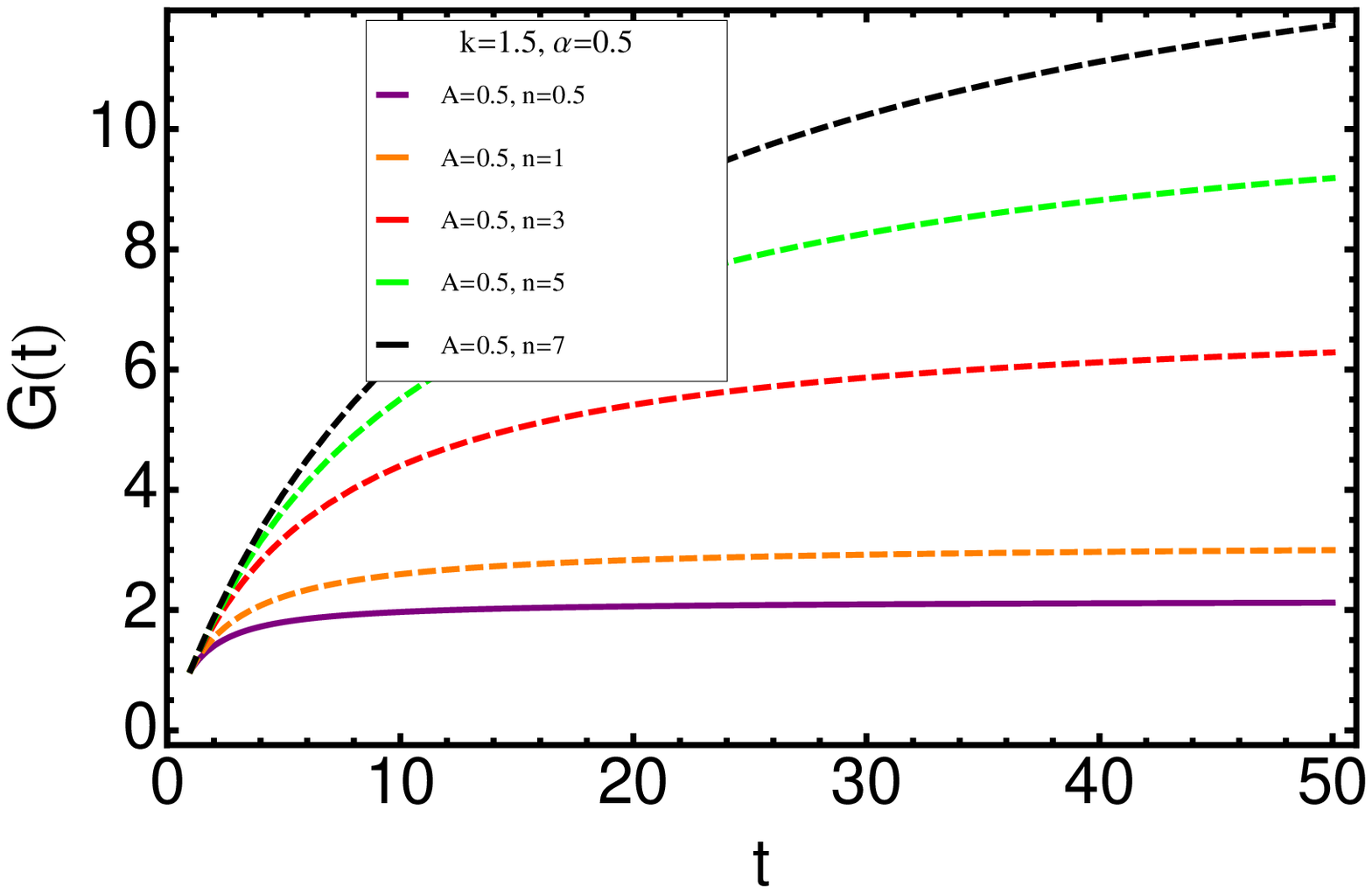}
 \end{array}$
 \end{center}
\caption{Behavior of $G$ against $t$ for non interacting components
where we choose $a_{0}=2$ and $\rho_{0}=1$.}
 \label{fig:1}
\end{figure}

\begin{figure}[h]
 \begin{center}$
 \begin{array}{cccc}
\includegraphics[width=48 mm]{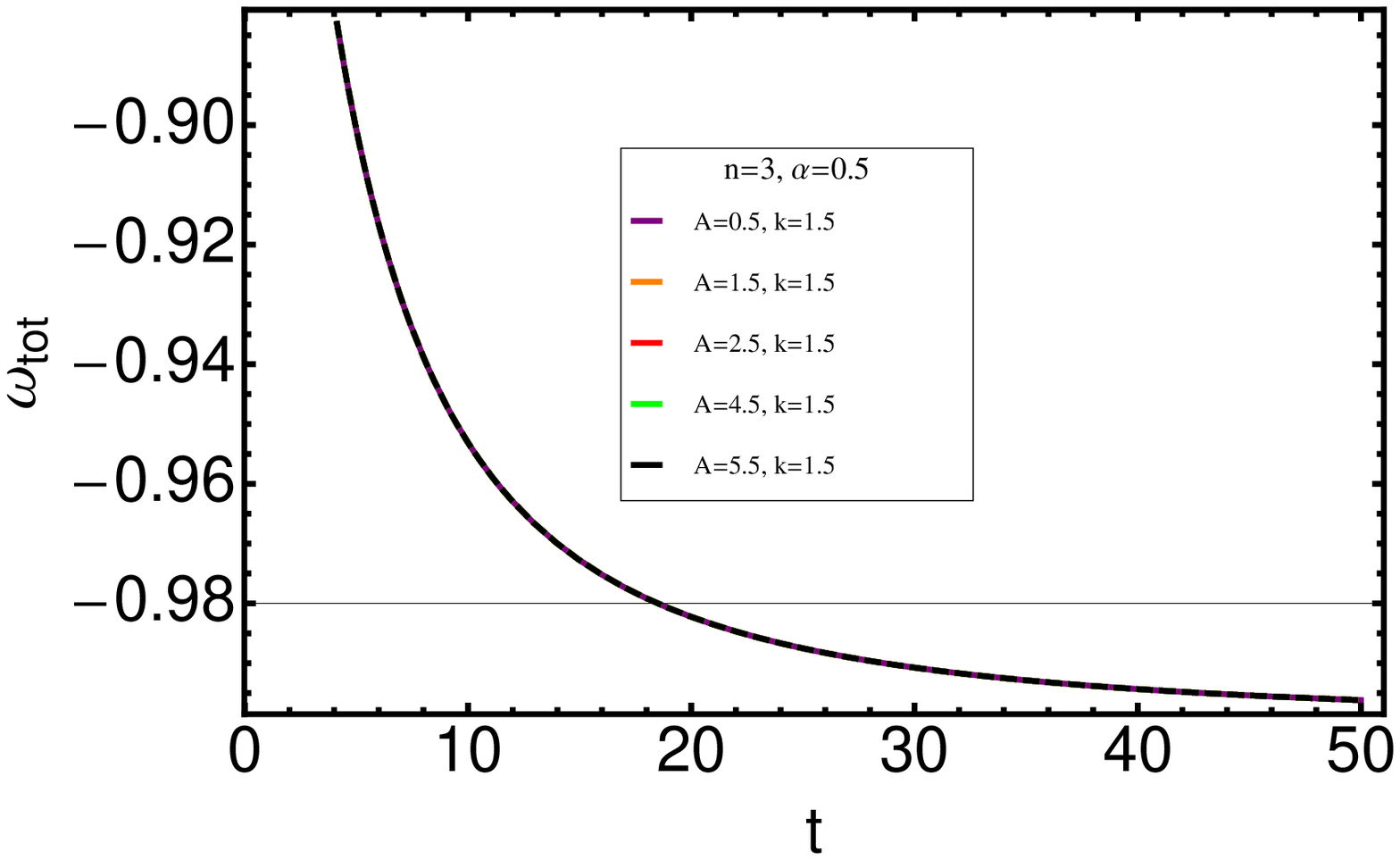} &
\includegraphics[width=48 mm]{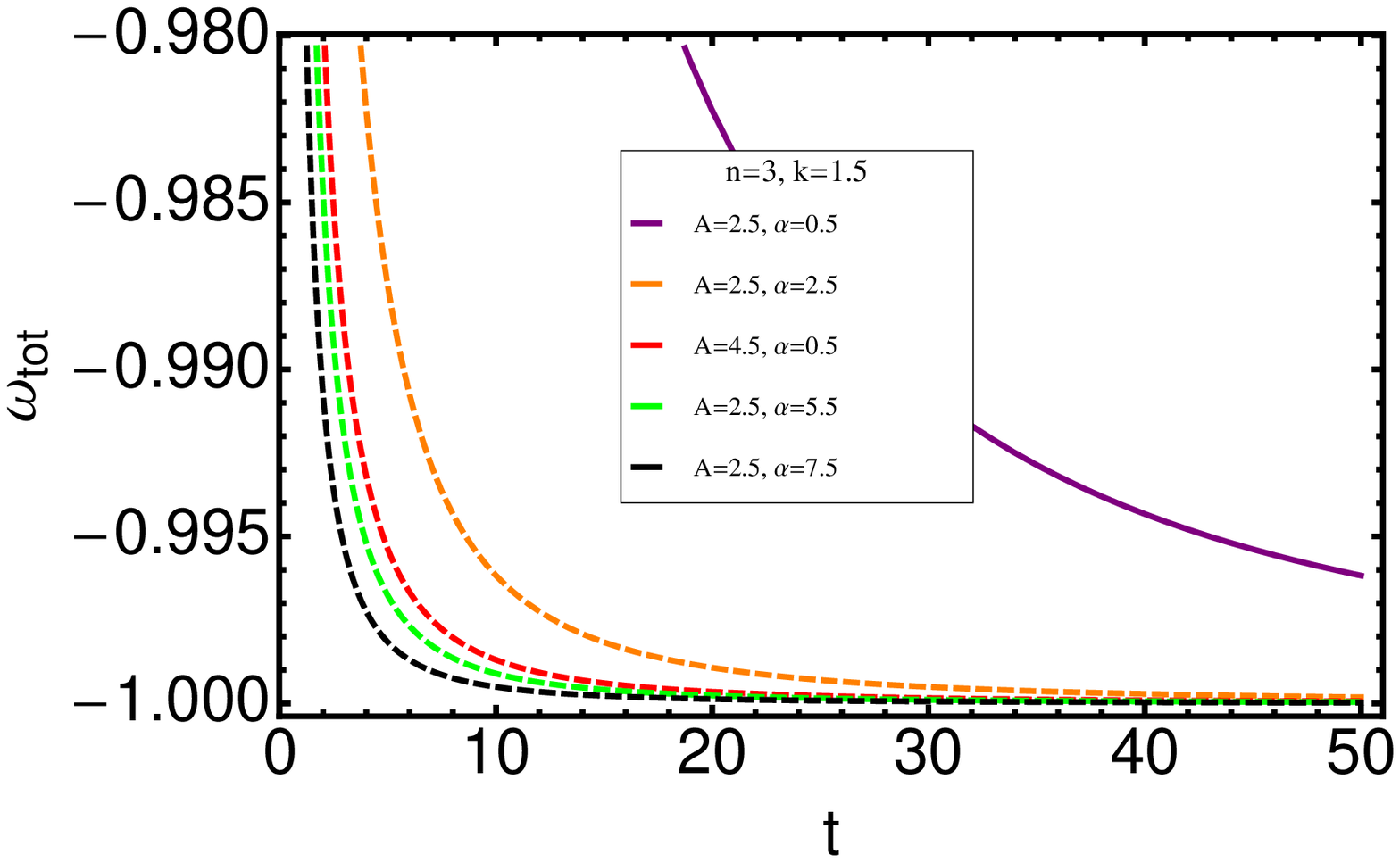}\\
\includegraphics[width=48 mm]{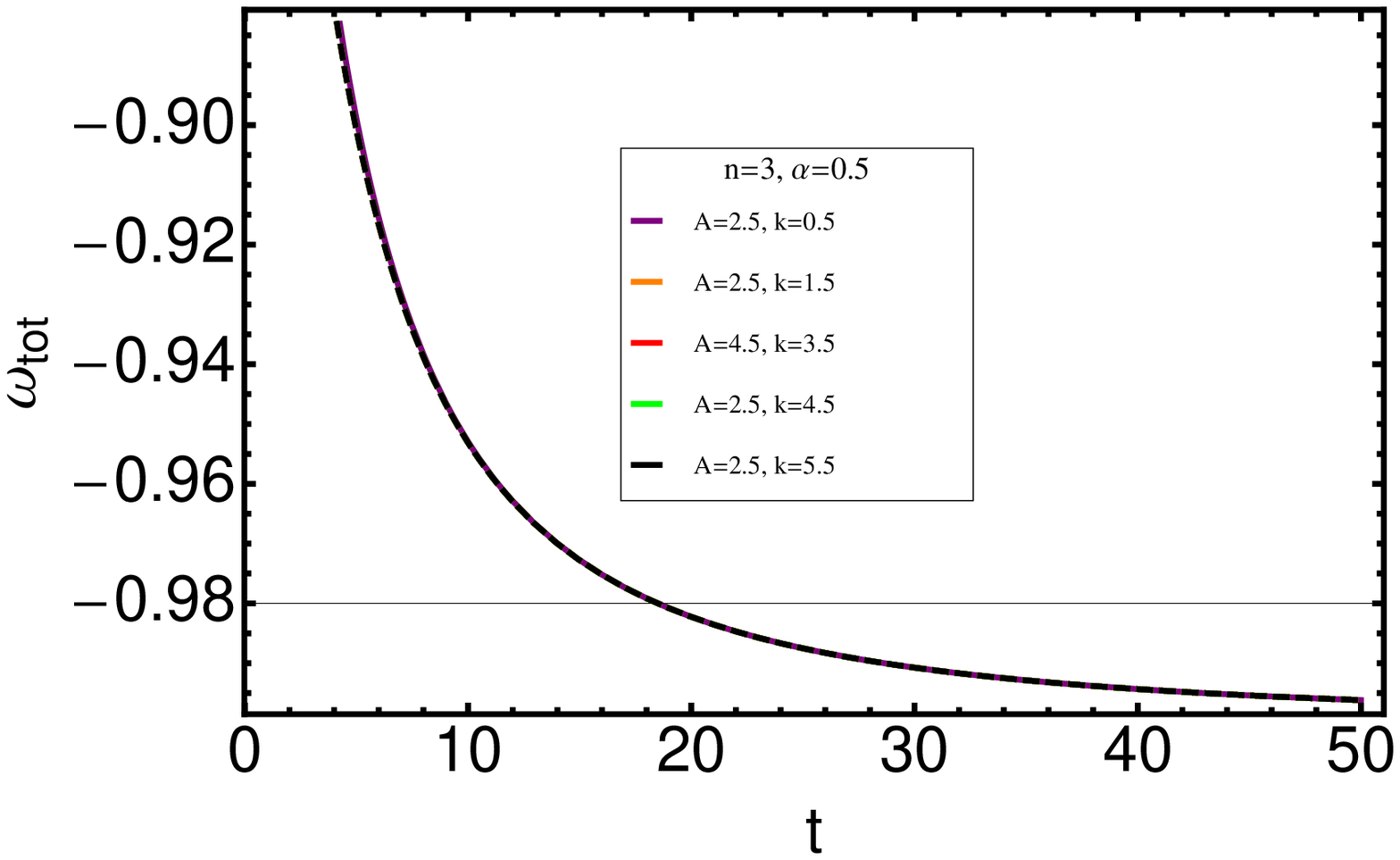} &
\includegraphics[width=48 mm]{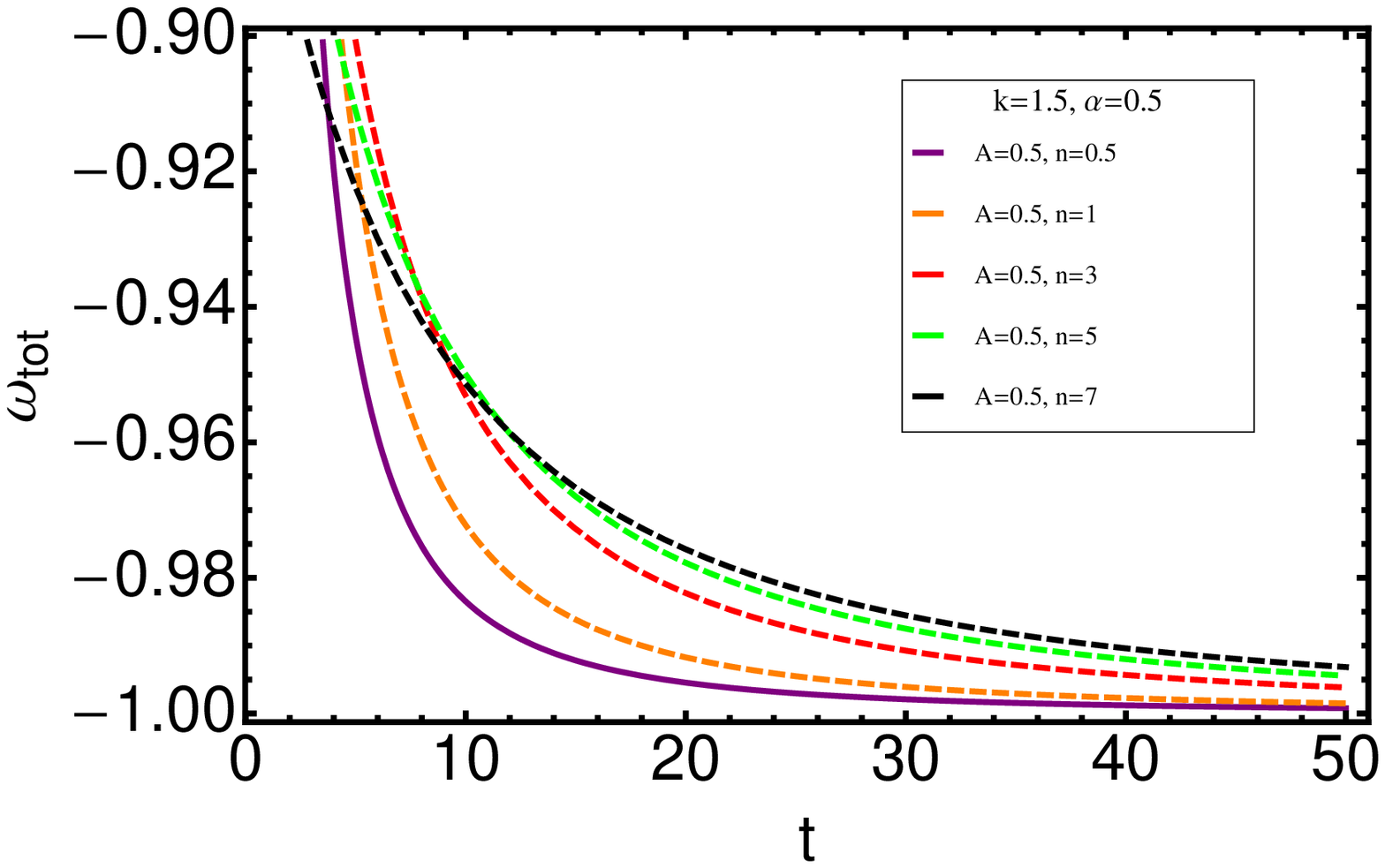}
 \end{array}$
 \end{center}
\caption{Behavior of $\omega_{tot}$ against $t$ for non interacting
components where we choose  $a_{0}=2$ and $\rho_{0}=1$.}
 \label{fig:2}
\end{figure}

\begin{figure}[h]
 \begin{center}$
 \begin{array}{cccc}
\includegraphics[width=48 mm]{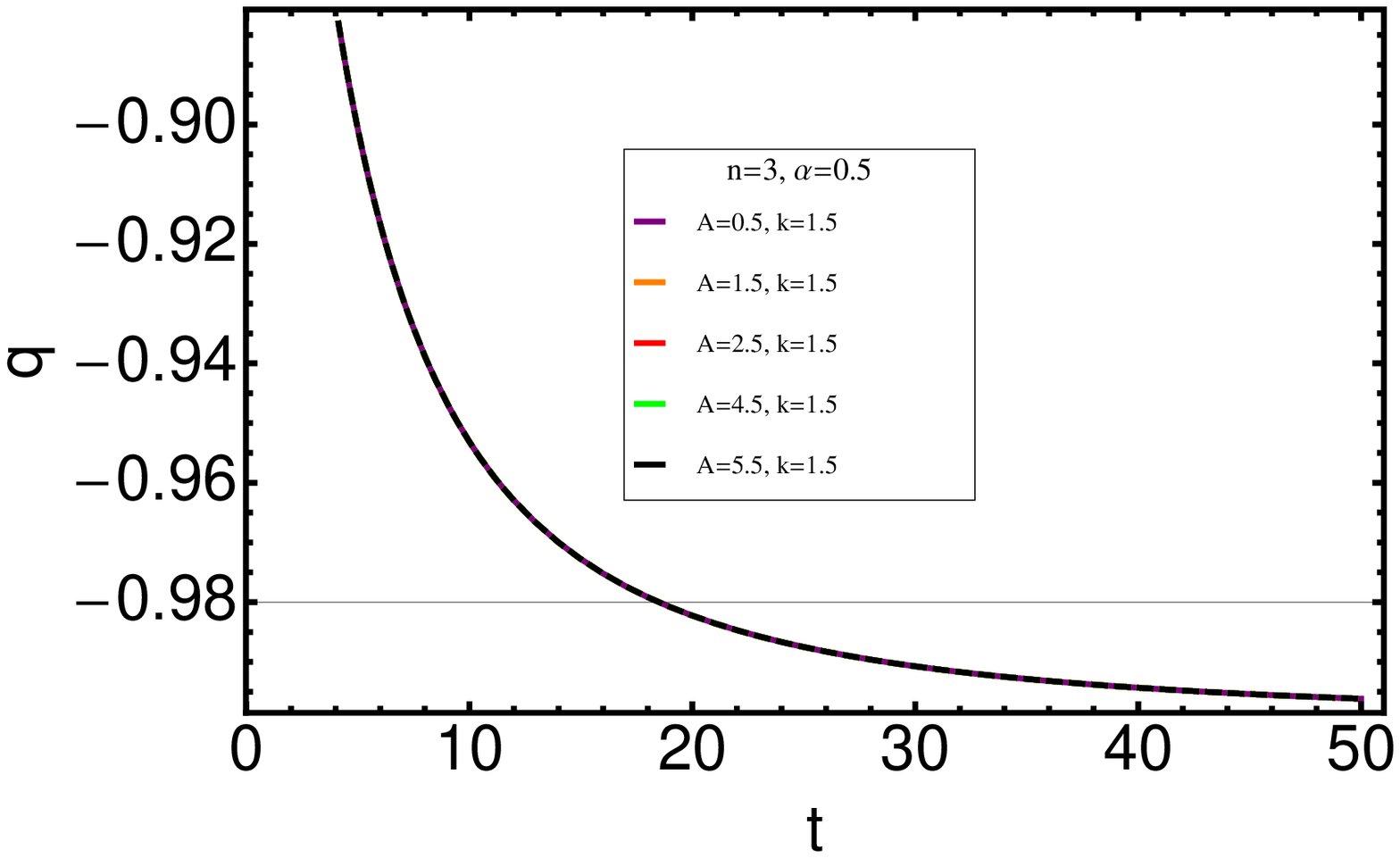} &
\includegraphics[width=48 mm]{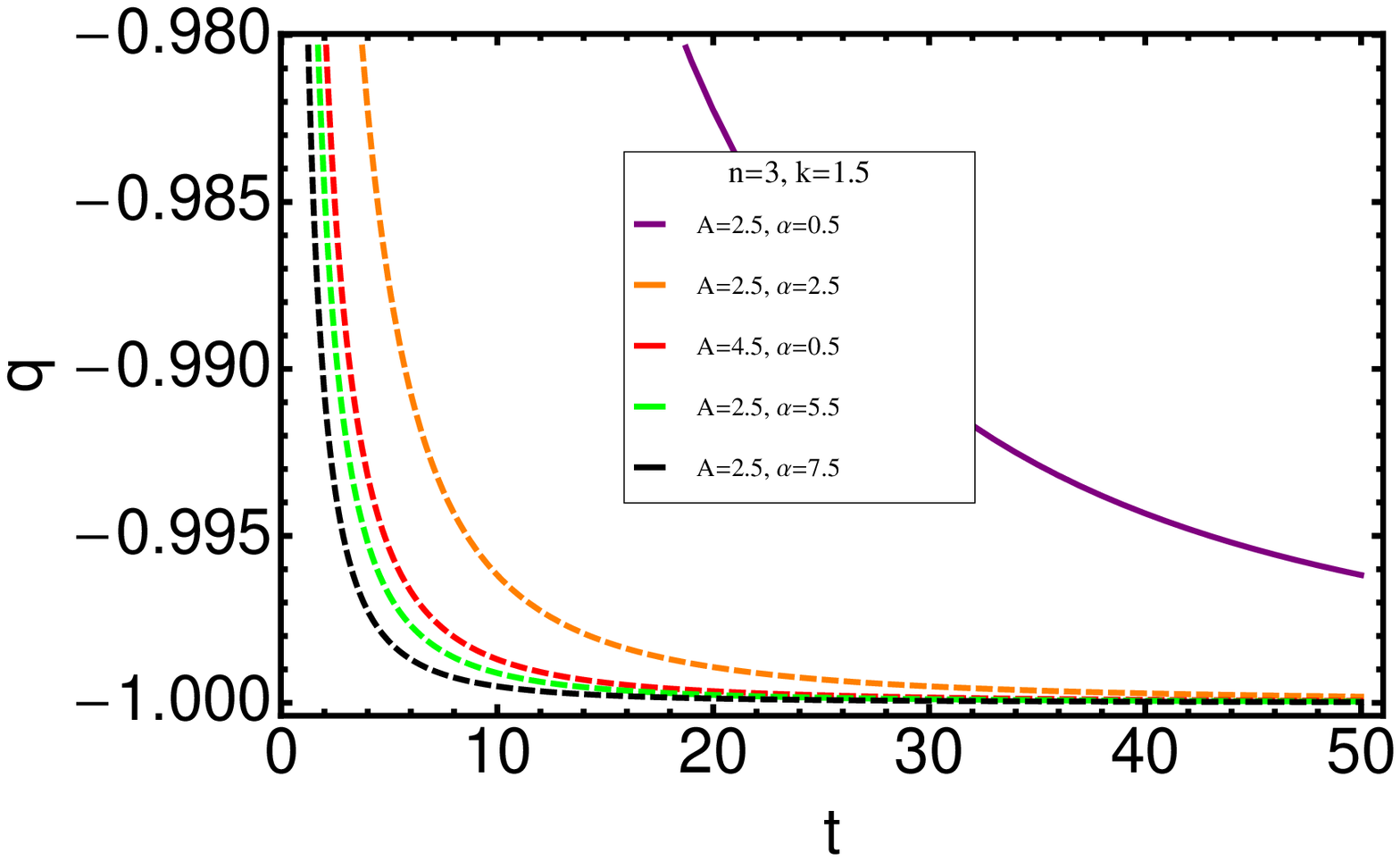}\\
\includegraphics[width=48 mm]{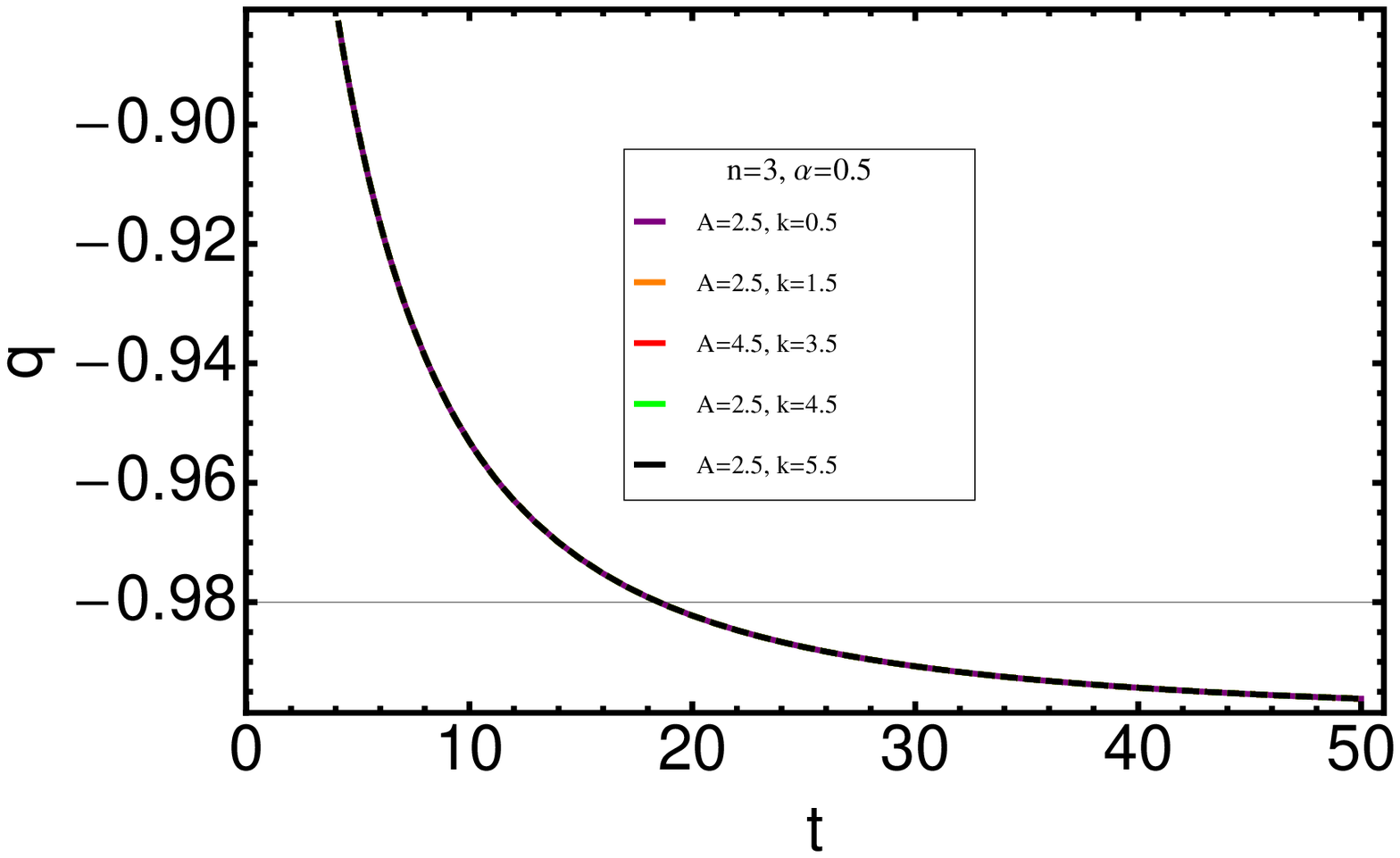} &
\includegraphics[width=48 mm]{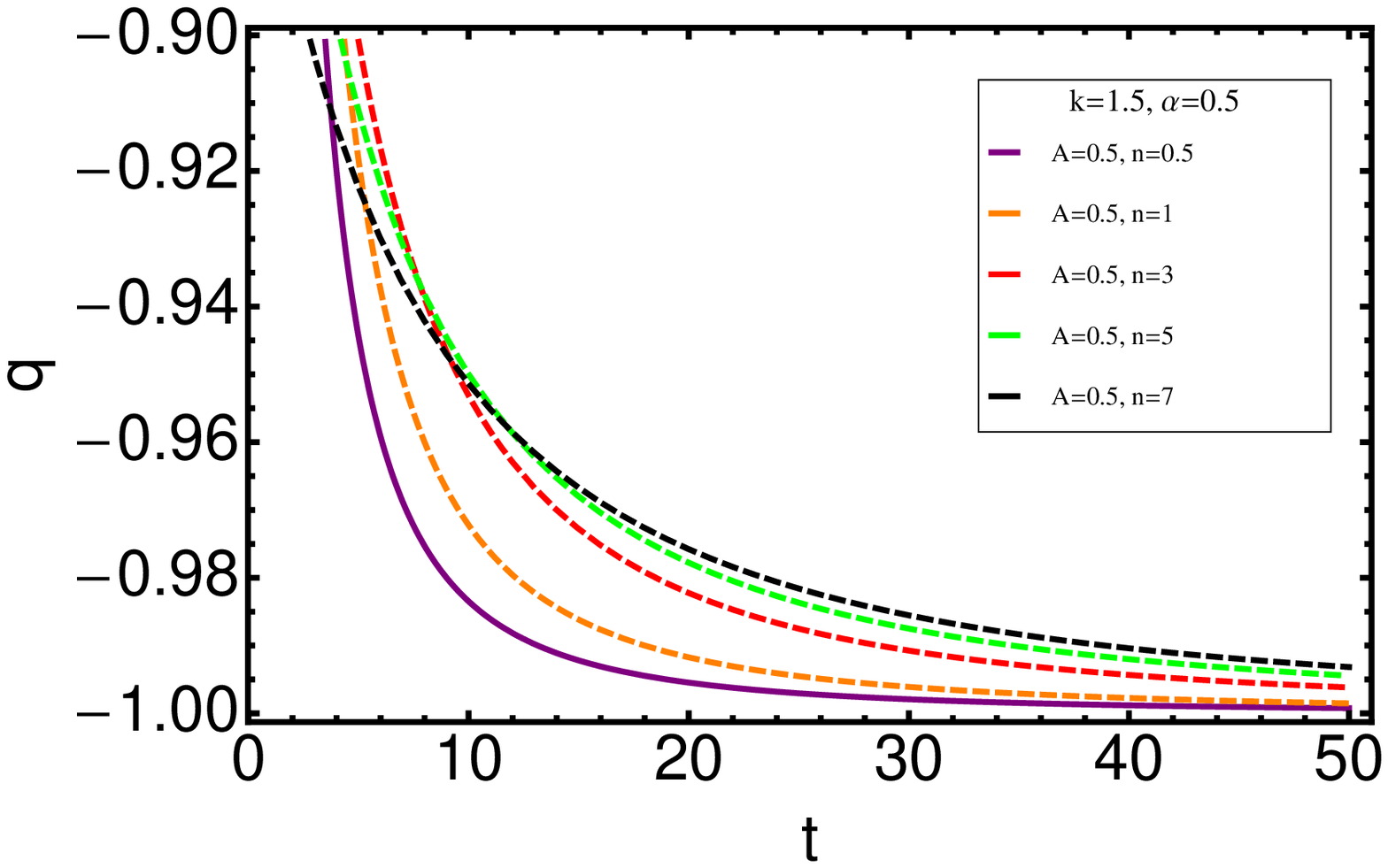}
 \end{array}$
 \end{center}
\caption{Behavior of $q$ against $t$ for non interacting components
where we choose  $a_{0}=2$ and $\rho_{0}=1$.}
 \label{fig:3}
\end{figure}

\begin{figure}[h]
 \begin{center}$
 \begin{array}{cccc}
\includegraphics[width=48 mm]{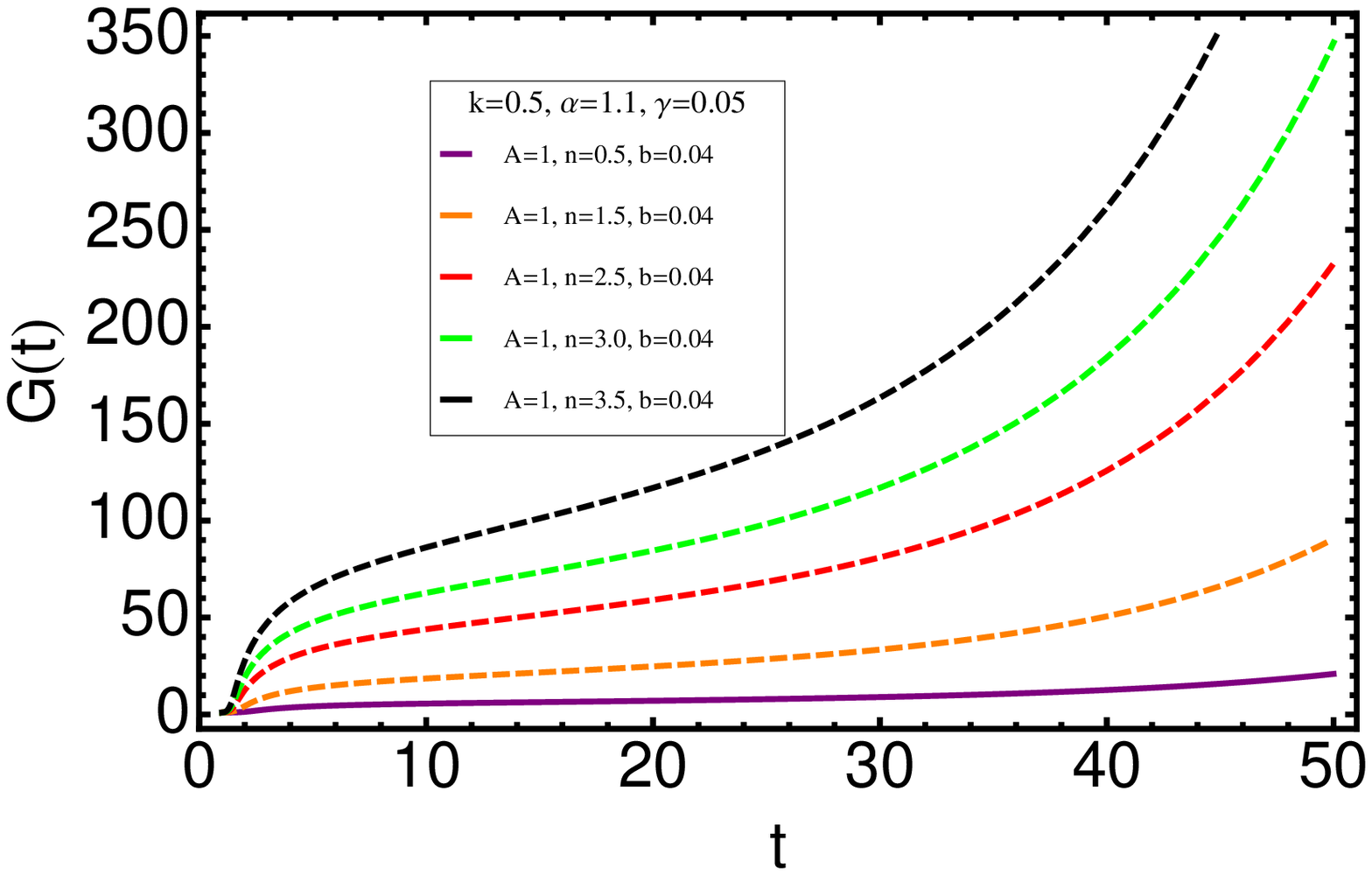} &
\includegraphics[width=48 mm]{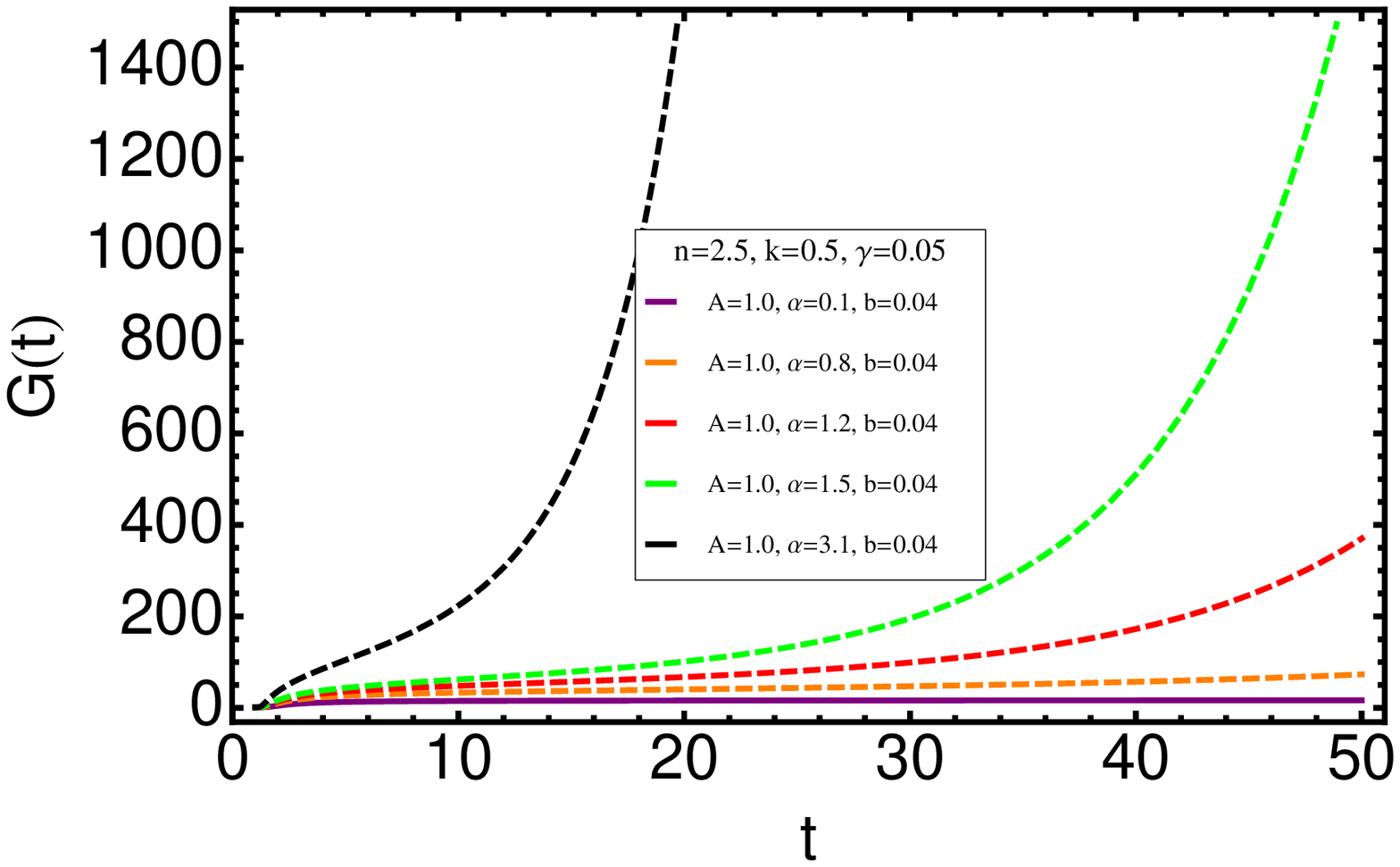}\\
\includegraphics[width=48 mm]{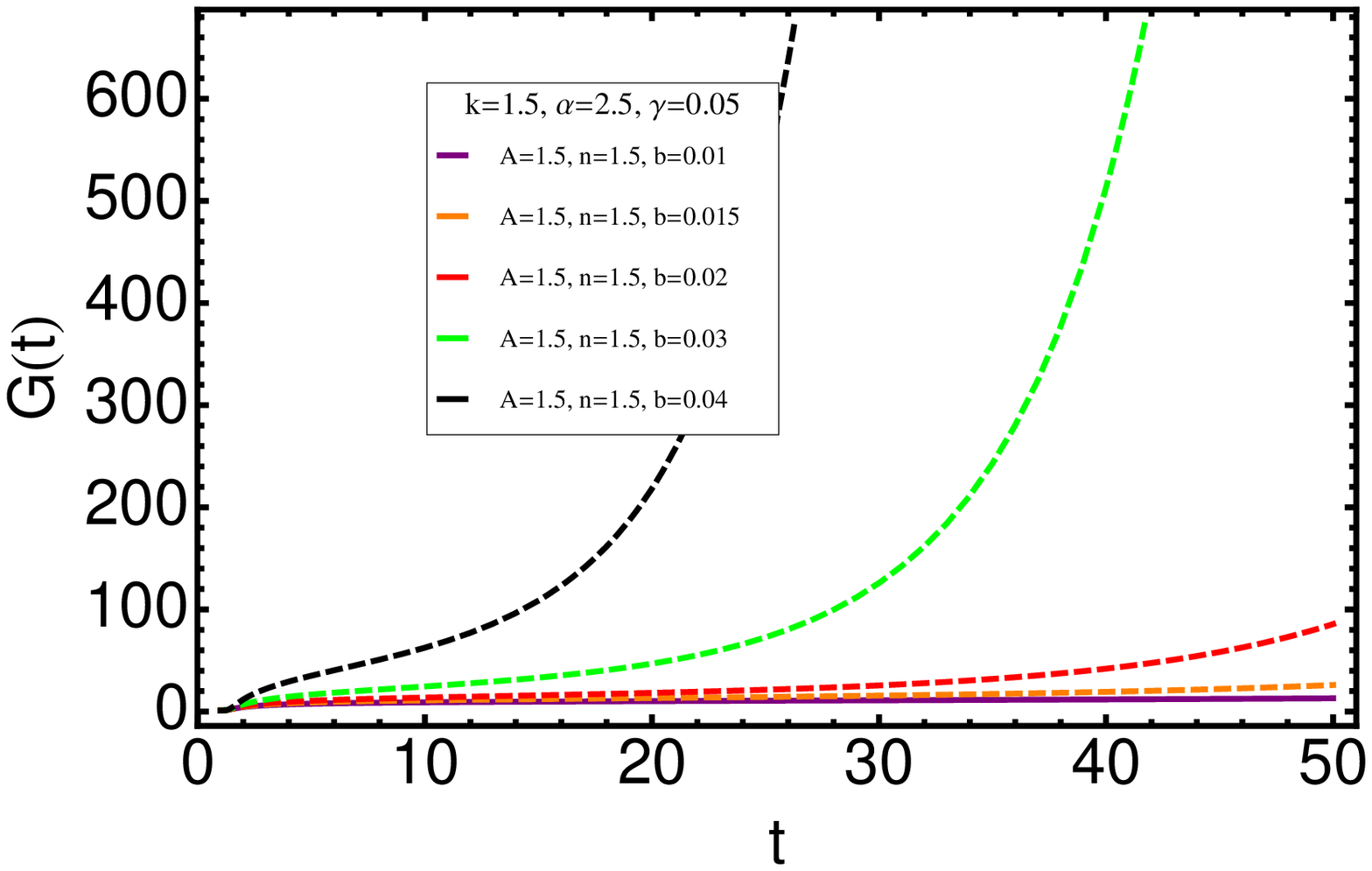} &
\includegraphics[width=48 mm]{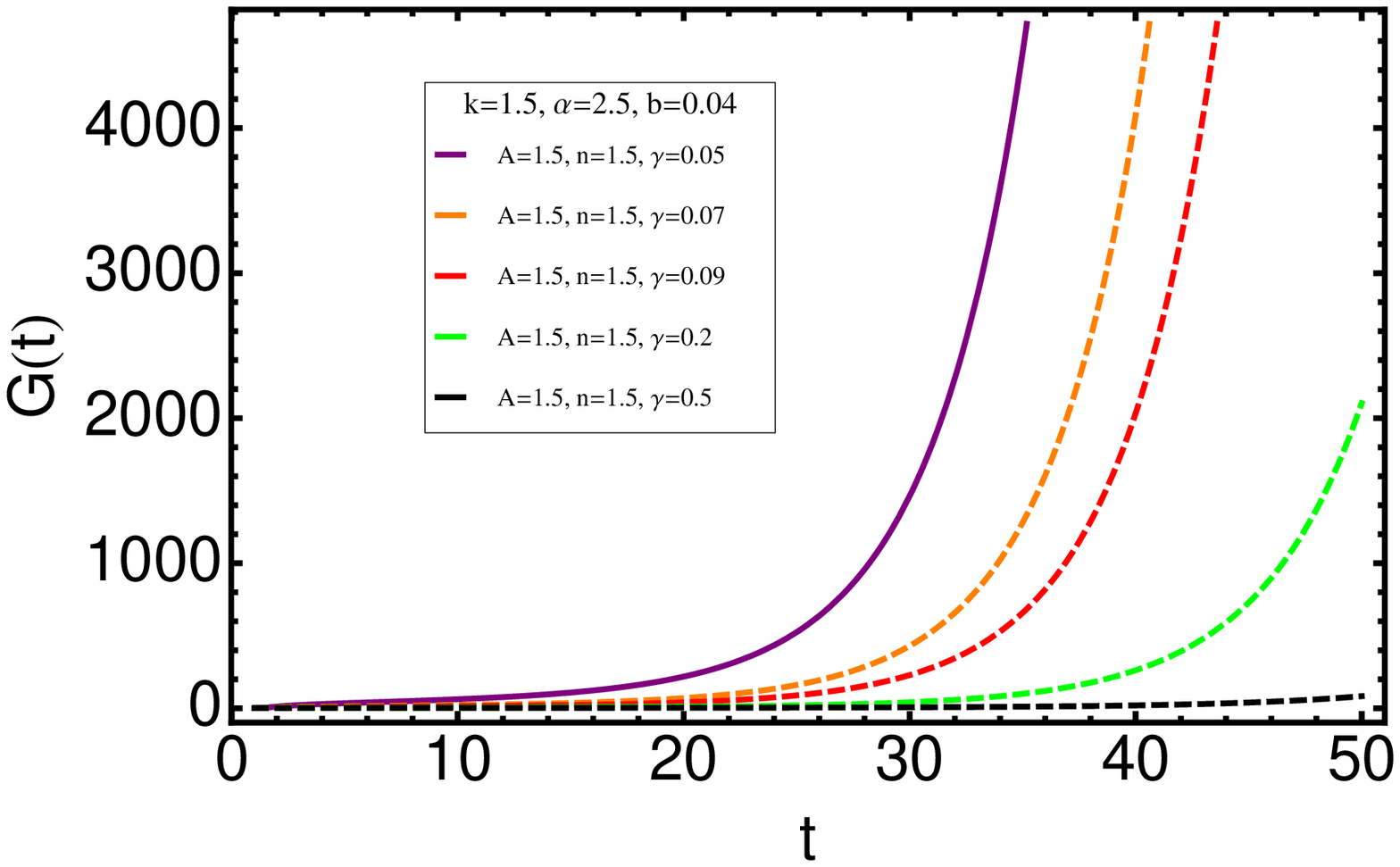}
 \end{array}$
 \end{center}
\caption{Behavior of $G(t)$ against $t$ for interacting components
where we choose $a_{0}=1$.}
 \label{fig:4}
\end{figure}

\begin{figure}[h]
 \begin{center}$
 \begin{array}{cccc}
\includegraphics[width=48 mm]{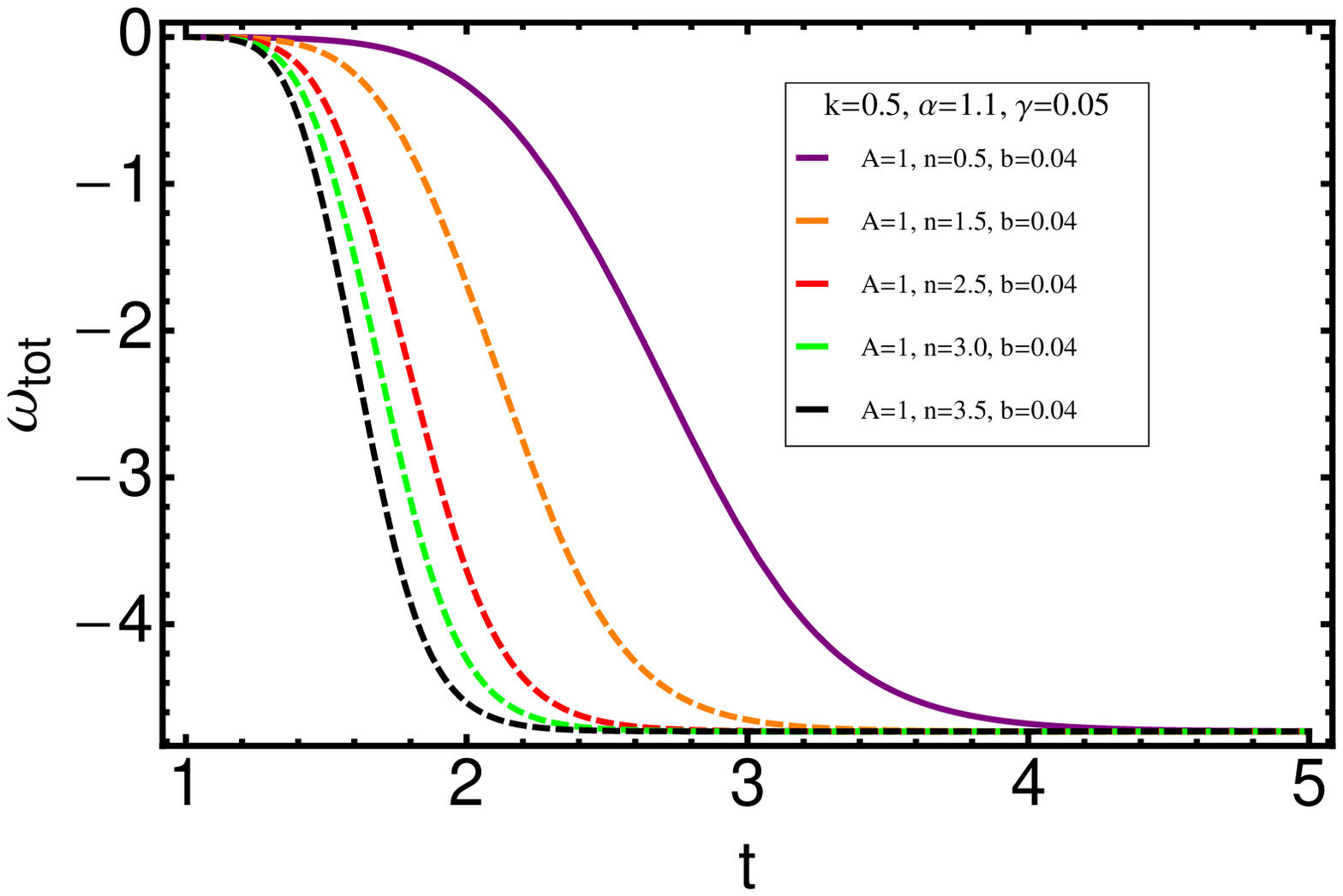} &
\includegraphics[width=48 mm]{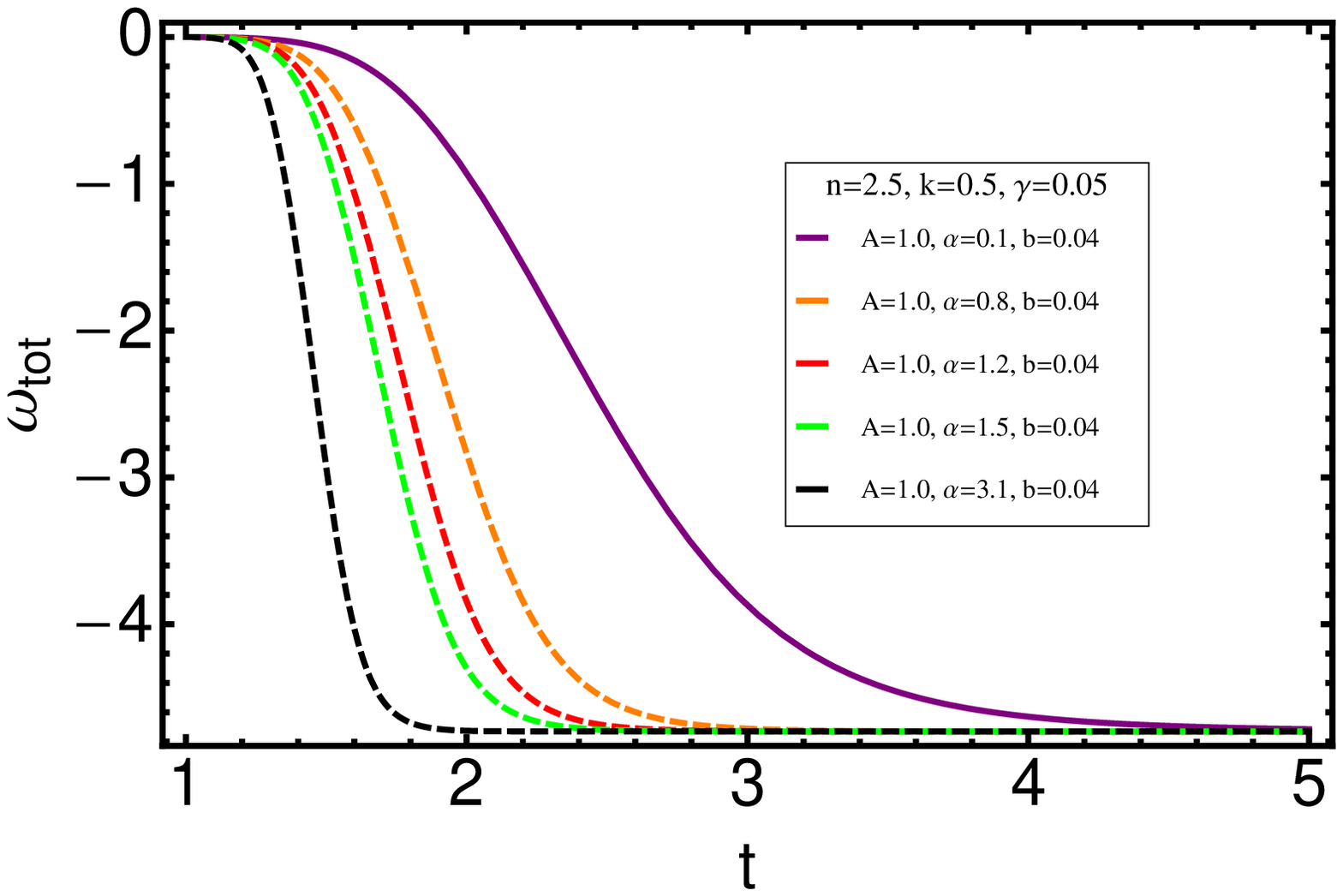}\\
\includegraphics[width=48 mm]{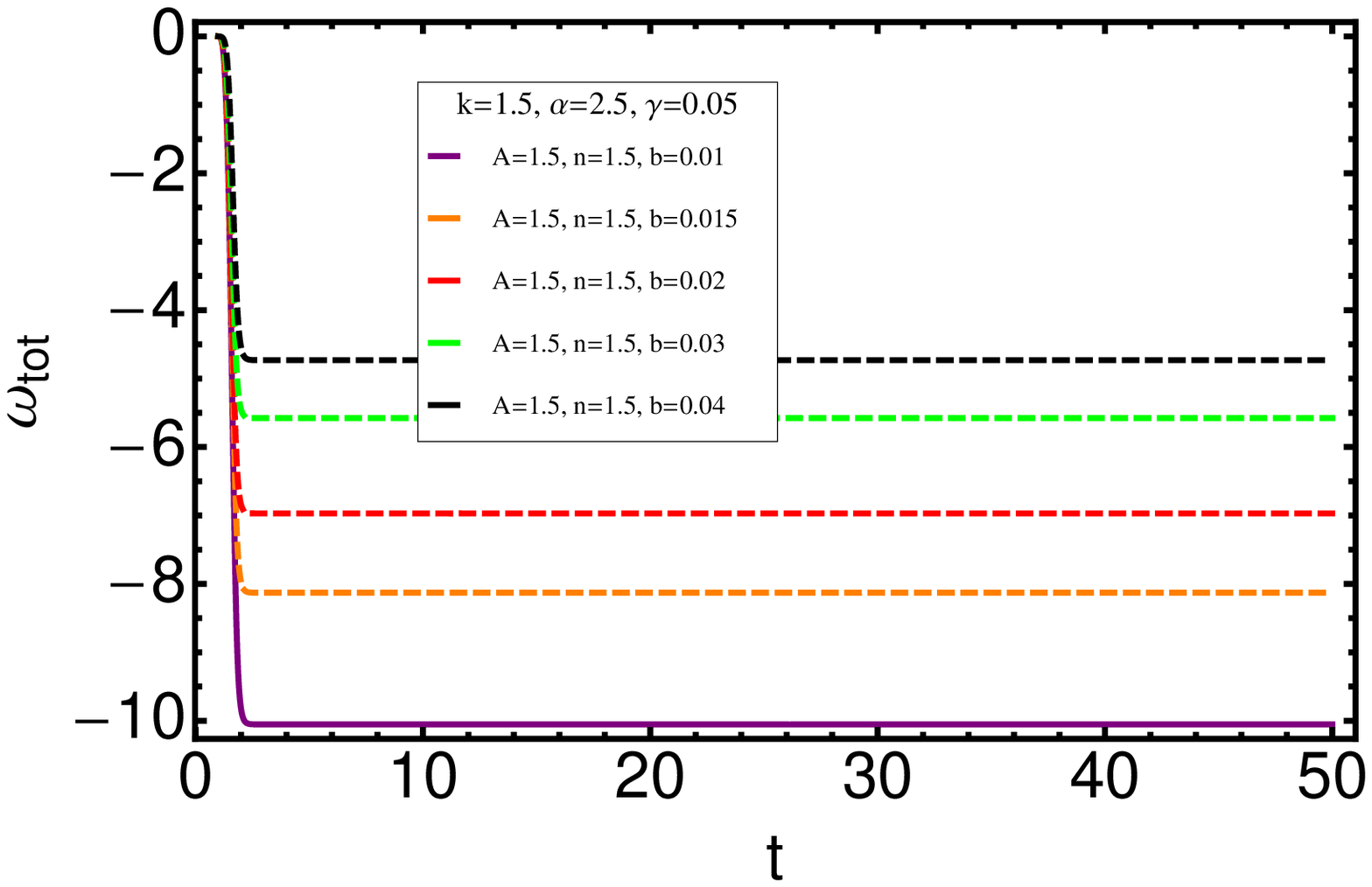} &
\includegraphics[width=48 mm]{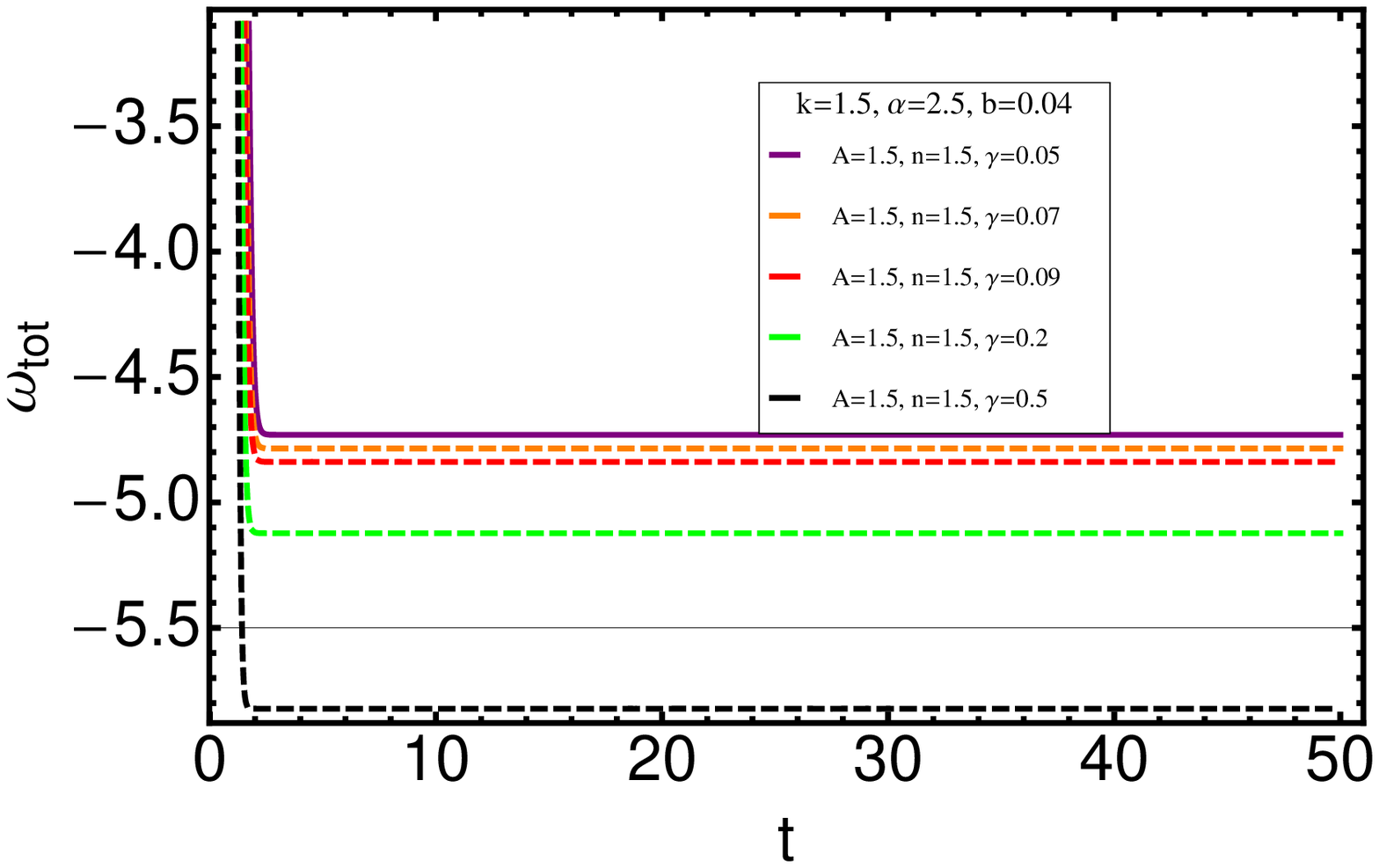}
 \end{array}$
 \end{center}
\caption{Behavior of $\omega_{tot}$ against $t$ for interacting
components where we choose $a_{0}=1$.}
 \label{fig:5}
\end{figure}

\begin{figure}[h]
 \begin{center}$
 \begin{array}{cccc}
\includegraphics[width=48 mm]{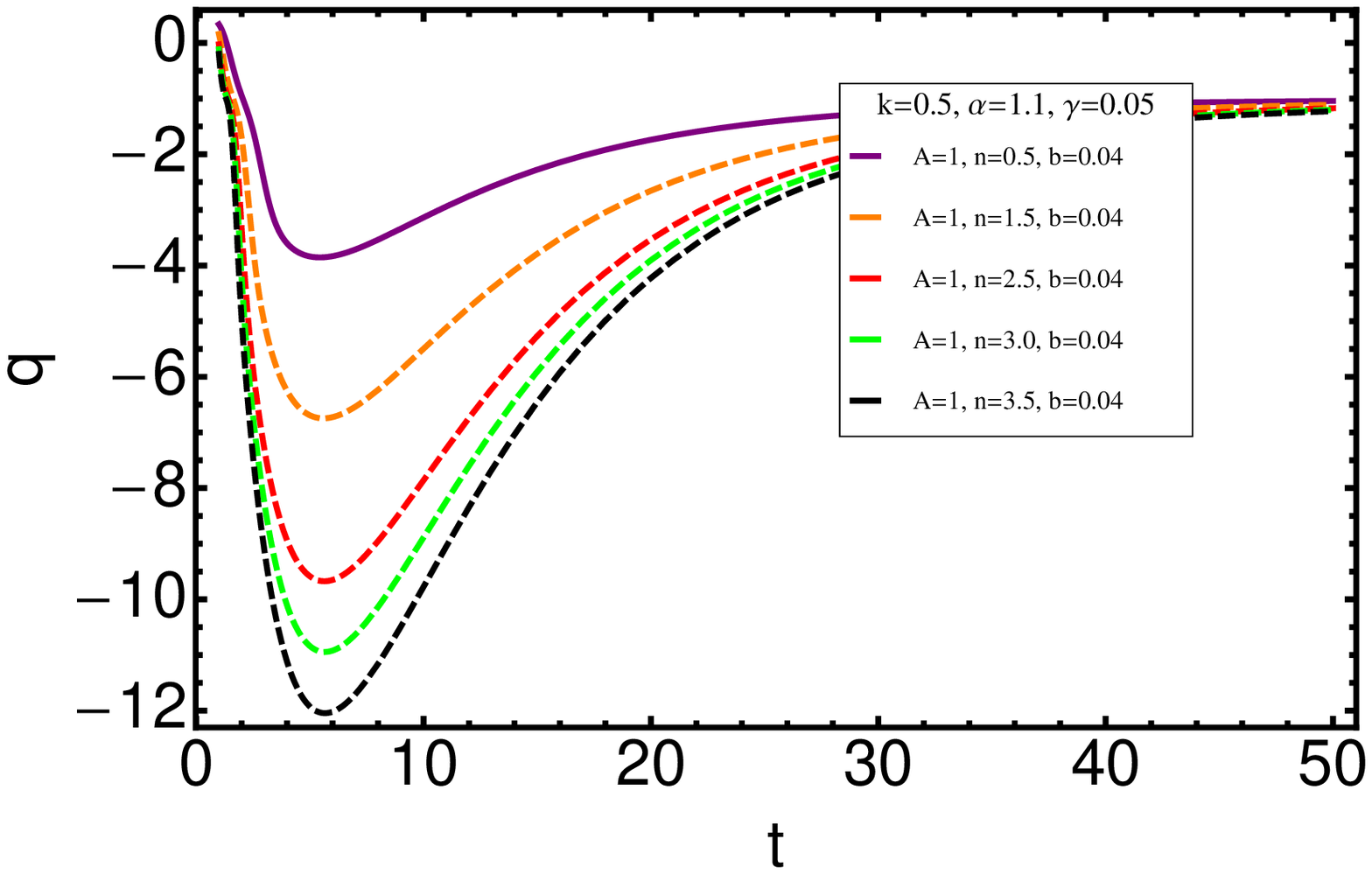} &
\includegraphics[width=48 mm]{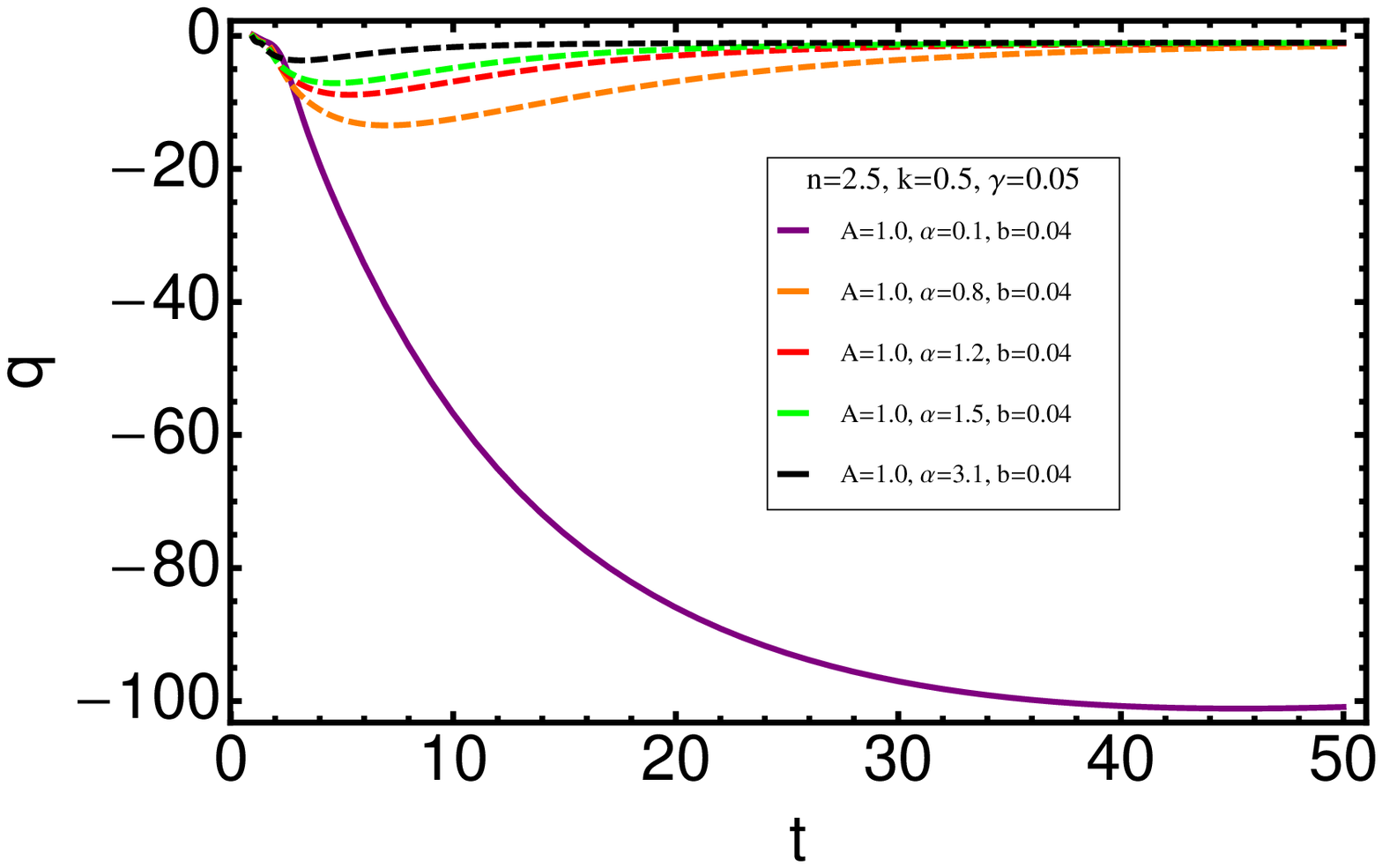}\\
\includegraphics[width=48 mm]{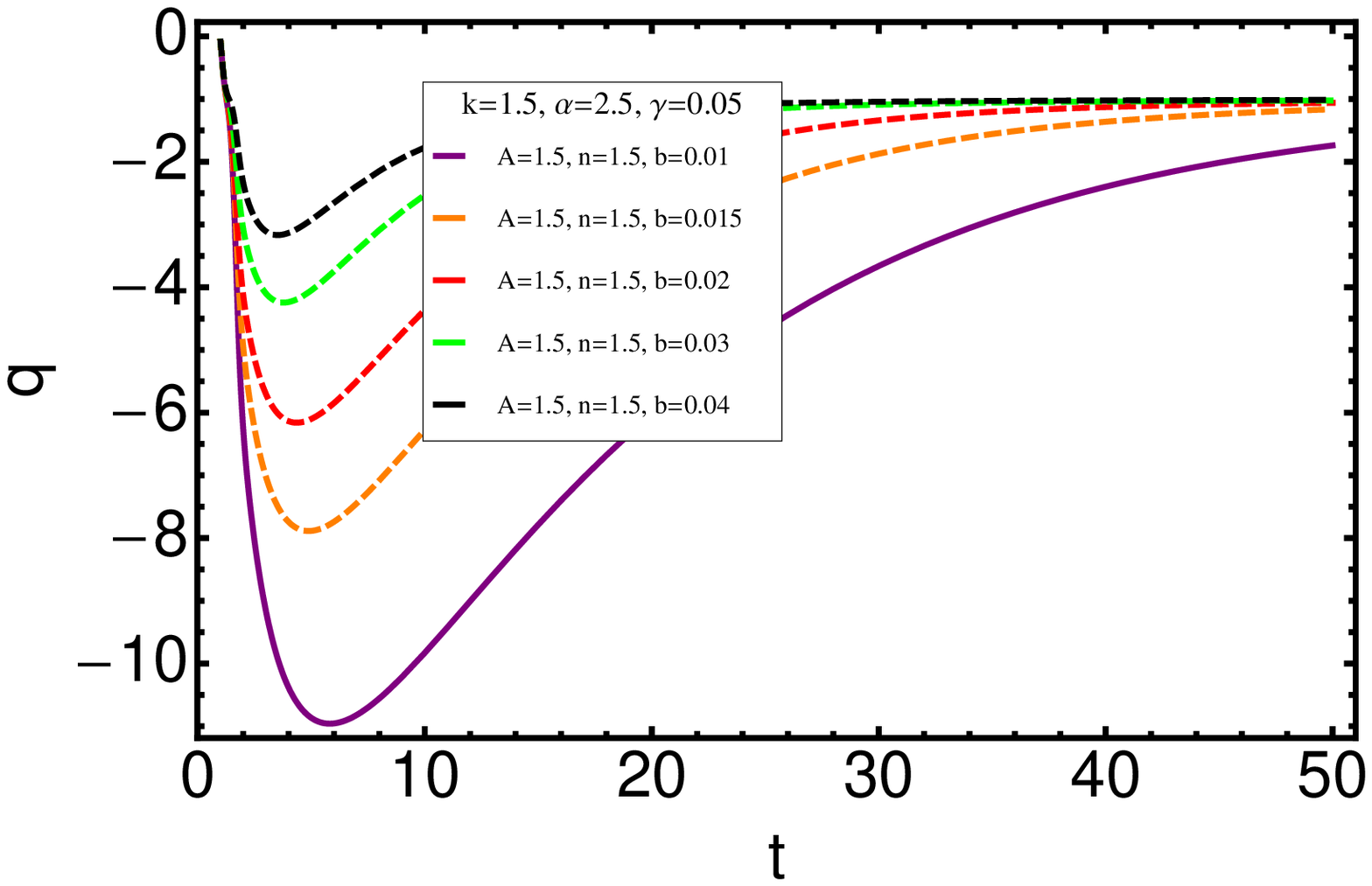} &
\includegraphics[width=48 mm]{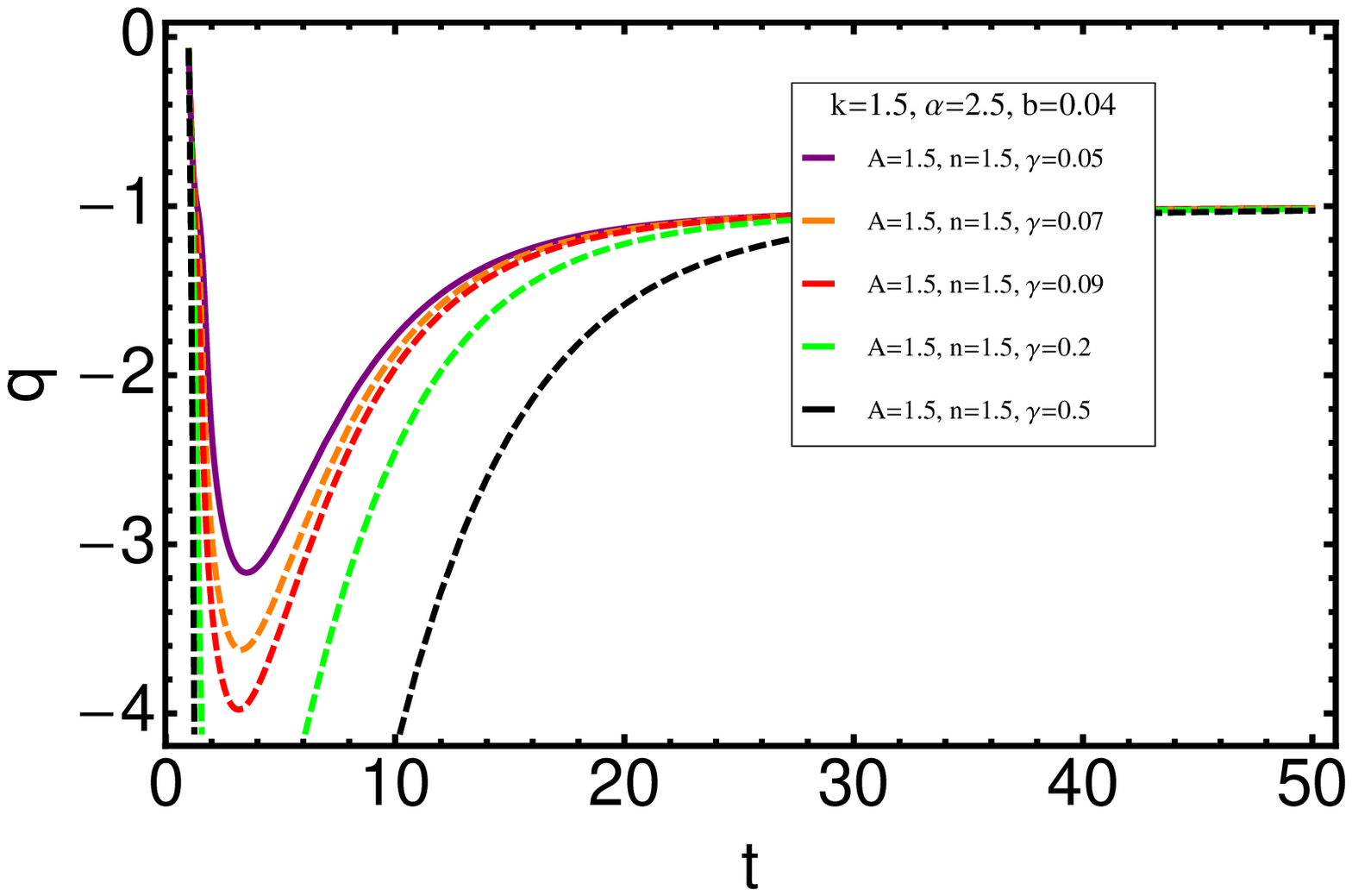}
 \end{array}$
 \end{center}
\caption{Behavior of $q$ against $t$ for interacting components
where we choose $a_{0}=1$.}
 \label{fig:6}
\end{figure}

\begin{figure}[h]
 \begin{center}$
 \begin{array}{cccc}
\includegraphics[width=48 mm]{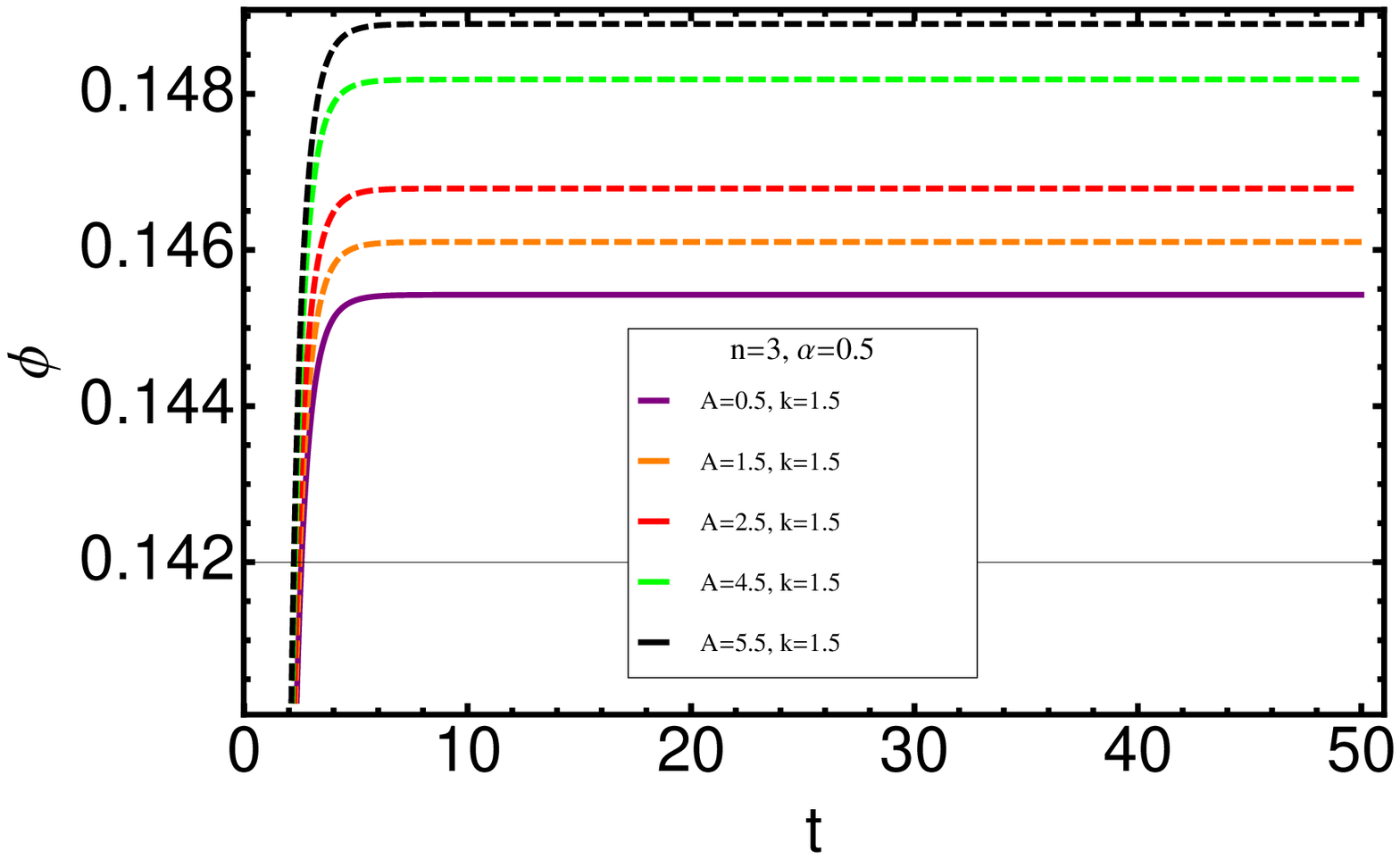} &
\includegraphics[width=48 mm]{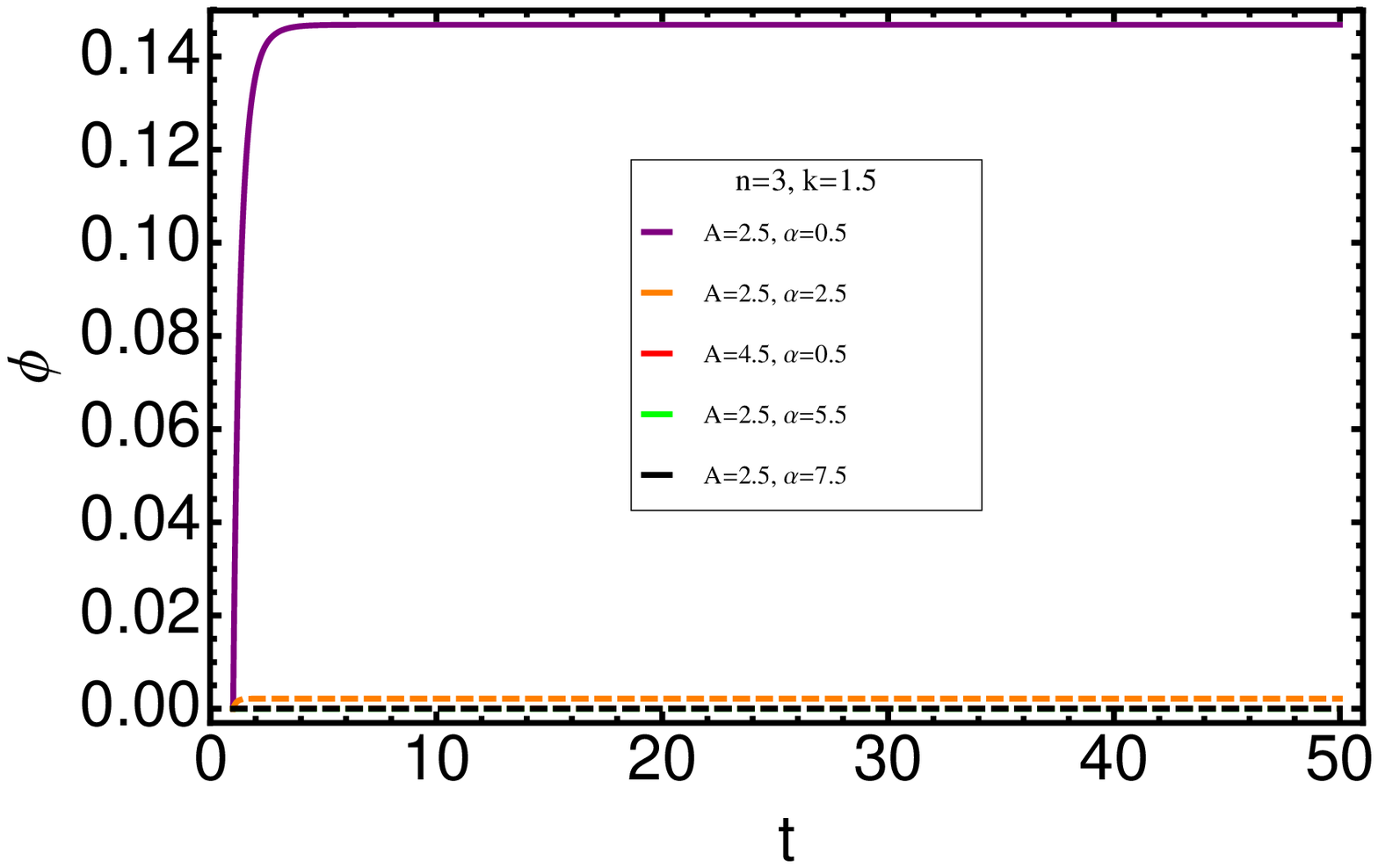}\\
\includegraphics[width=48 mm]{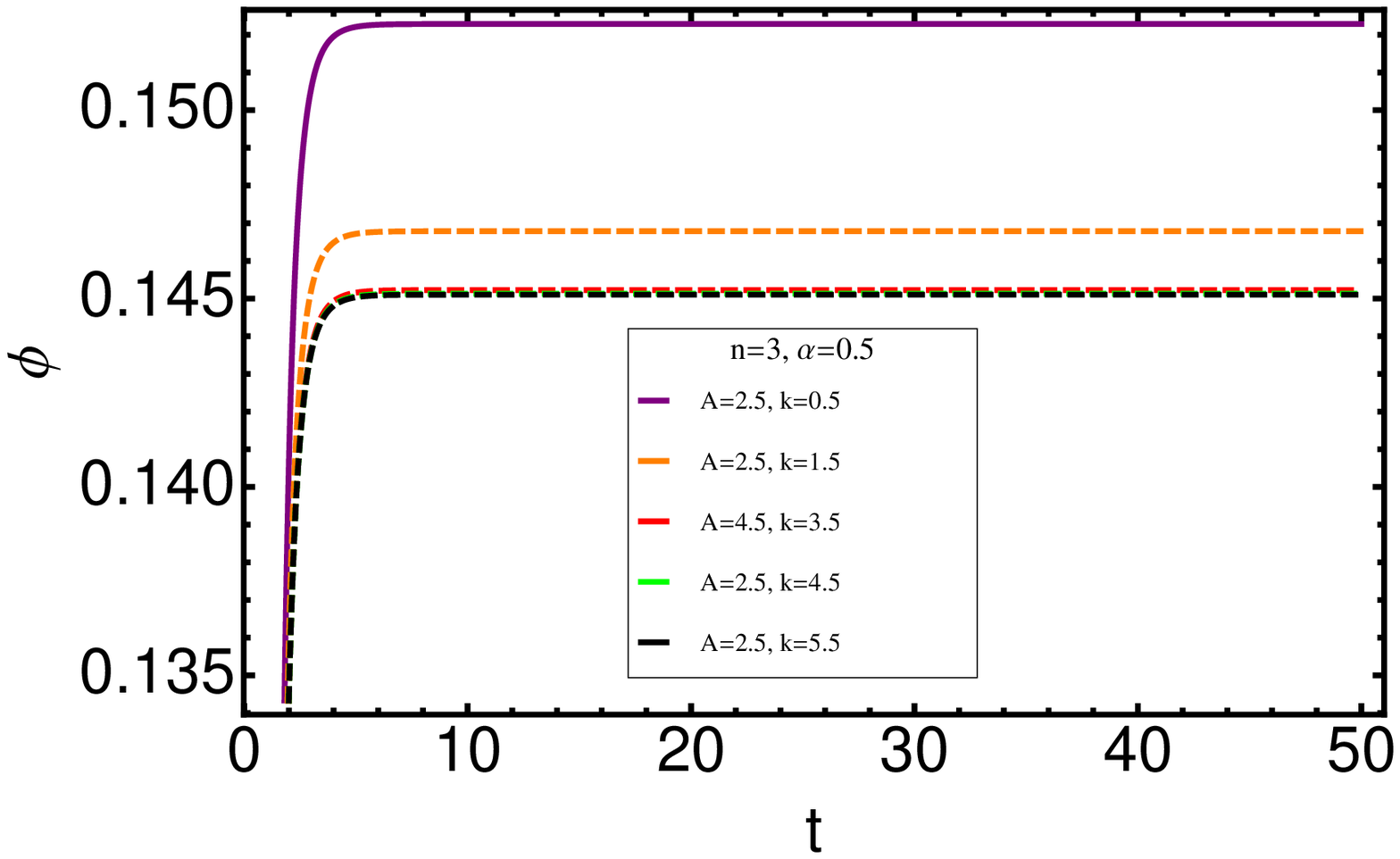} &
\includegraphics[width=48 mm]{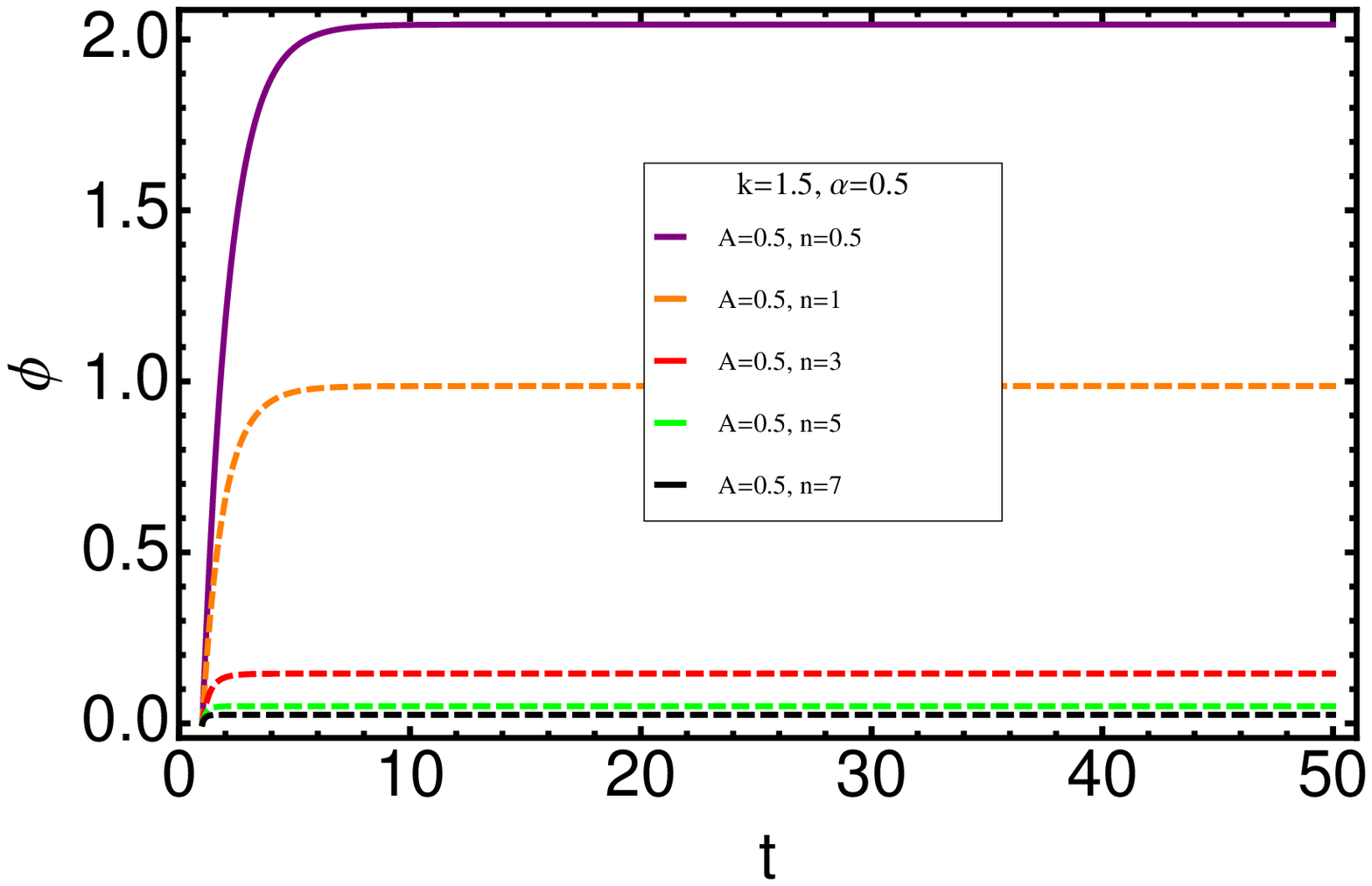}
 \end{array}$
 \end{center}
\caption{Behavior of $\phi$ against $t$ for non interacting
components where we choose $a_{0}=2$ and $\rho_{0}=1$.}
 \label{fig:7}
\end{figure}

\begin{figure}[h]
 \begin{center}$
 \begin{array}{cccc}
\includegraphics[width=48 mm]{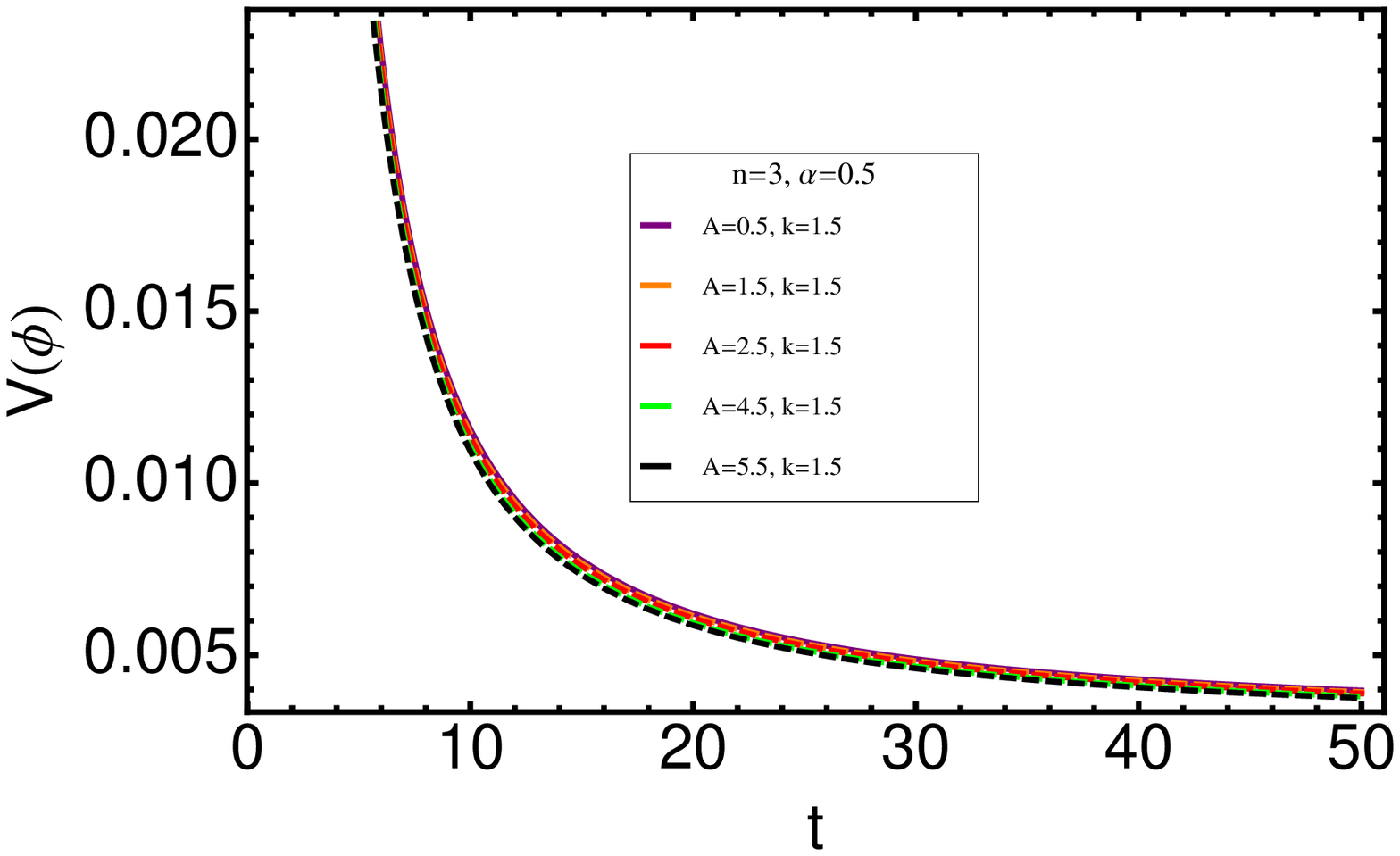} &
\includegraphics[width=48 mm]{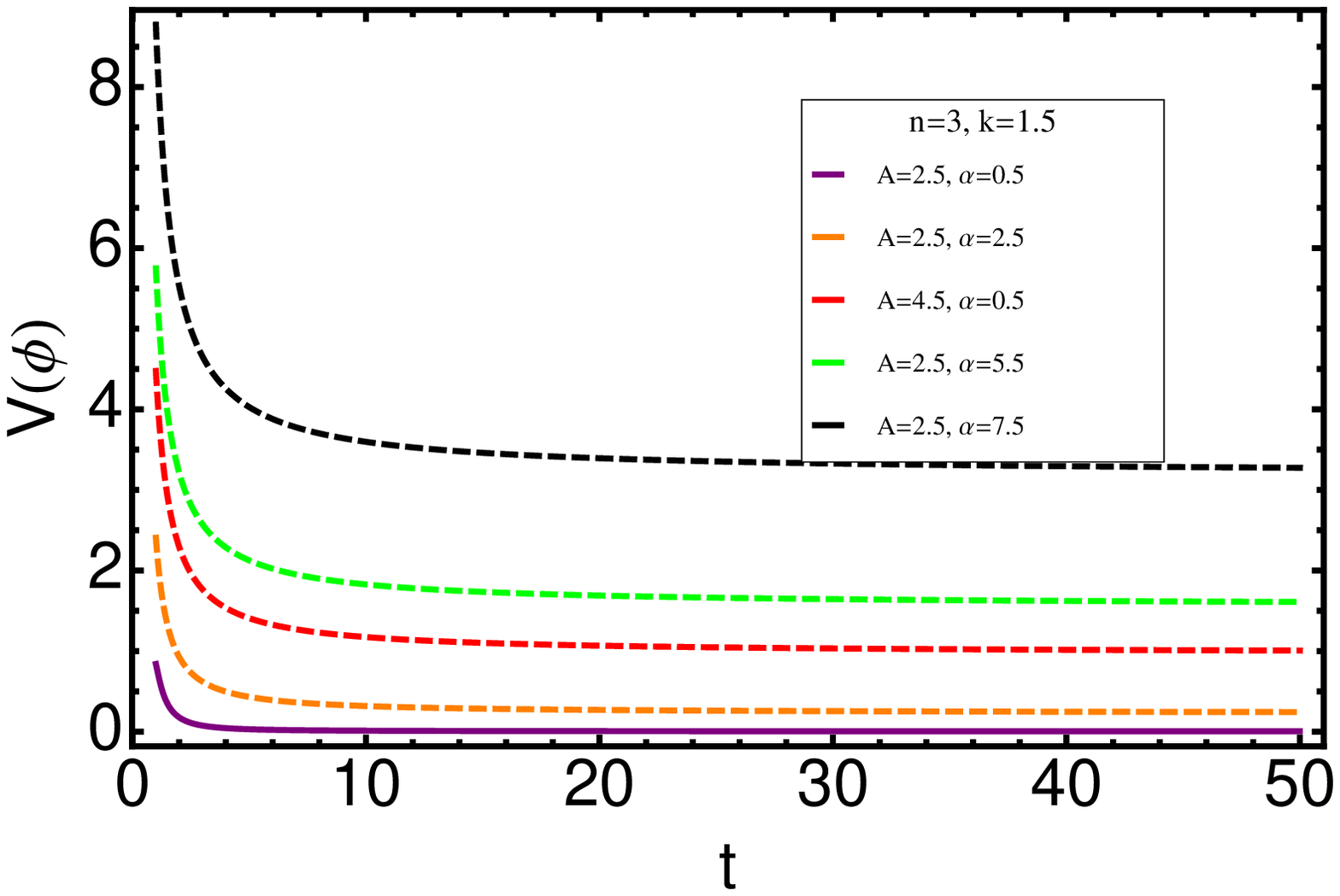}\\
\includegraphics[width=48 mm]{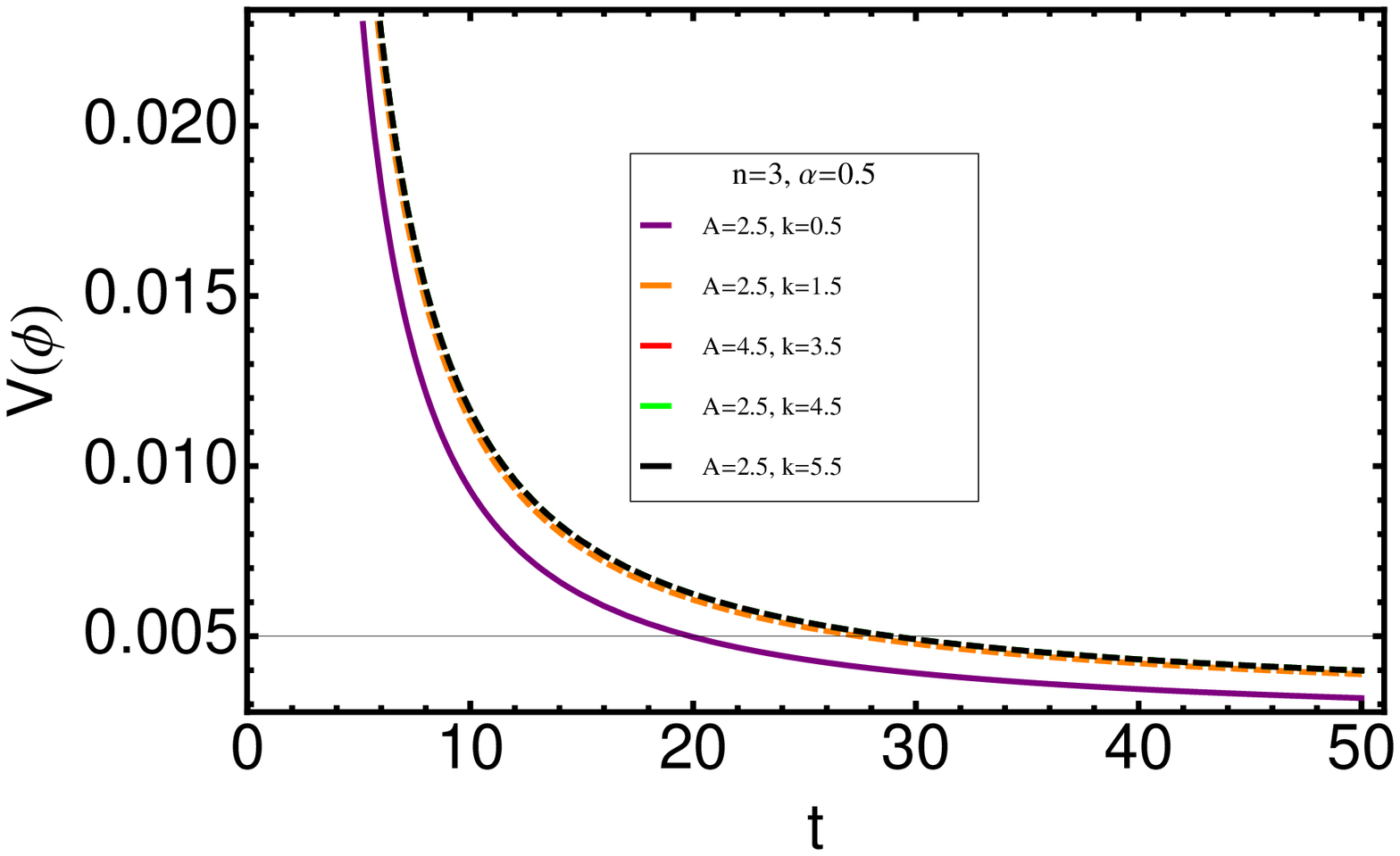} &
\includegraphics[width=48 mm]{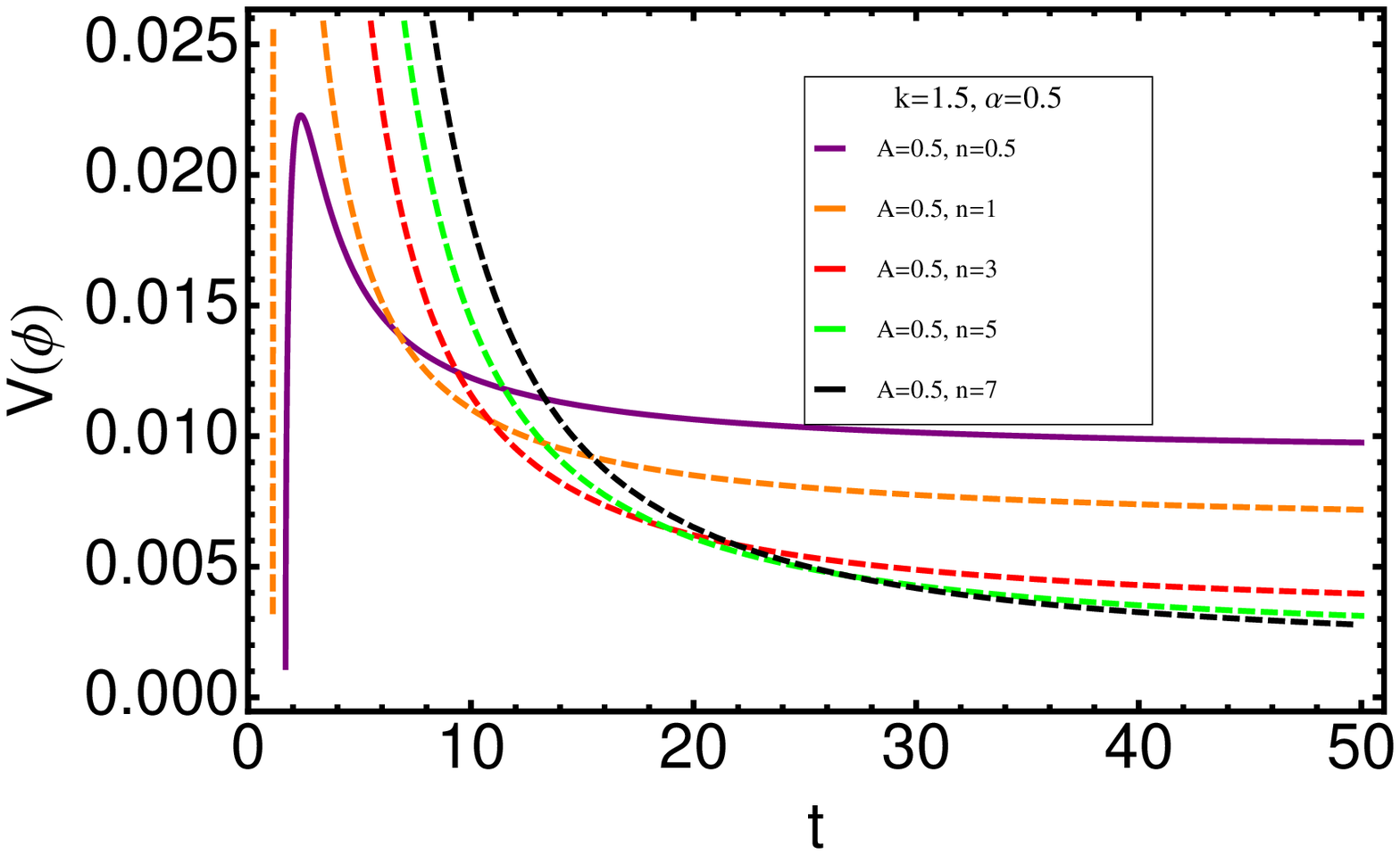}
 \end{array}$
 \end{center}
\caption{Behavior of $V$ against $t$ for non interacting components
where we choose $a_{0}=2$ and $\rho_{0}=1$.}
 \label{fig:8}
\end{figure}

\begin{figure}[h]
 \begin{center}$
 \begin{array}{cccc}
\includegraphics[width=48 mm]{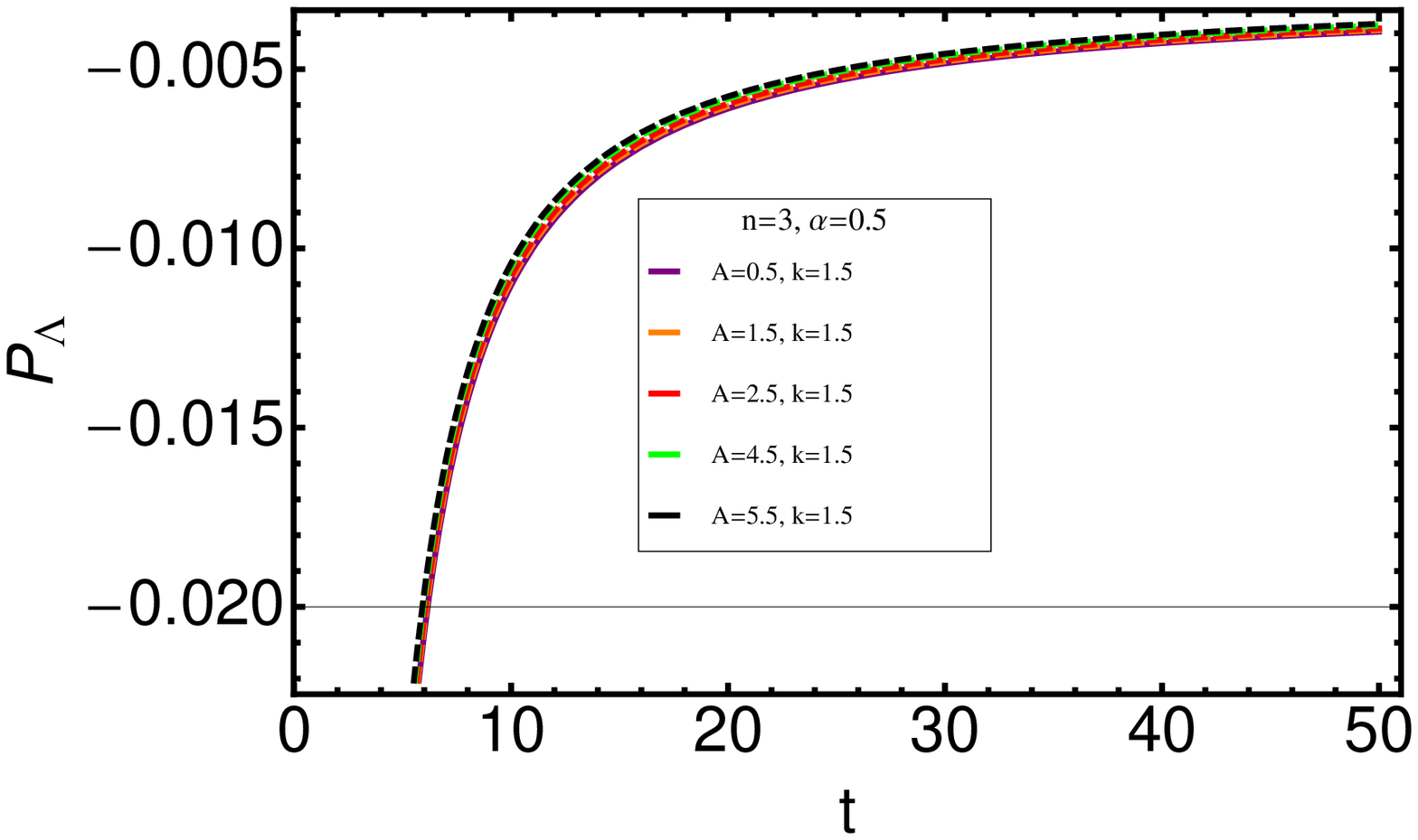} &
\includegraphics[width=48 mm]{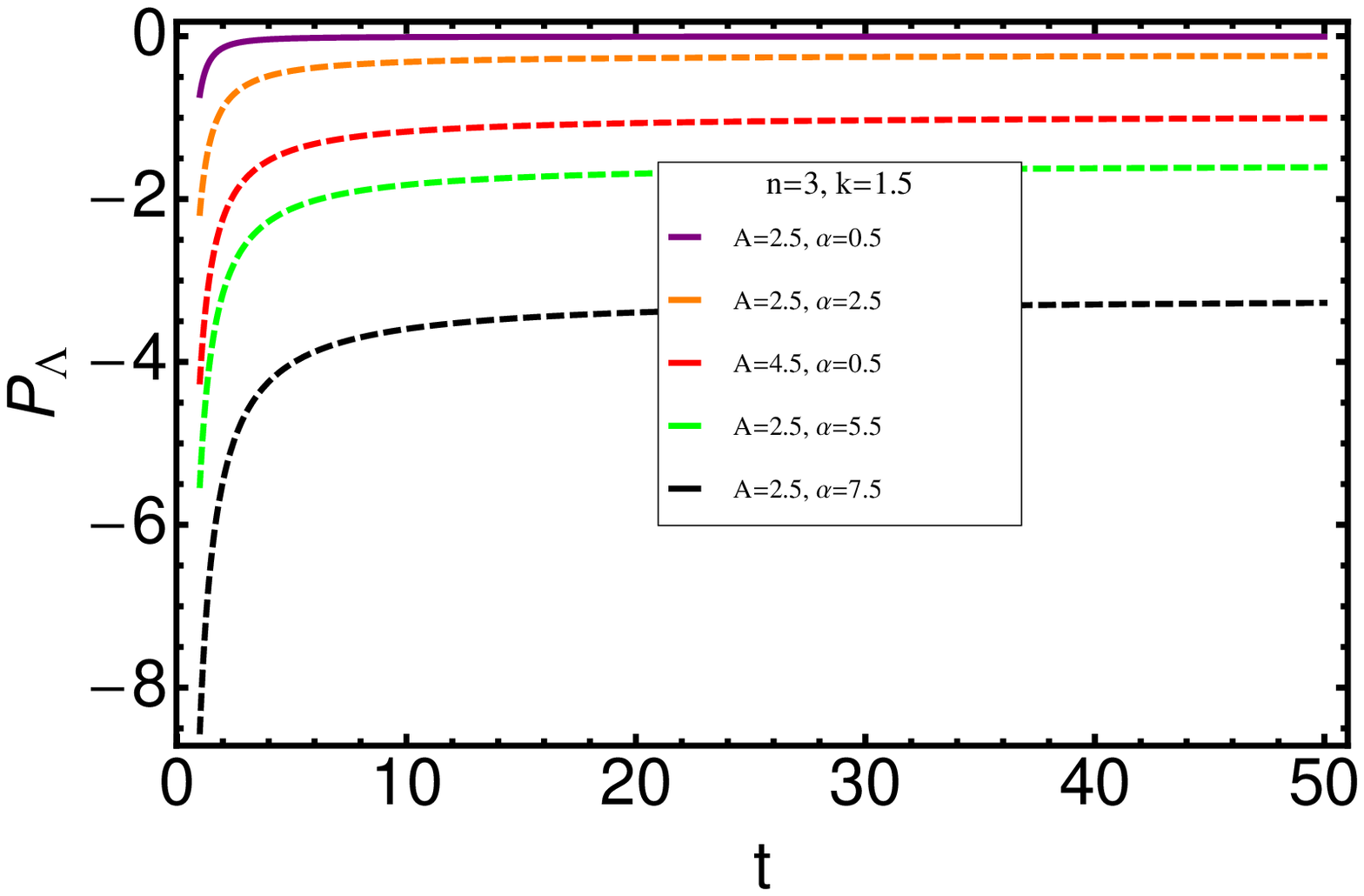}\\
\includegraphics[width=48 mm]{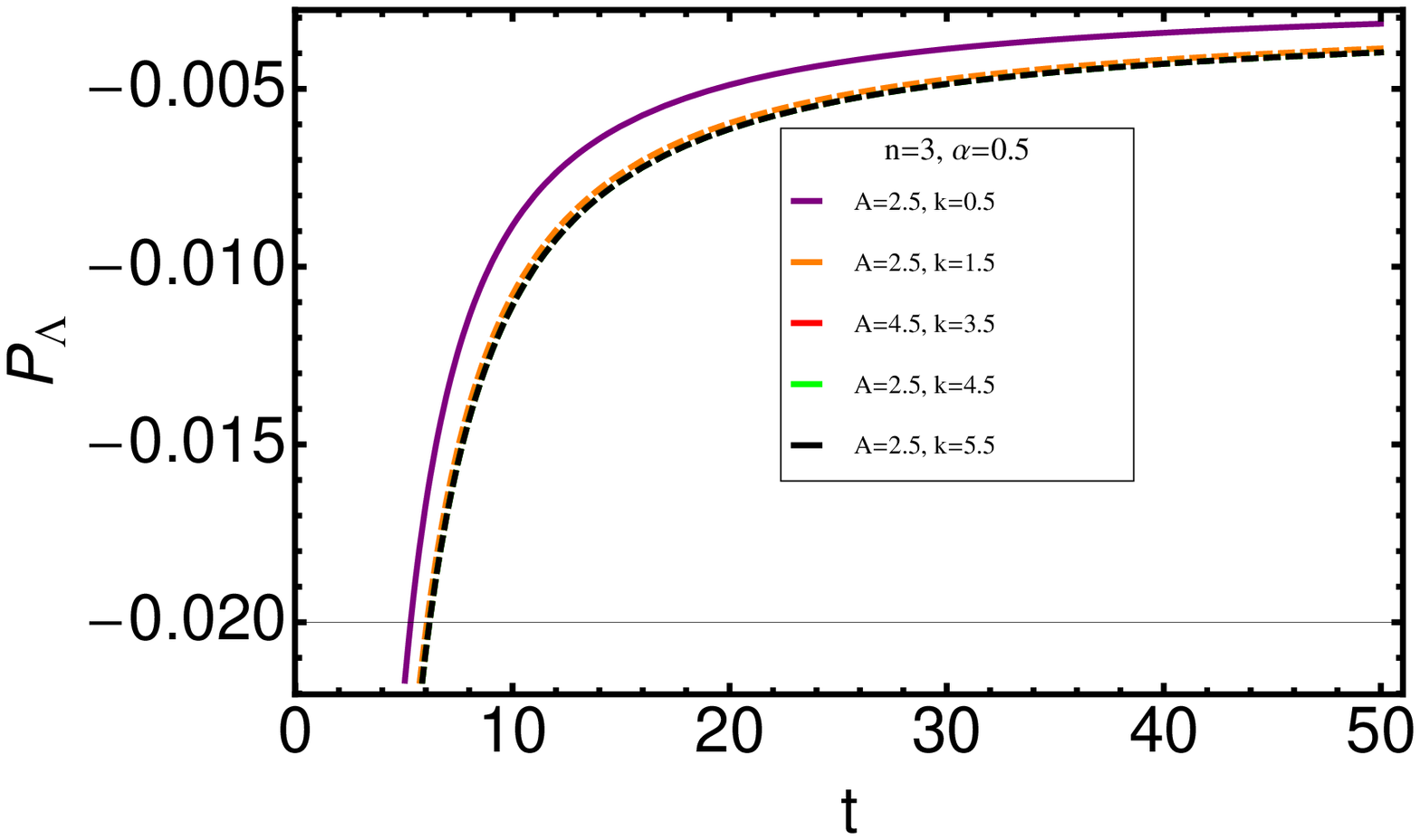} &
\includegraphics[width=48 mm]{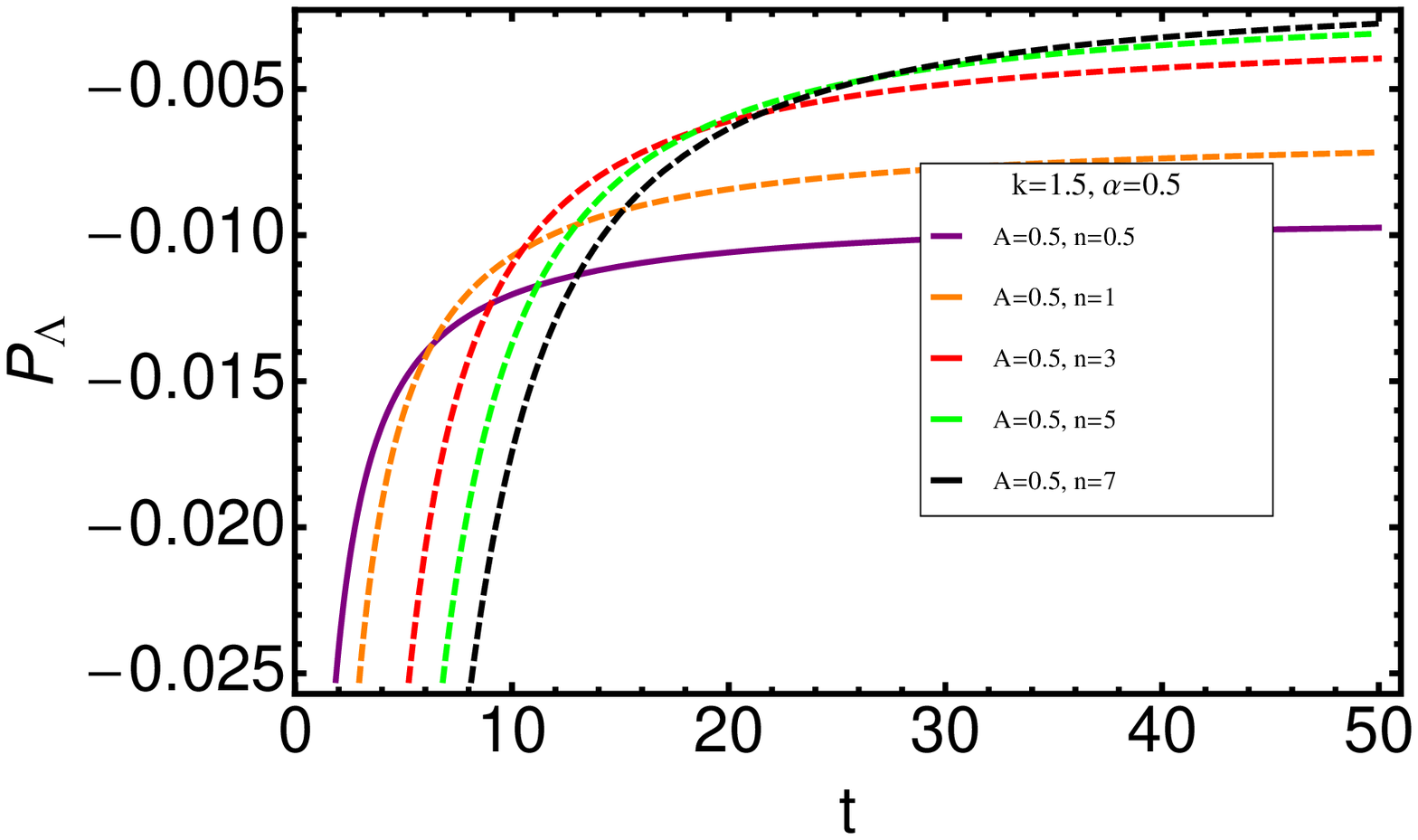}
 \end{array}$
 \end{center}
\caption{Behavior of $P_{\Lambda}$ against $t$ for non interacting
components where we choose $a_{0}=2$ and $\rho_{0}=1$.}
 \label{fig:9}
\end{figure}

\begin{figure}[h]
 \begin{center}$
 \begin{array}{cccc}
\includegraphics[width=48 mm]{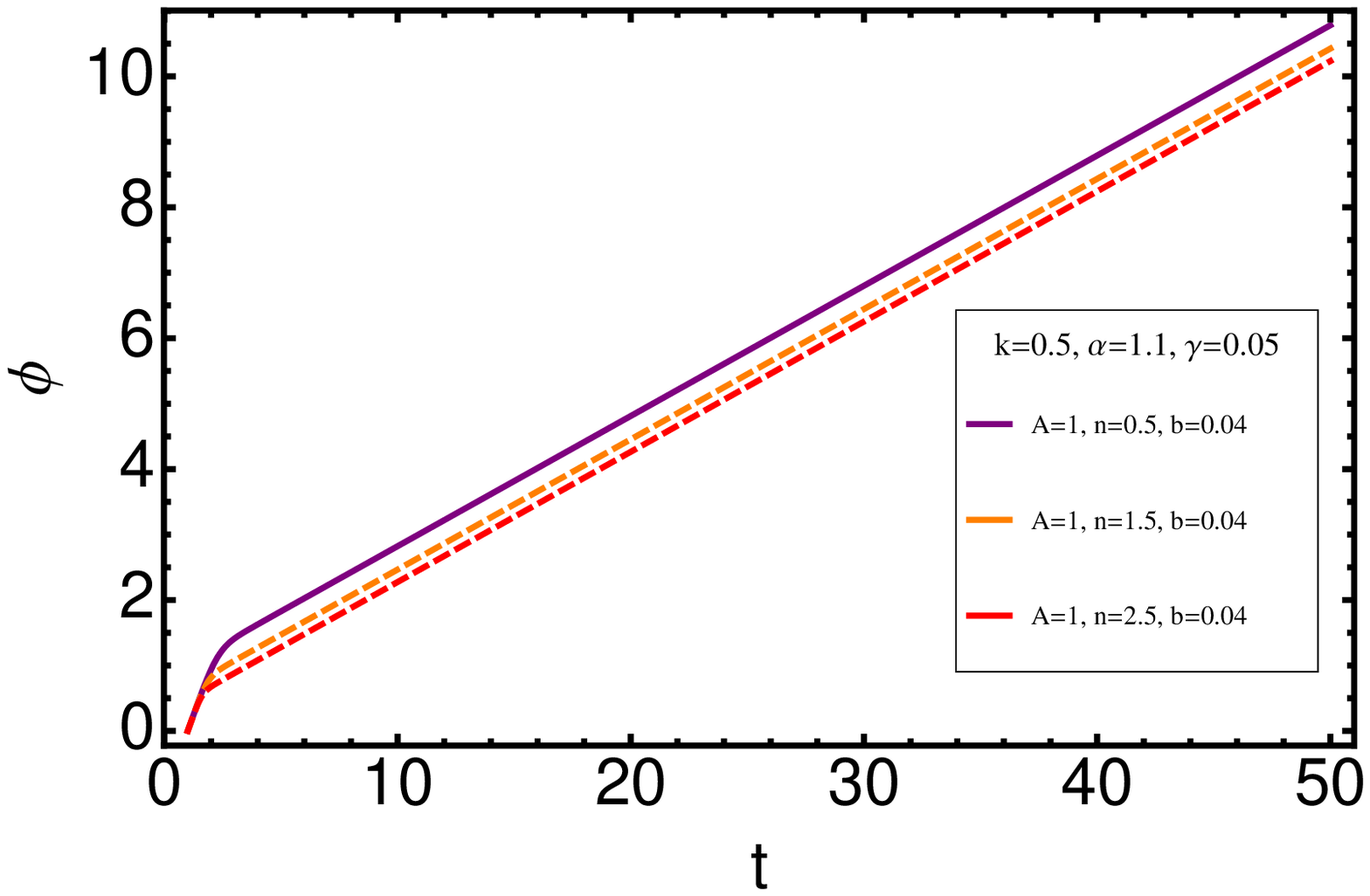} &
\includegraphics[width=48 mm]{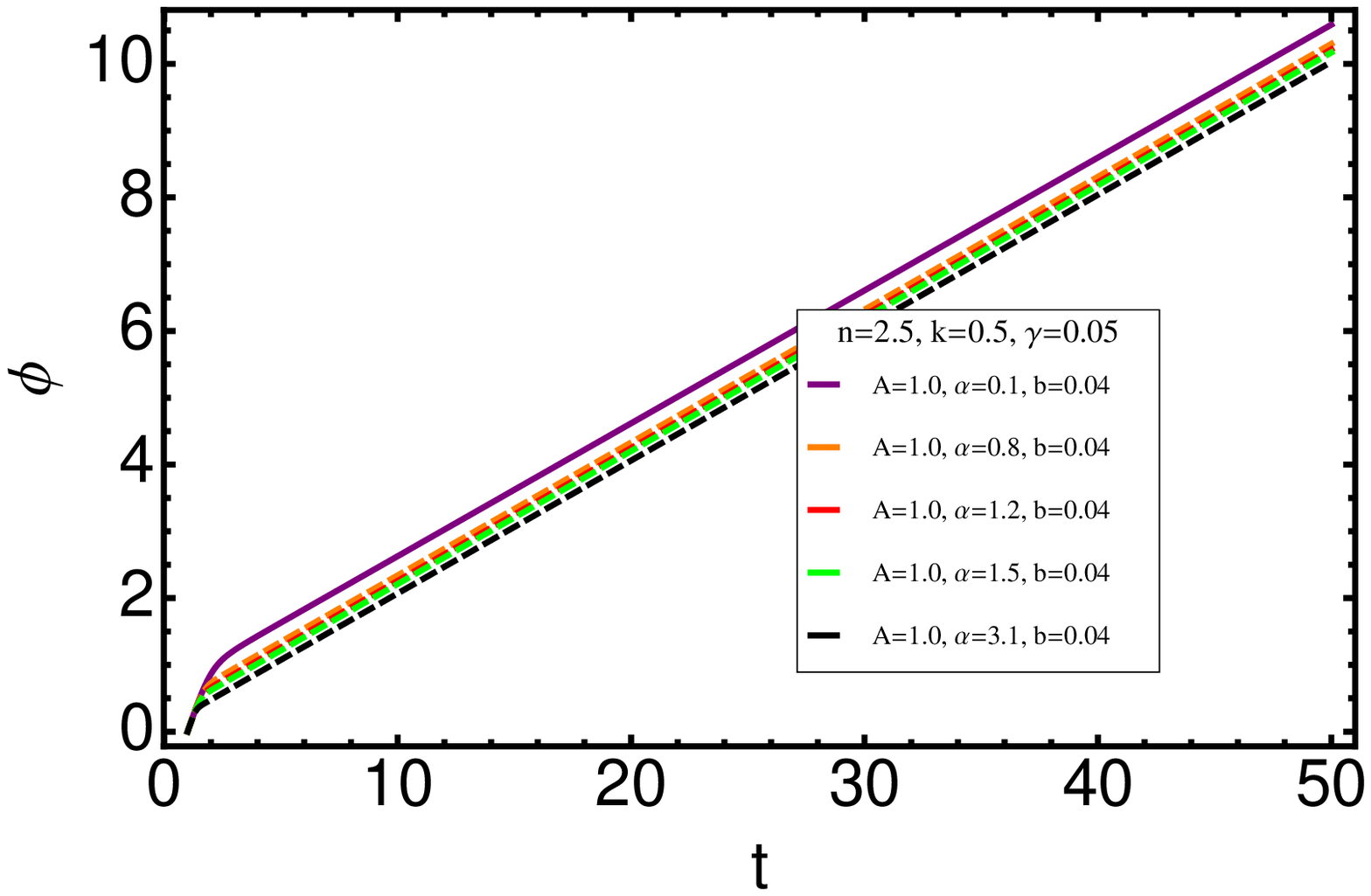}\\
\includegraphics[width=48 mm]{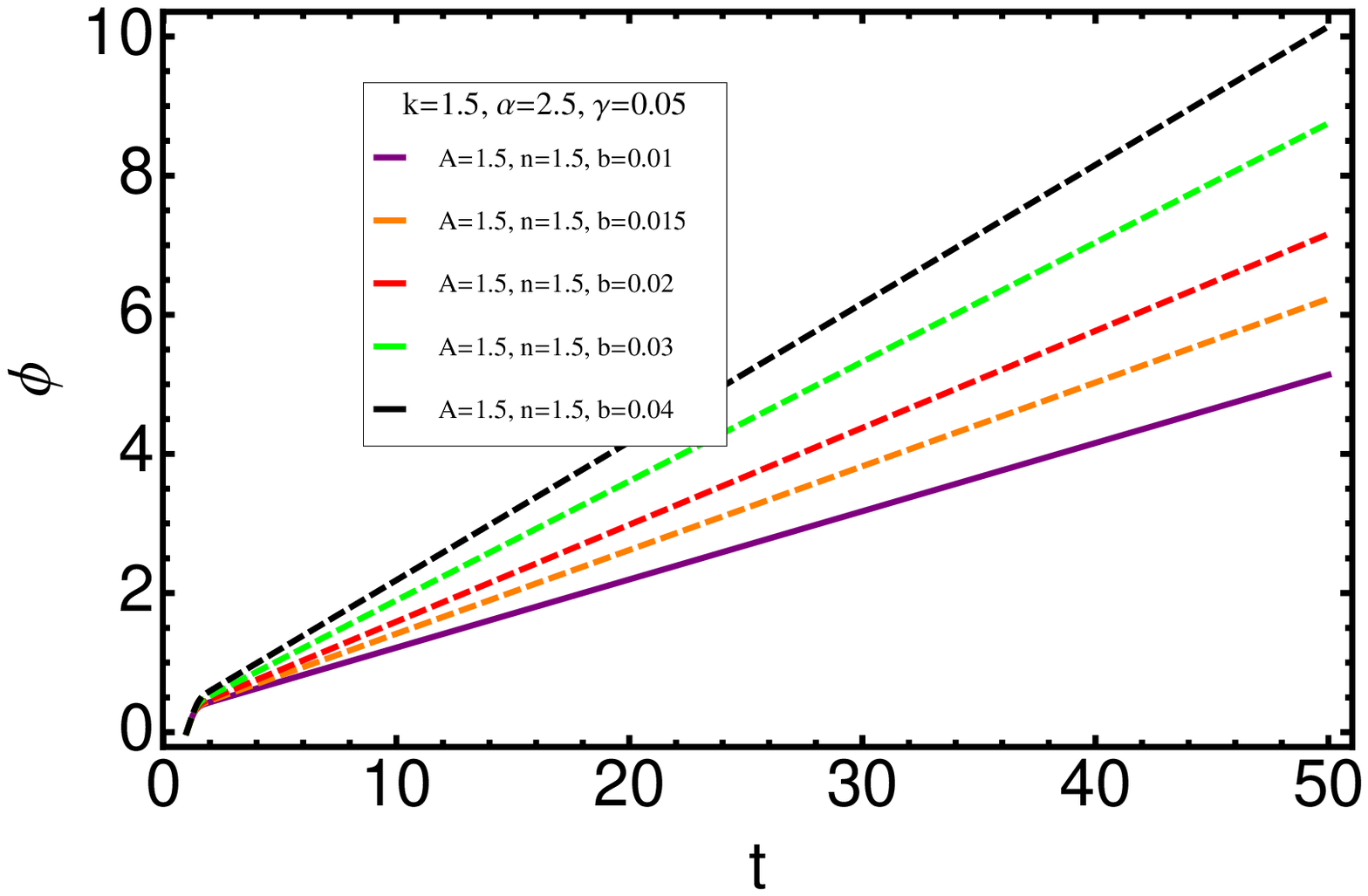} &
\includegraphics[width=48 mm]{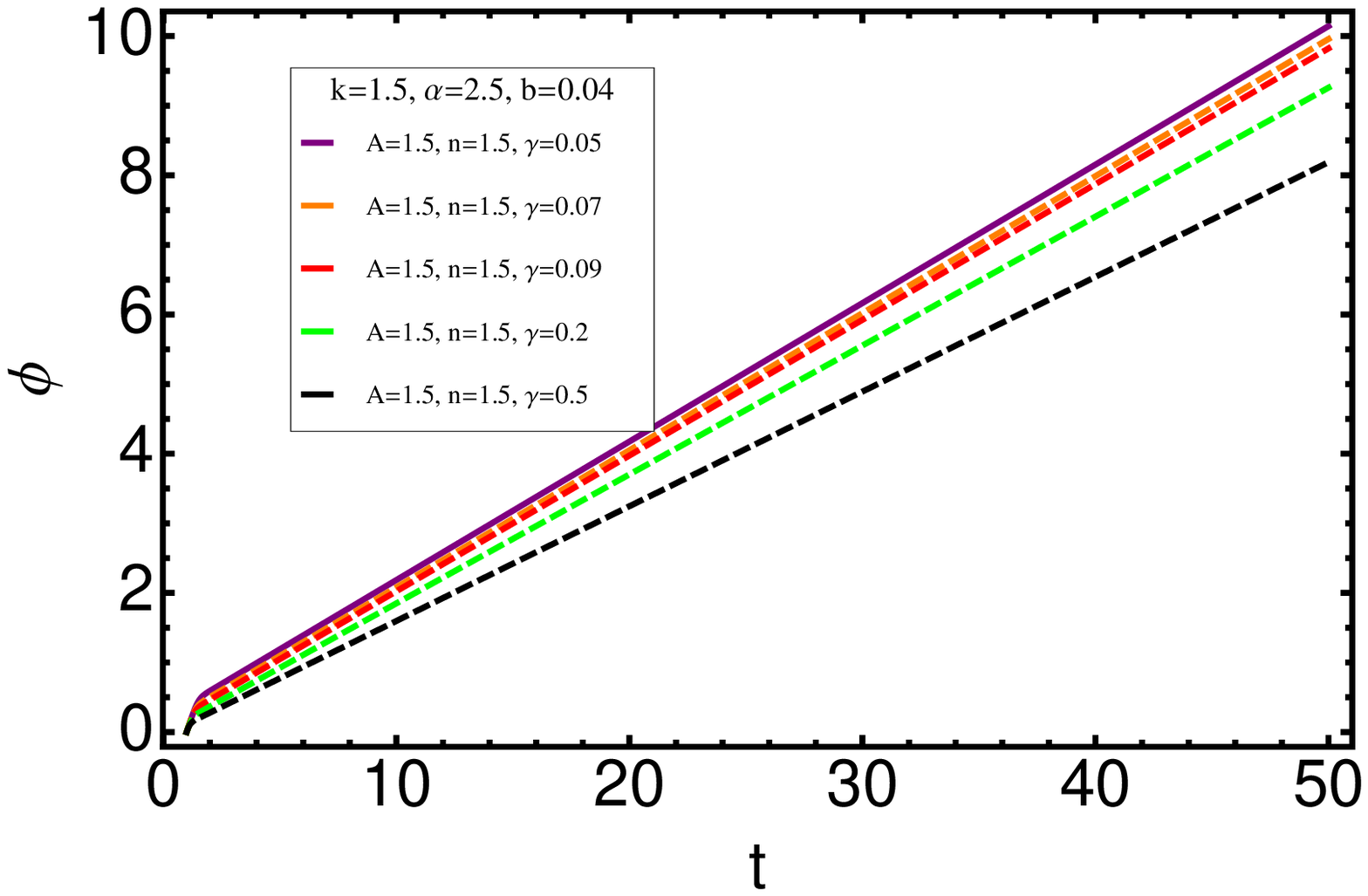}
 \end{array}$
 \end{center}
\caption{Behavior of $\phi$ against $t$ for interacting components
where we choose $a_{0}=1$.}
 \label{fig:10}
\end{figure}

\begin{figure}[h]
 \begin{center}$
 \begin{array}{cccc}
\includegraphics[width=48 mm]{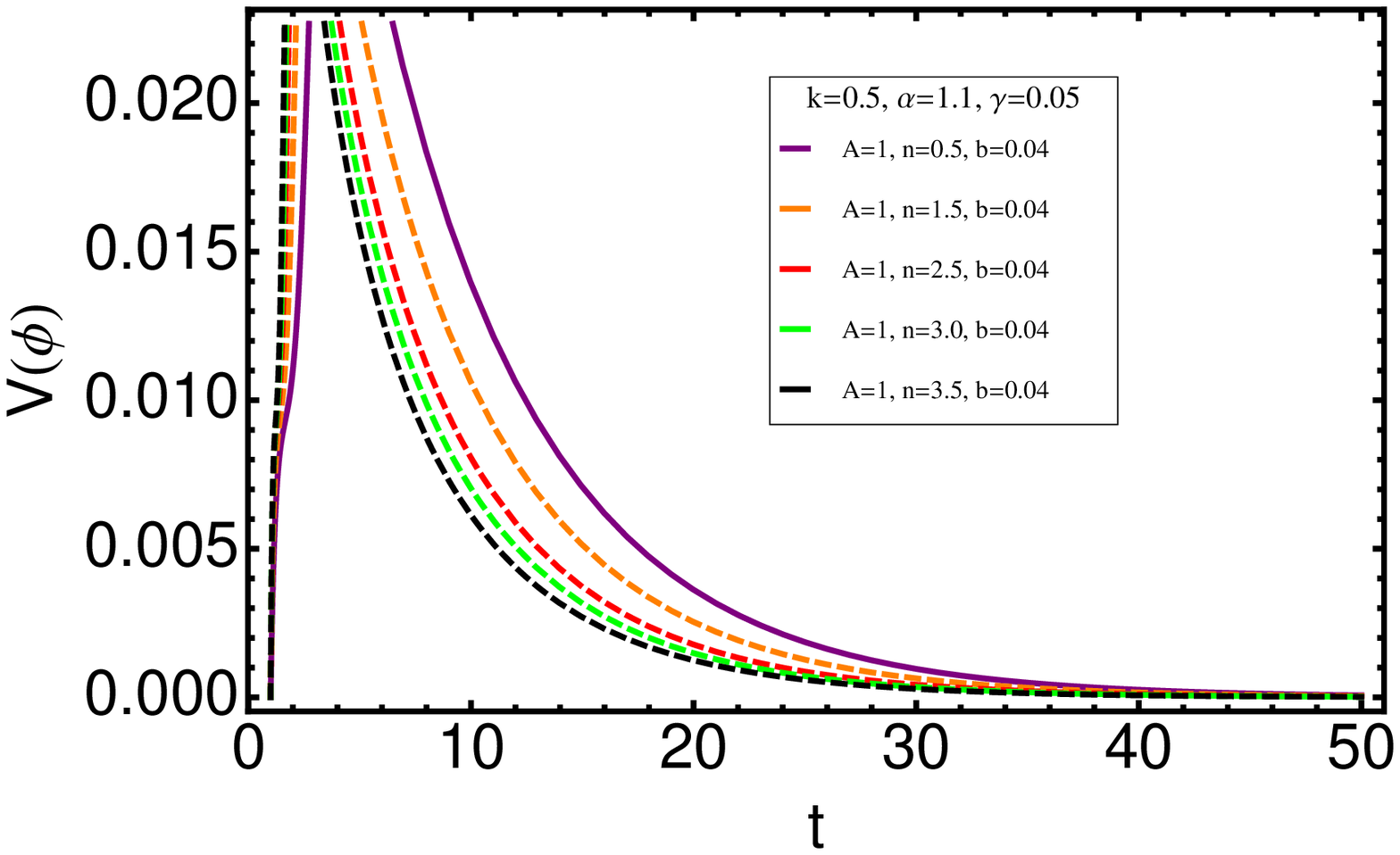} &
\includegraphics[width=48 mm]{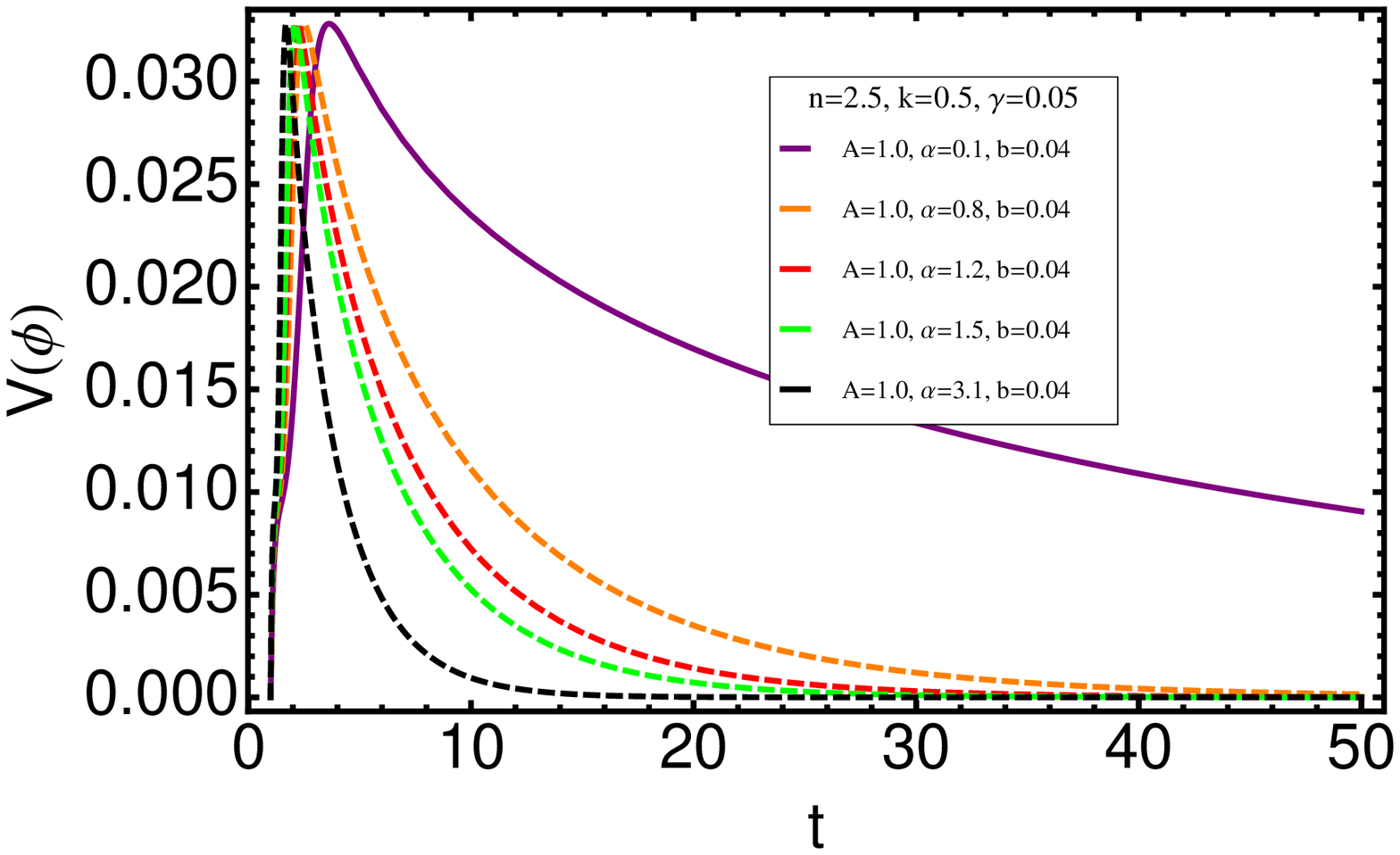}\\
\includegraphics[width=48 mm]{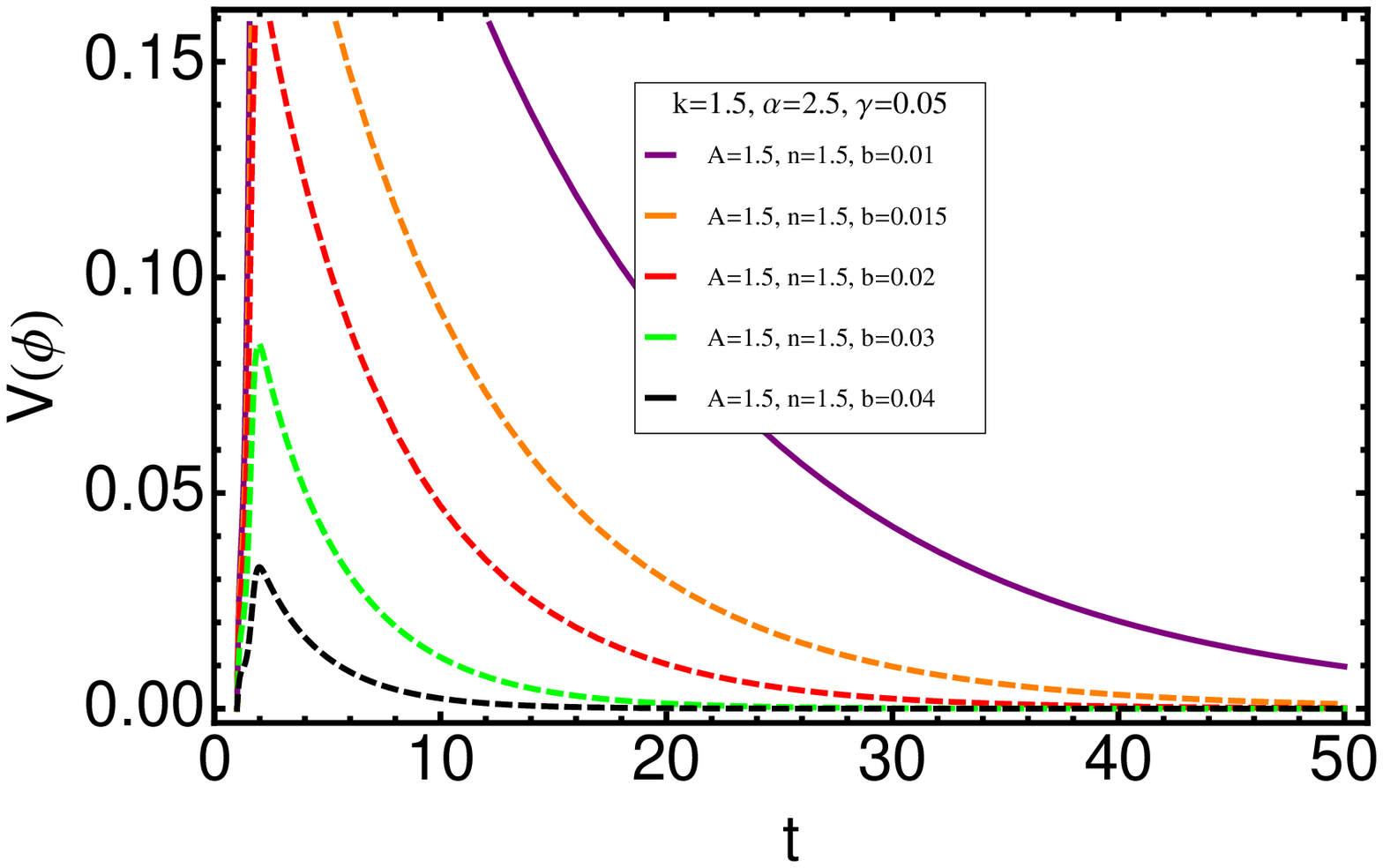} &
\includegraphics[width=48 mm]{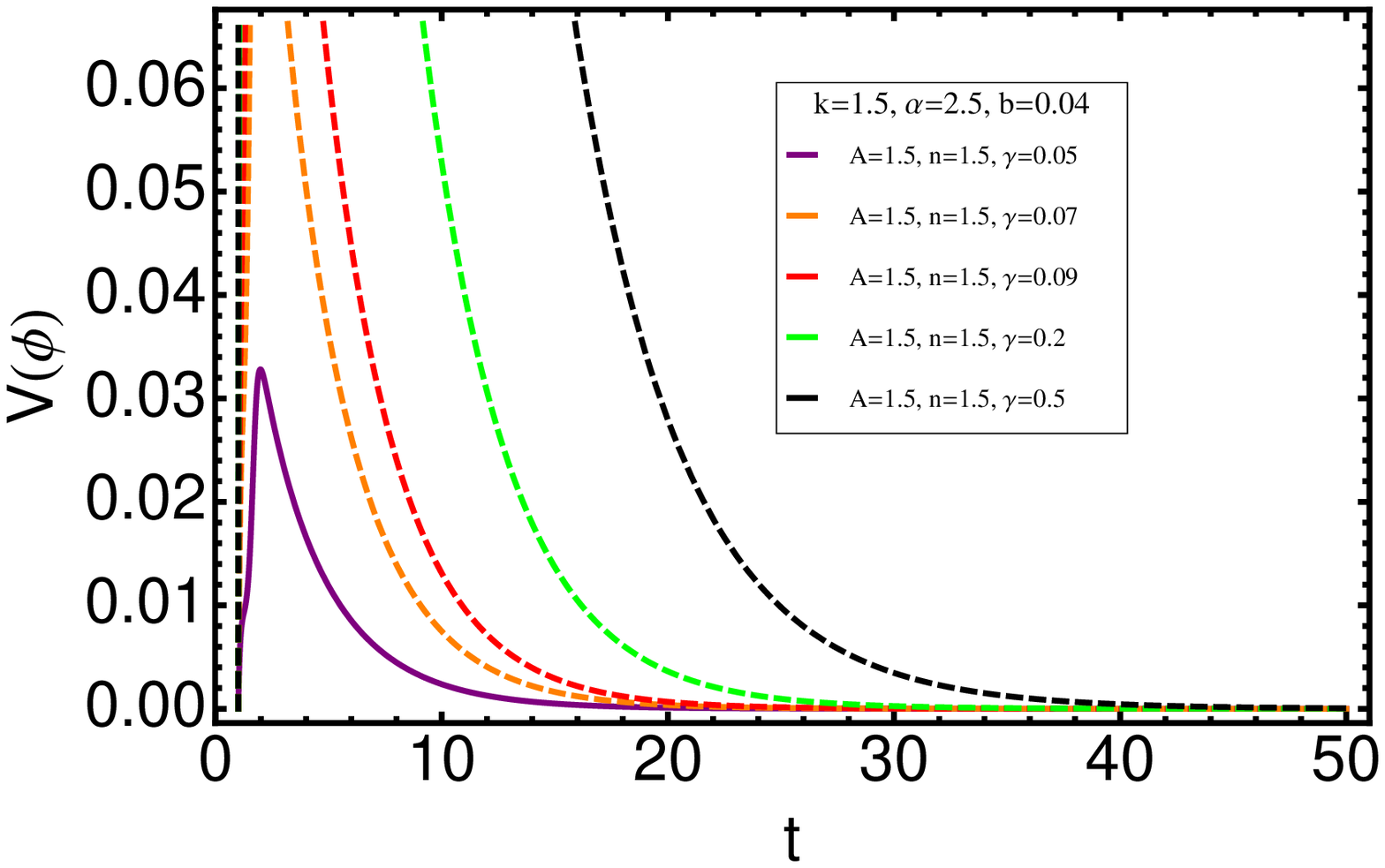}
 \end{array}$
 \end{center}
\caption{Behavior of $V$ against $t$ for interacting components
where we choose $a_{0}=1$.}
 \label{fig:11}
\end{figure}

\begin{figure}[h]
 \begin{center}$
 \begin{array}{cccc}
\includegraphics[width=48 mm]{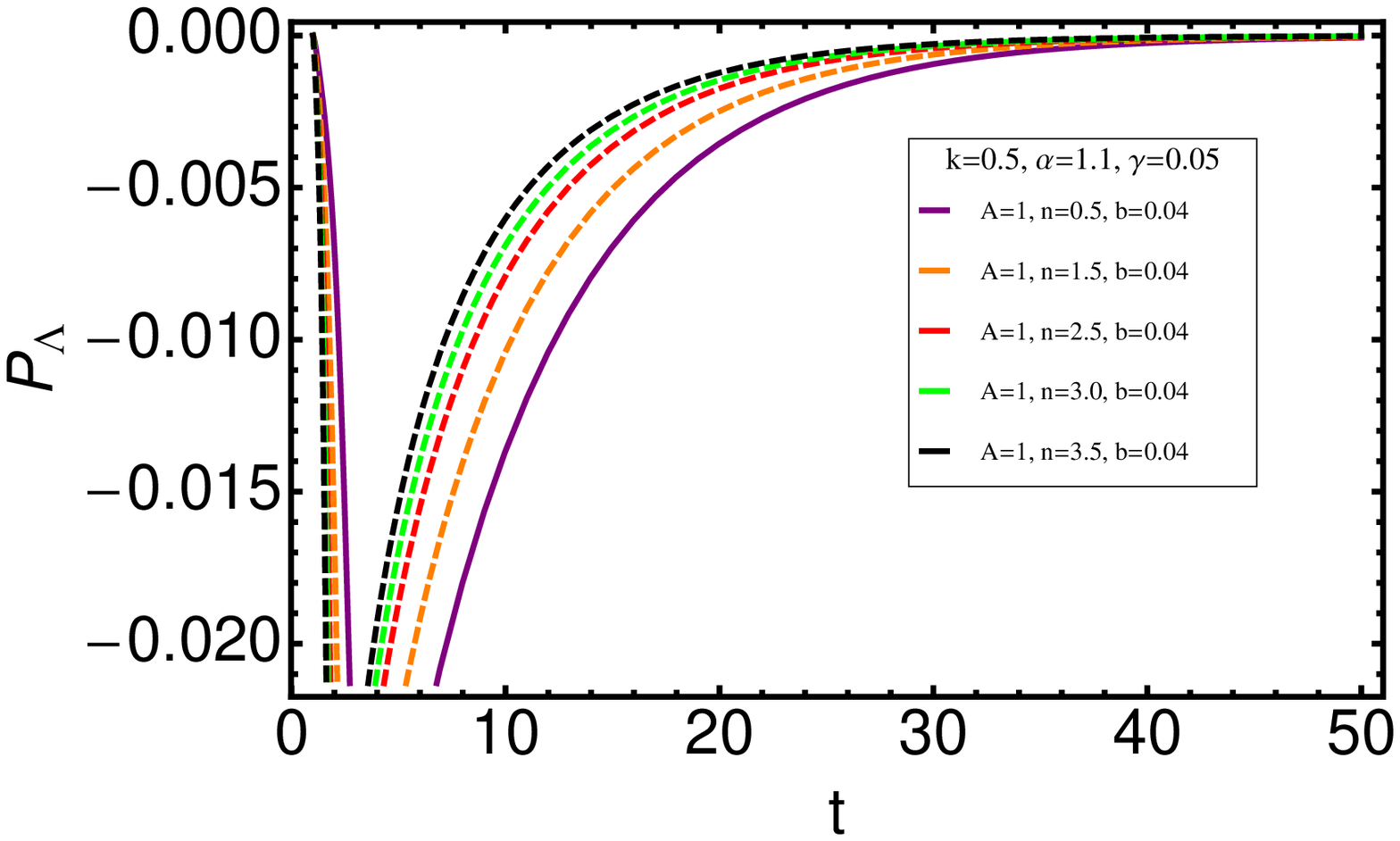} &
\includegraphics[width=48 mm]{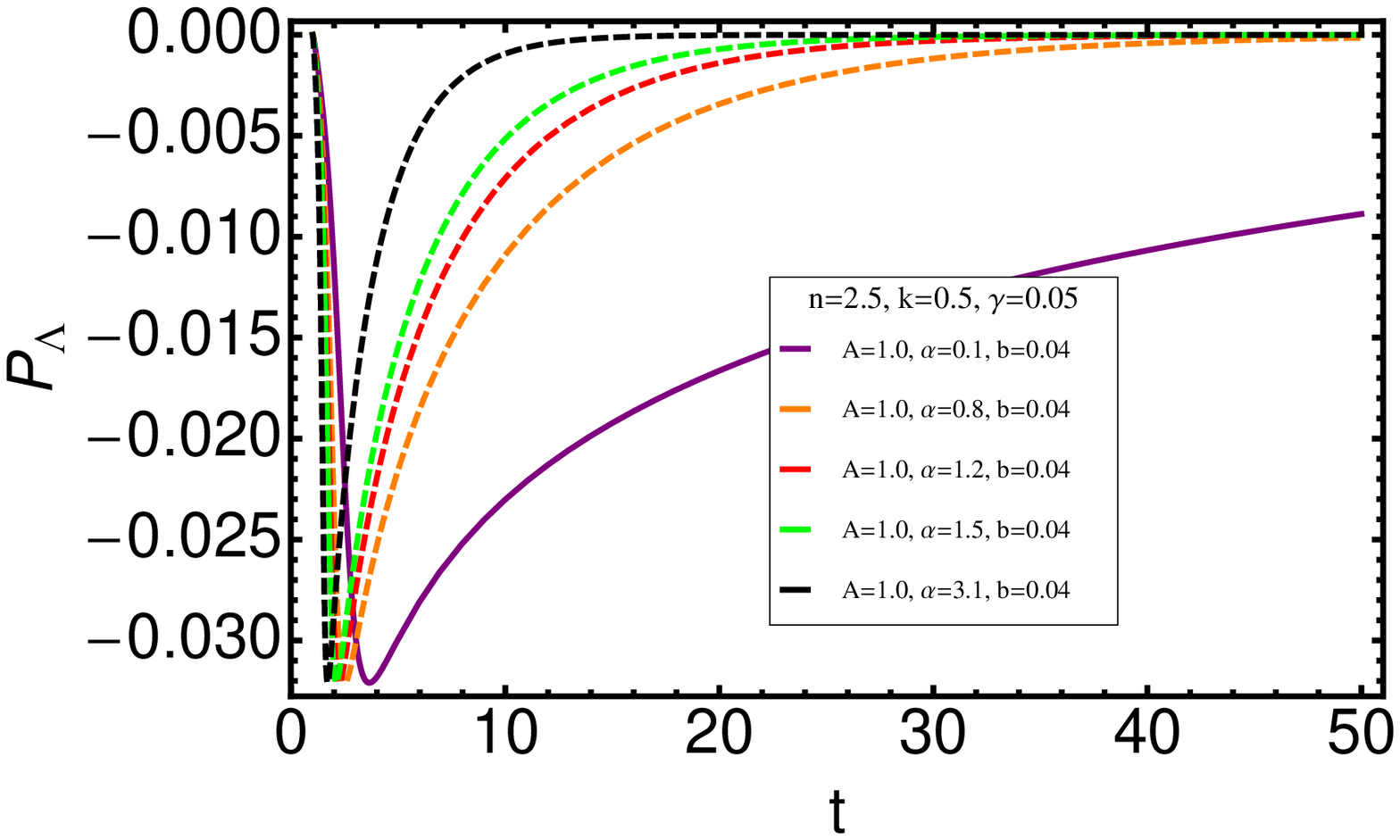}\\
\includegraphics[width=48 mm]{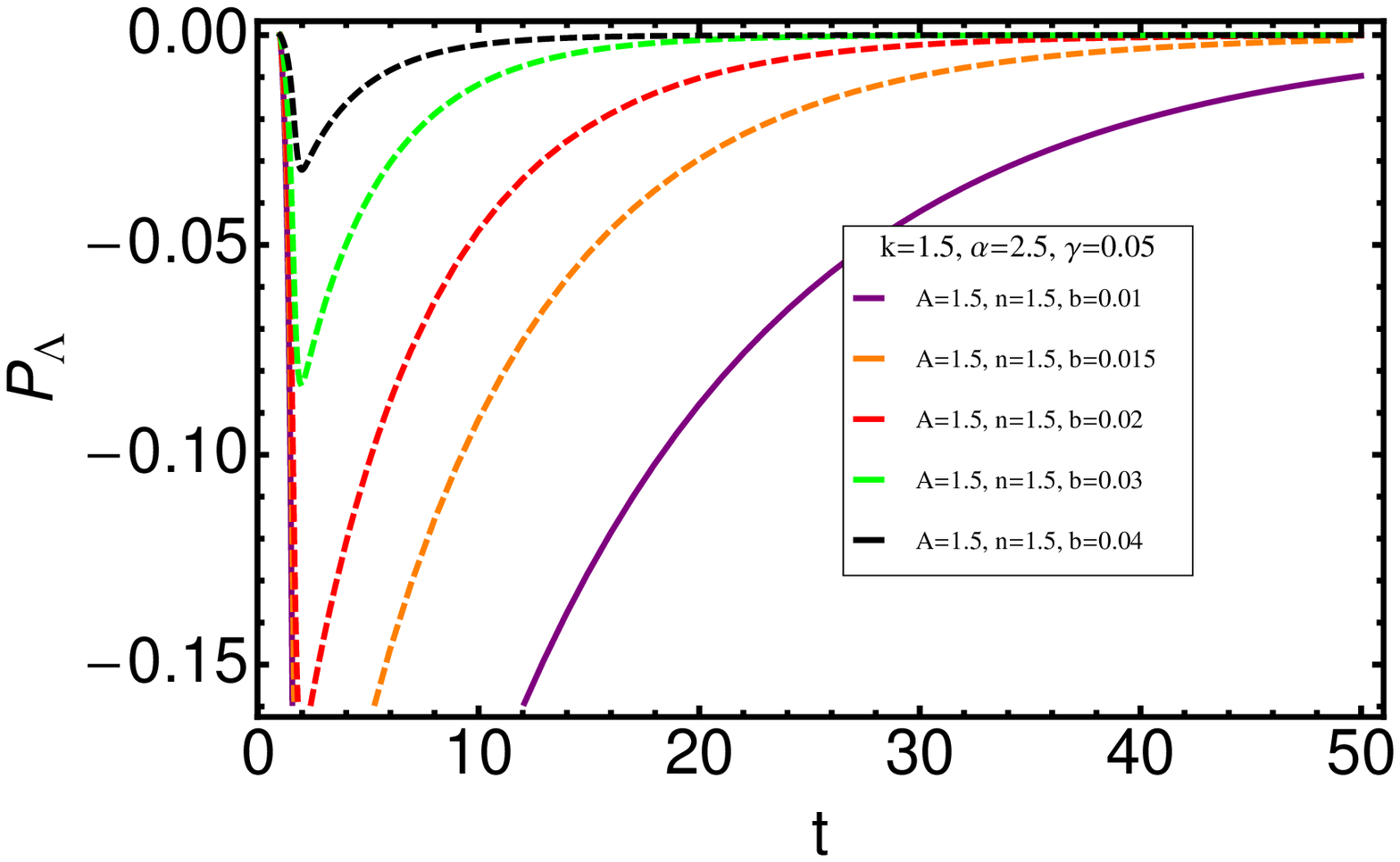} &
\includegraphics[width=48 mm]{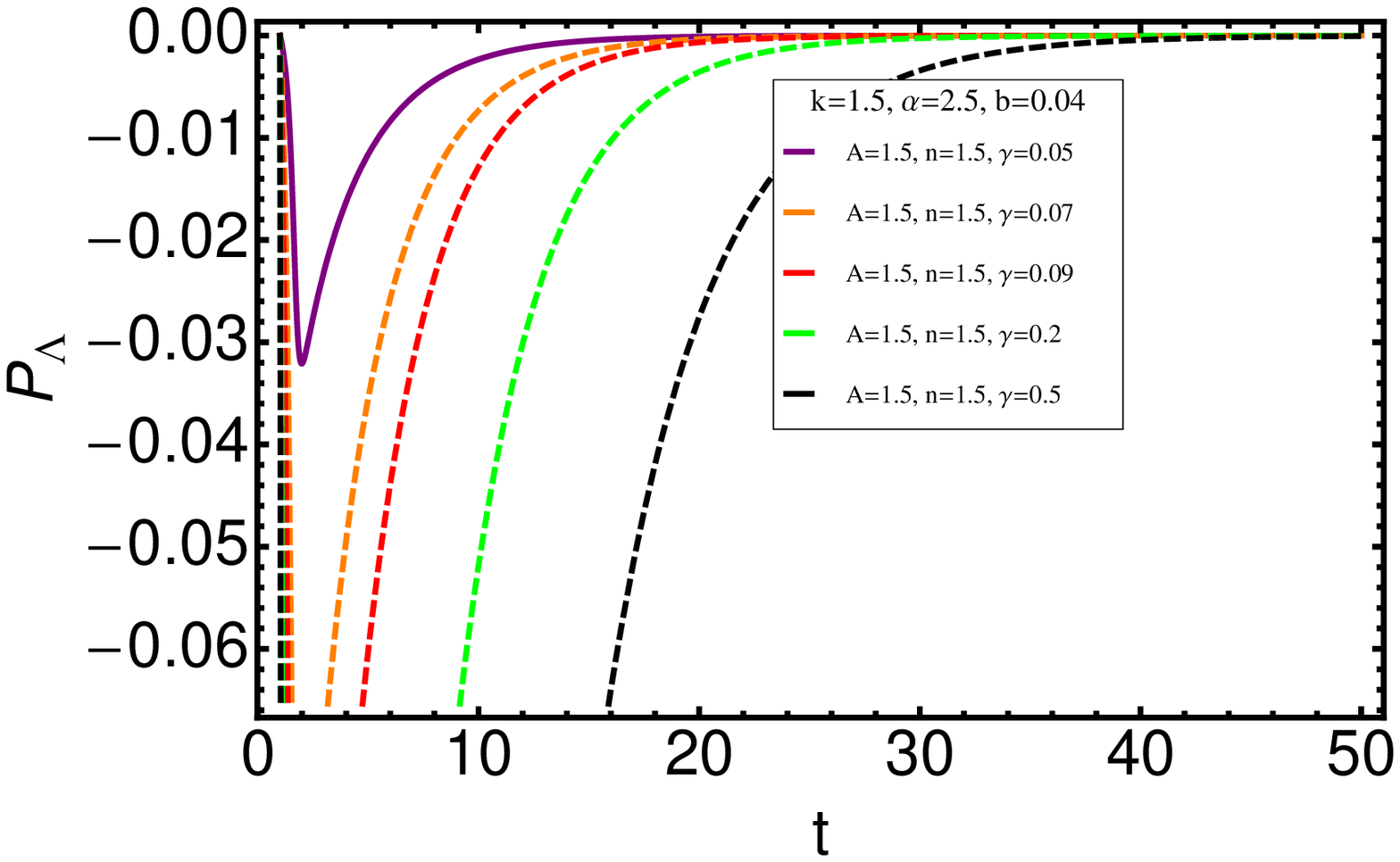}
 \end{array}$
 \end{center}
\caption{Behavior of $P_{\Lambda}$ against $t$ for interacting
components where we choose $a_{0}=1$.}
 \label{fig:12}
\end{figure}

\begin{figure}[h]
 \begin{center}$
 \begin{array}{cccc}
\includegraphics[width=48 mm]{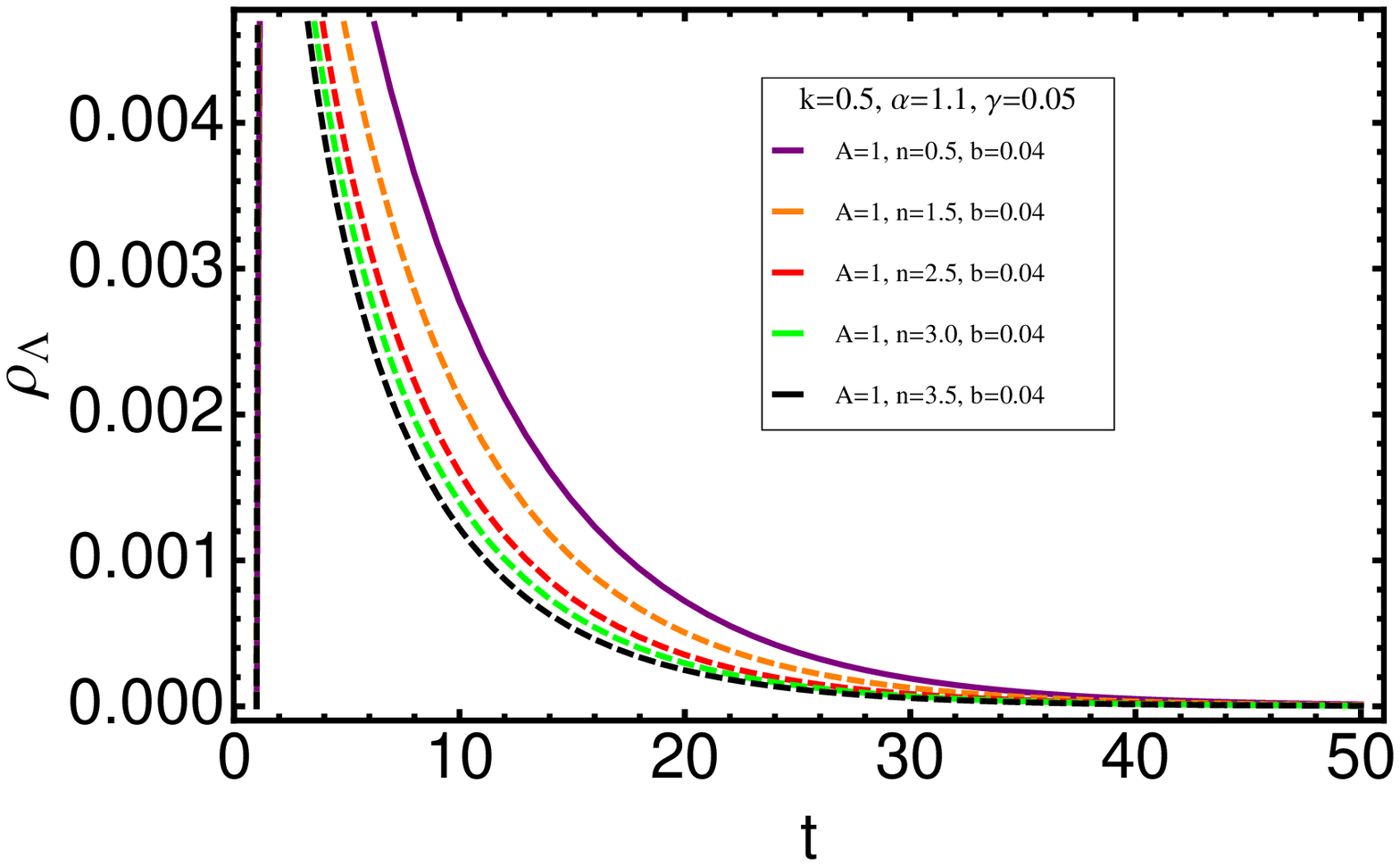} &
\includegraphics[width=48 mm]{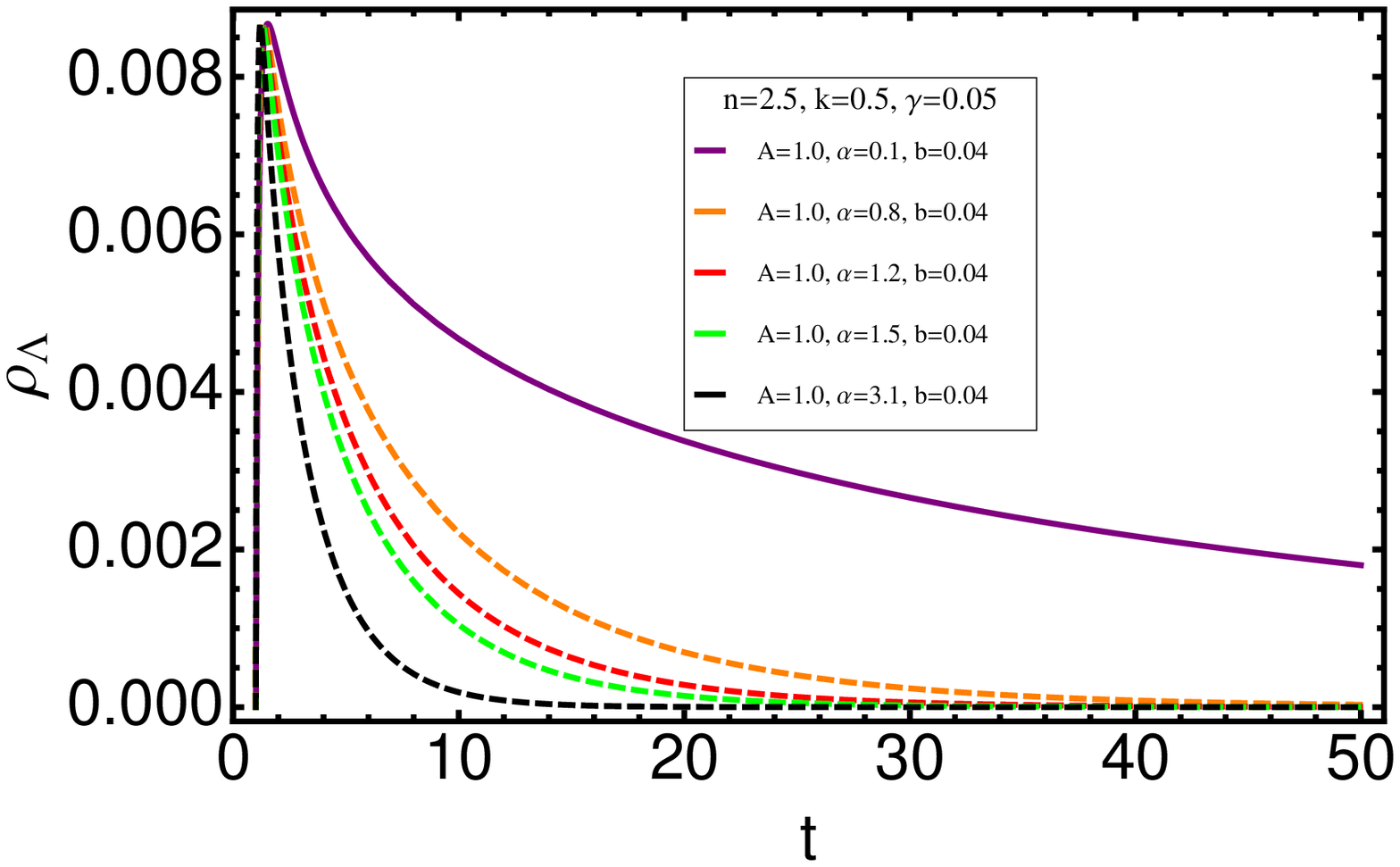}\\
\includegraphics[width=48 mm]{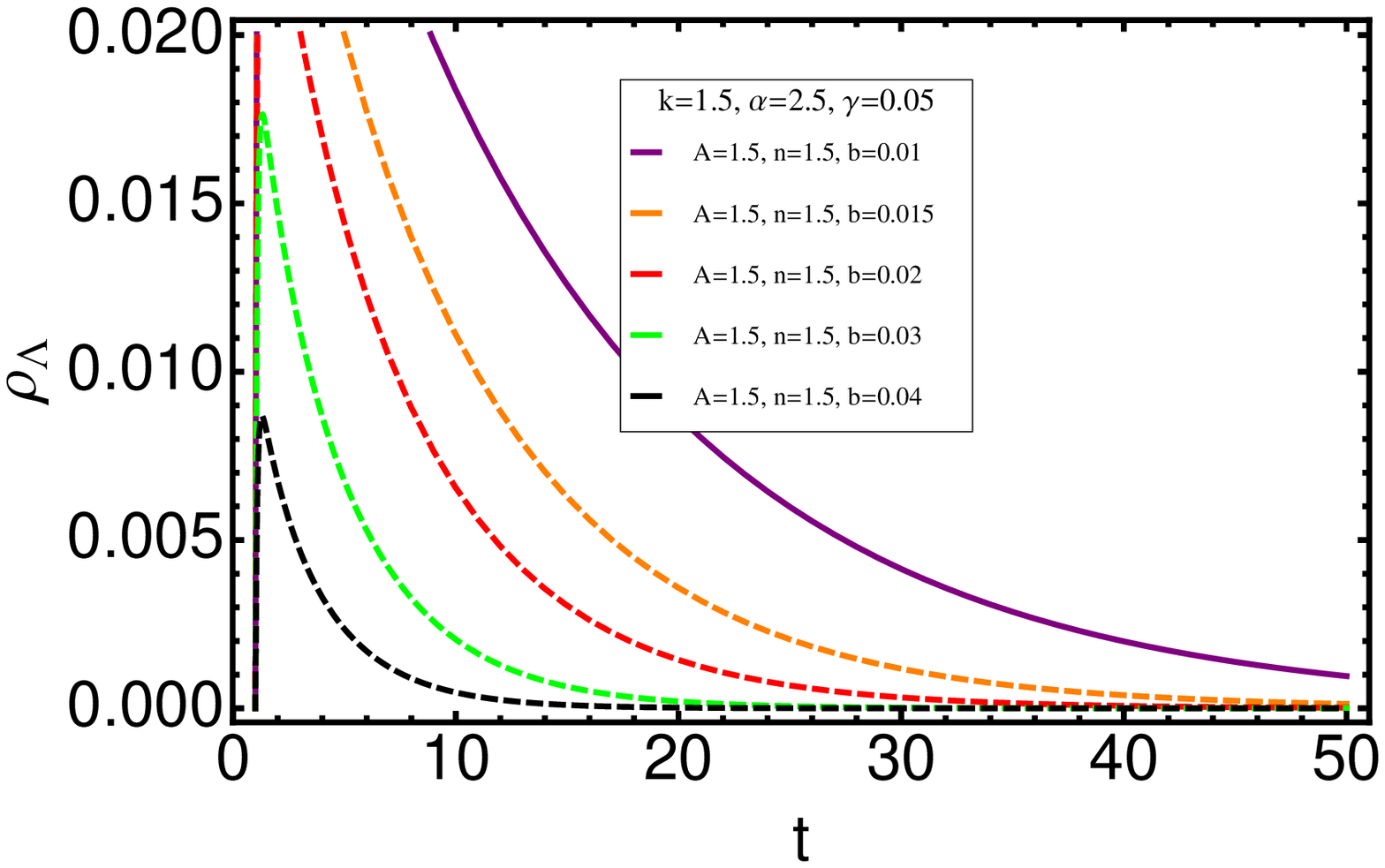} &
\includegraphics[width=48 mm]{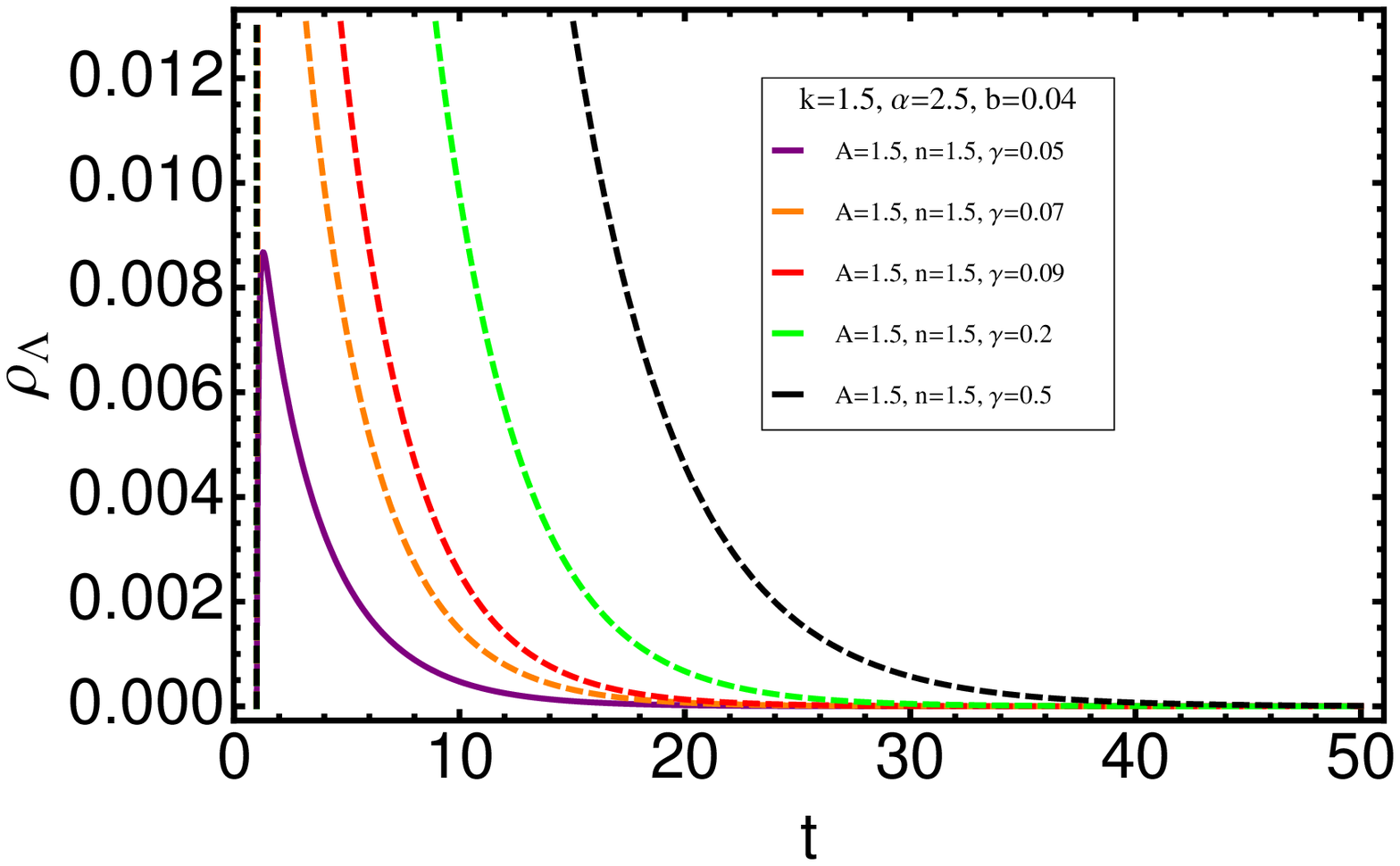}
 \end{array}$
 \end{center}
\caption{Behavior of $\rho_{\Lambda}$ against $t$ for interacting
components where we choose $a_{0}=1$.}
 \label{fig:13}
\end{figure}

\begin{figure}[h]
 \begin{center}$
 \begin{array}{cccc}
\includegraphics[width=48 mm]{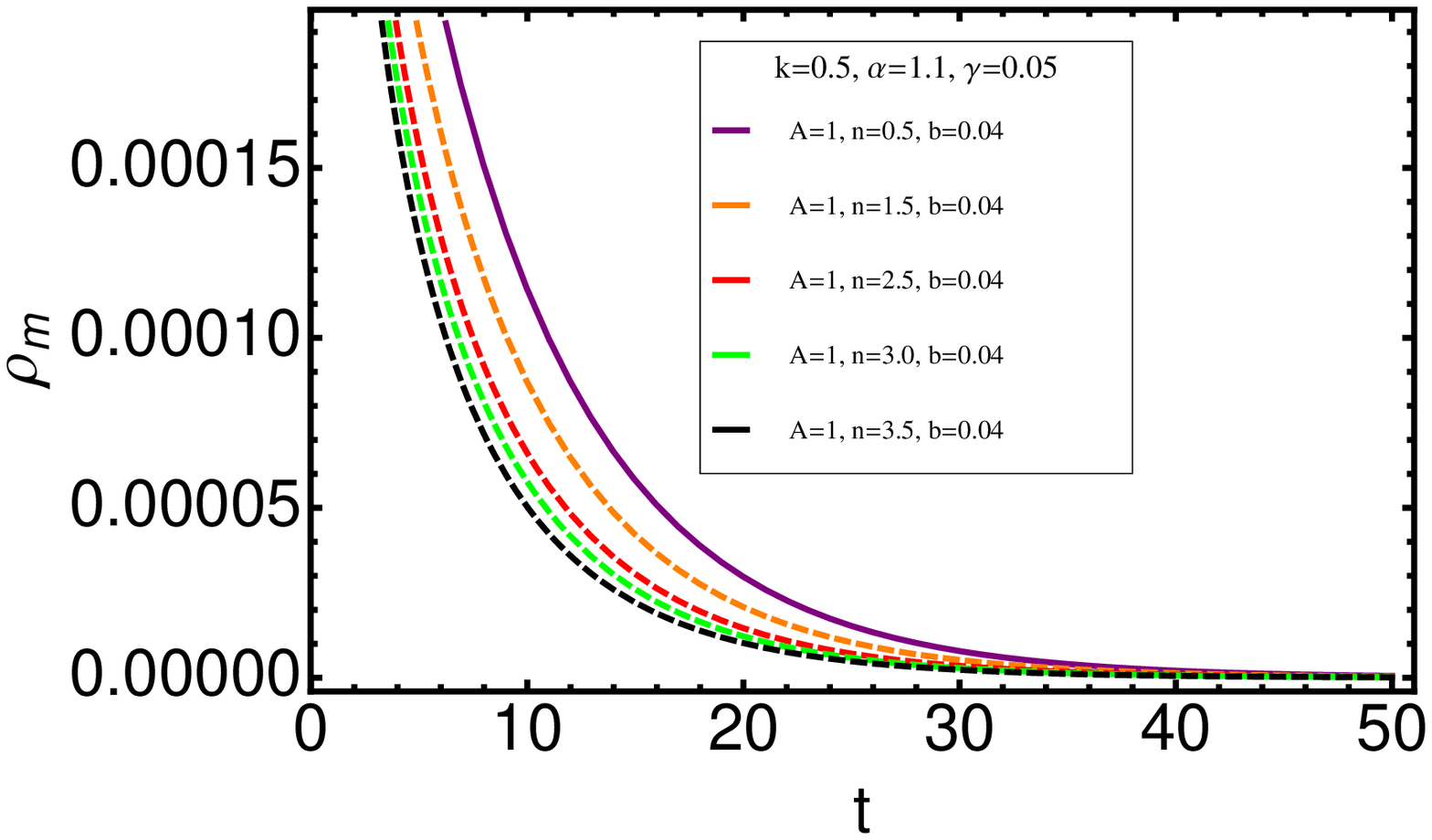} &
\includegraphics[width=48 mm]{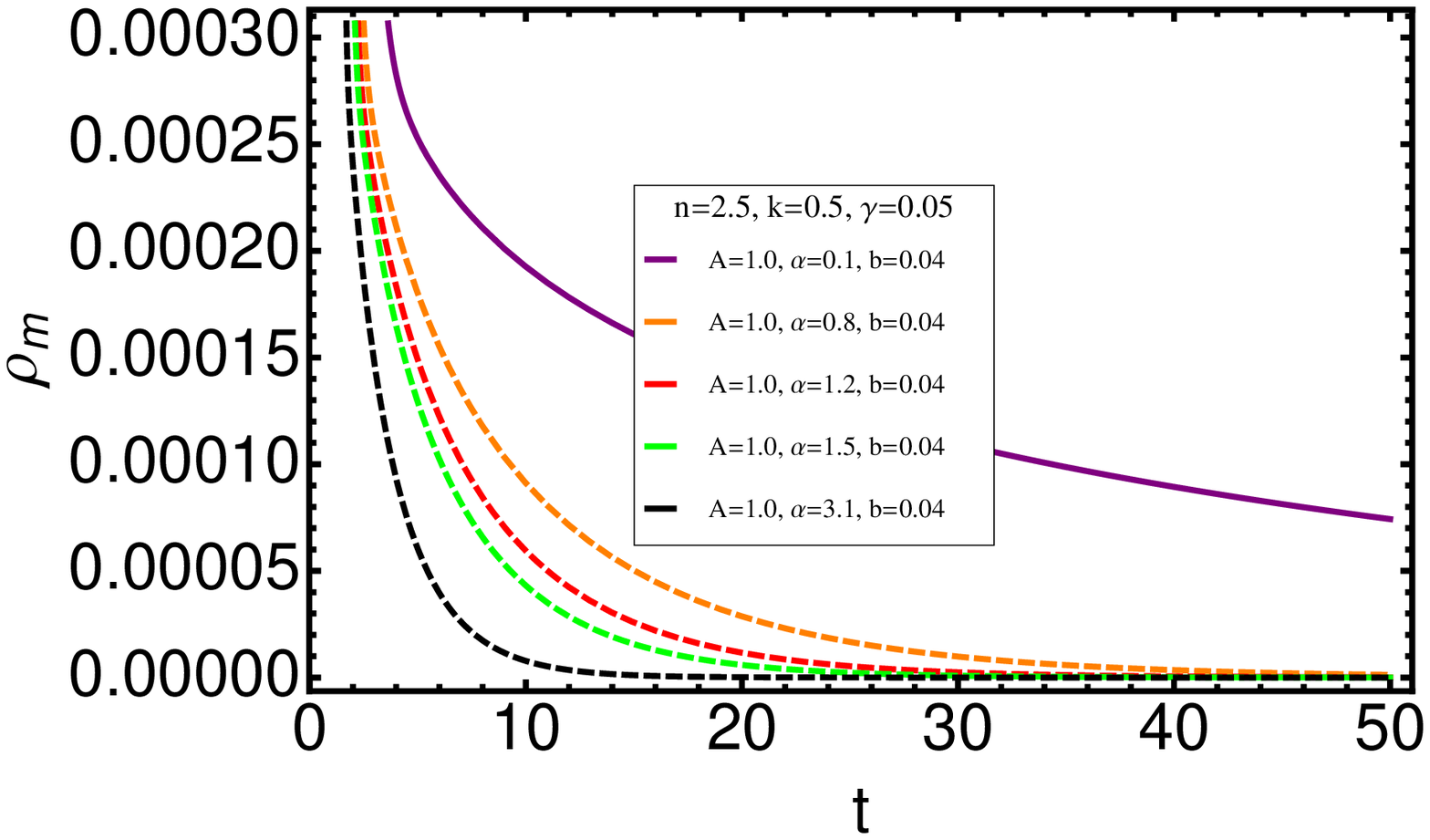}\\
\includegraphics[width=48 mm]{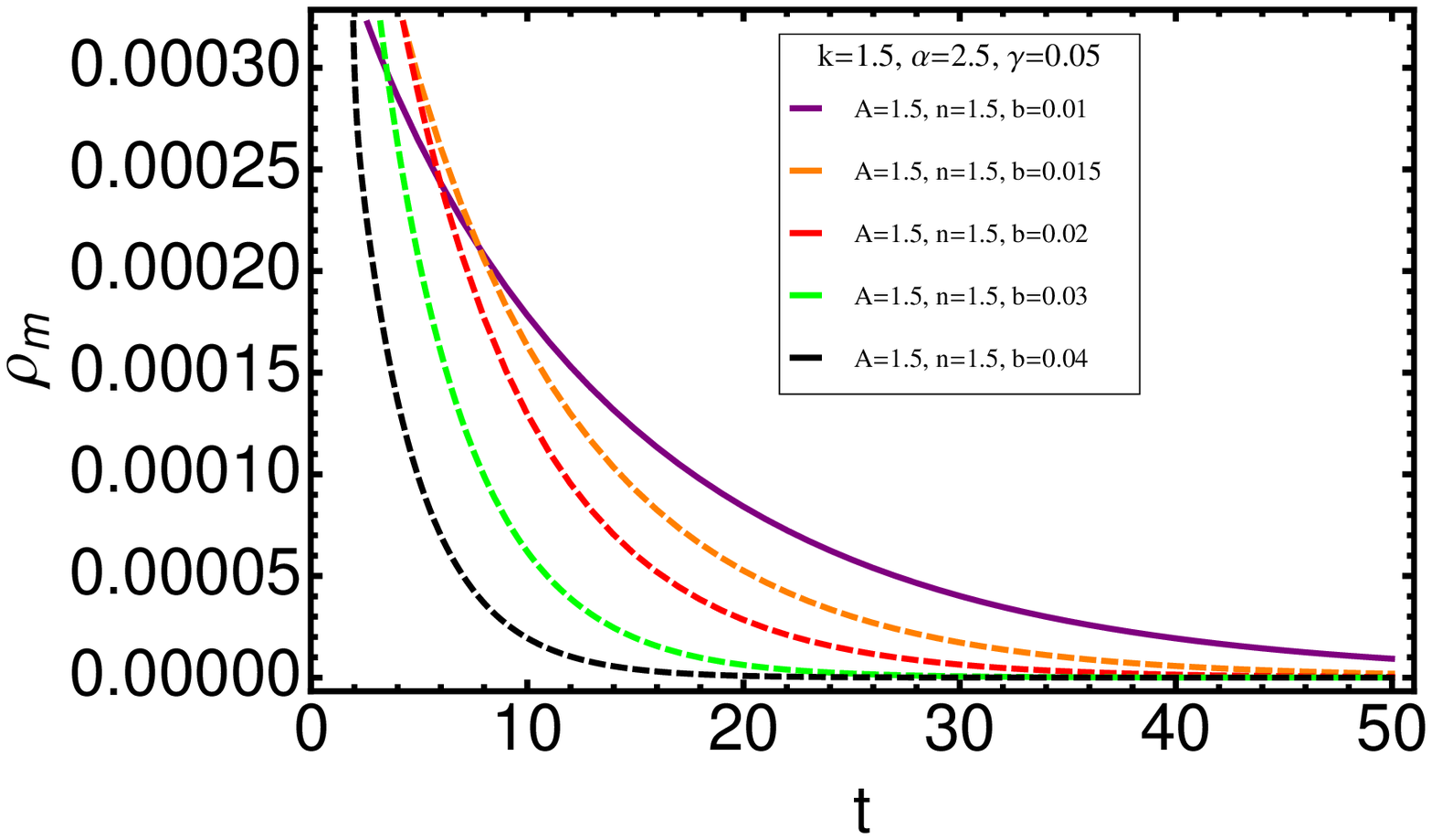} &
\includegraphics[width=48 mm]{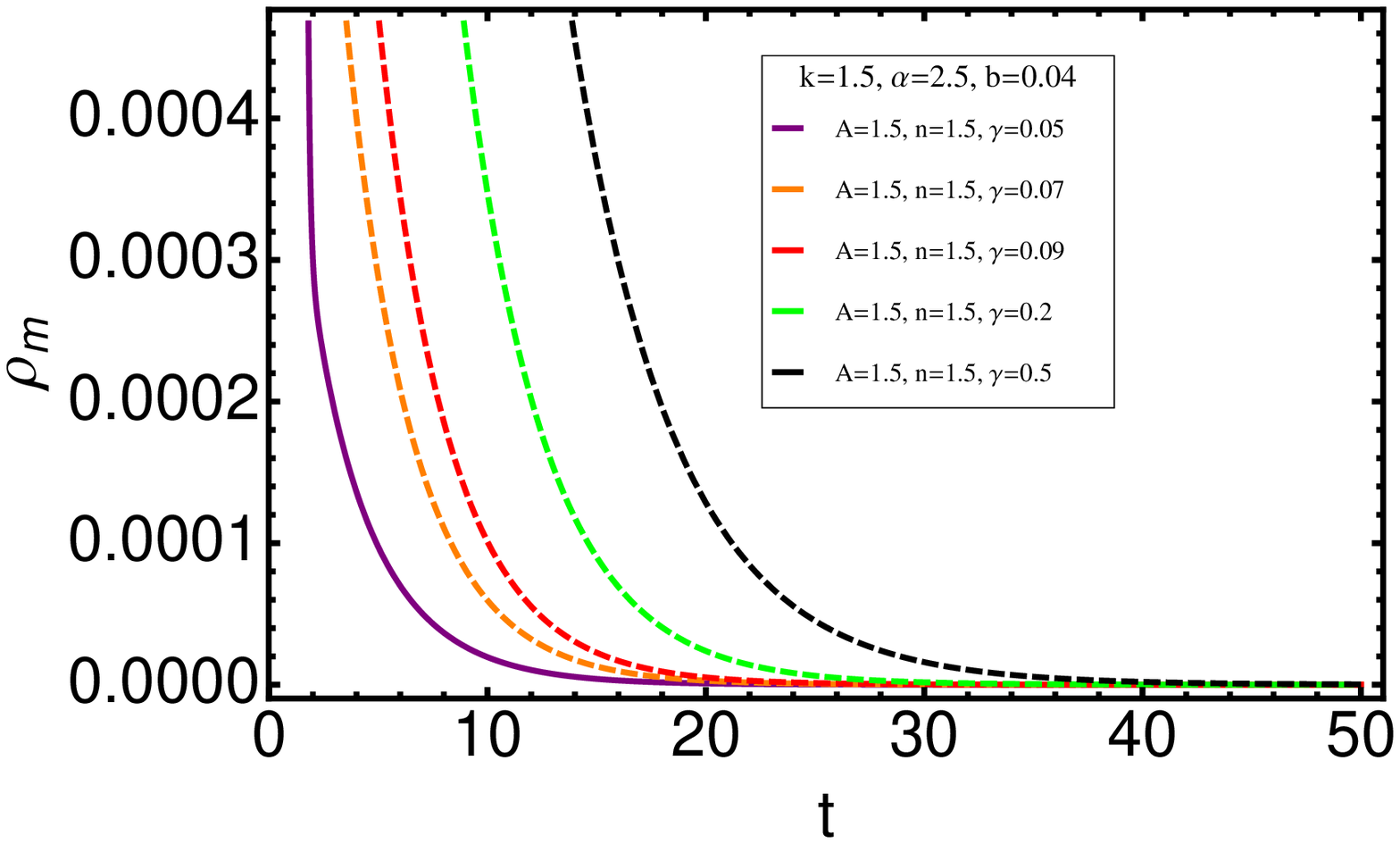}
 \end{array}$
 \end{center}
\caption{Behavior of $\rho_{m}$ against $t$ for interacting
components where we choose $a_{0}=1$.}
 \label{fig:14}
\end{figure}

\begin{figure}[h]
 \begin{center}$
 \begin{array}{cccc}
\includegraphics[width=50 mm]{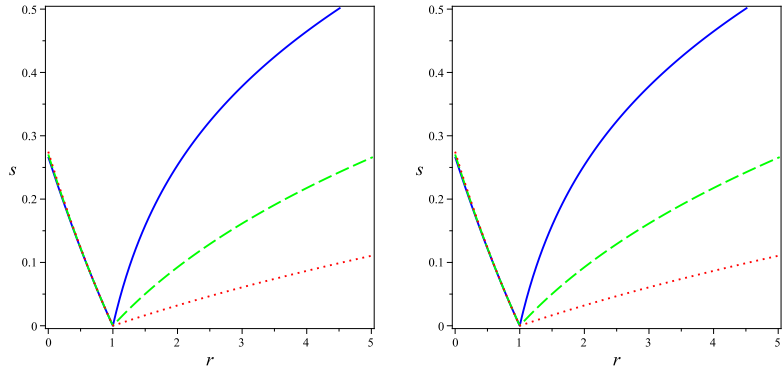}&
\includegraphics[width=50 mm]{MS1781fig15-1.eps}
 \end{array}$
 \end{center}
\caption{Behavior of $s$ against $r$ by choosing $a_{0}=1$. Left:
$n=1$ (solid line), $n=2$ (dashed line) and $n=5$ (dotted line).
Right: $n=0.6$ (solid line), $n=0.5$ (dashed line).}
 \label{fig:15}
\end{figure}

\begin{figure}[h]
 \begin{center}$
 \begin{array}{cccc}
\includegraphics[width=50 mm]{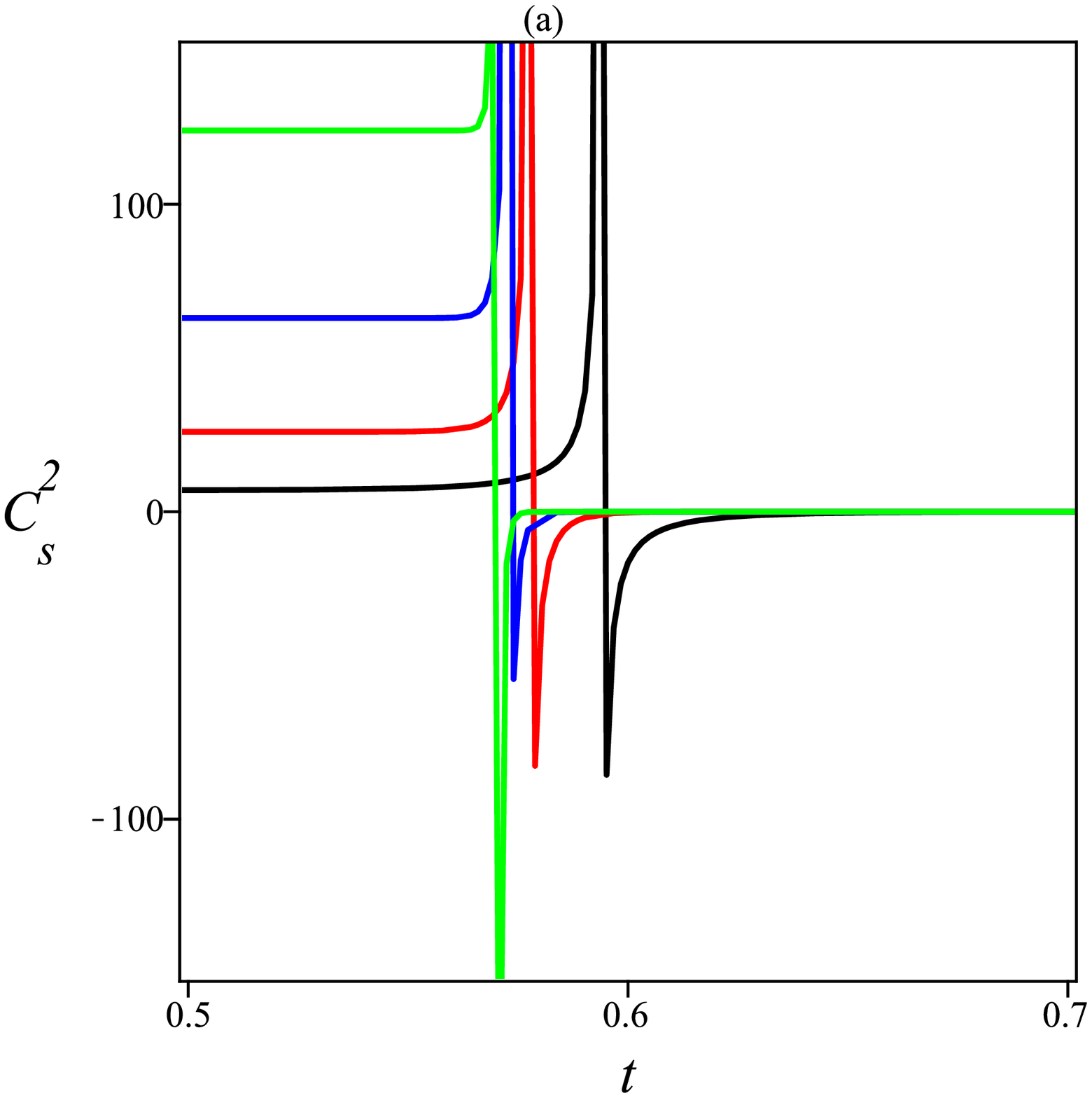}
 \end{array}$
 \end{center}
\caption{Behavior of $C_{s}^{2}$ against $r$, for $n=1$ and
$\alpha=1$. $b=\gamma=2$ (black), $b=\gamma=3$ (red), $b=\gamma=4$
(blue), $b=\gamma=5$ (green).}
 \label{fig:16}
\end{figure}

\section*{Discussion}
In this paper, we considered mutually interacting Tachyon dark energy
and extended it to the case of variable $G$ and $\Lambda$. We obtained
behavior of some cosmological quantities by using analytical and numerical analysis. Under some assumptions we obtained analytical expressions for energy densities in terms of time, which yield us to obtain the tensor to scalar ratio. We fixed some parameters as unity and reduced free parameters of the models. We found that higher values of $n$ is more agreement with observational data.\\
Below, we give two steps to explain cosmological quantities which obtained numerically.
In the first step we deal with $G(t)$, $\omega_{tot}(t)$ and $q$,
and in the second step we deal with $\phi(t)$, $V$, $P_{\Lambda}$, $\rho_{\Lambda}$ and $\rho_{m}$. In the first step we are able to compare our results with observational data, while in the second step there are no measurement on parameters.\\
Plots of the Fig. 1 show behavior of $G$ versus $t$ with variation
of $A$, $\alpha$, $n$ and $k$. We found that $G$ is increasing
function of $t$ at the early step, and yield to a constant at the later
step. It is clear from the Fig. 2 that increasing $A$ and $n$
increase value of $G$, but increasing $\alpha$ and $k$ decrease
value of
$G$.\\
Plots of the Fig. 2 show behavior of $\omega_{tot}$ versus $t$ with
variation of $A$, $\alpha$, $n$ and $k$. We found that
$\omega_{tot}$ is totaly negative after initial time and yields to
-1 at the large $t$. It is clear from the Fig. 2 that increasing of parameter
 $n$ increases value of $\omega_{tot}$, but increasing $\alpha$
decreases the value of
$\omega_{tot}$. We also found that variation of $A$ and $k$ have no important effect on $\omega_{tot}$.
Black line of the last plot, which is corresponding to the $n=7$, has opposite behavior at the early stage which is more expected, therefore, we can restrict this parameter as
$n>6$.\\\\\\
Plots of the Fig. 3 show behavior of the declaration parameter $q$
versus $t$ with variation of $A$, $\alpha$, $n$ and $k$. We found
that $q$ is totaly negative after initial time which is
corresponding to accelerating expansion of Universe. Also it yields
to -1 after large time which suggest constant value for Hubble
expansion parameter and agree with current data. It is clear from
the Fig. 3 that increasing
 $n$ increases value of $q$, but increasing $\alpha$
decreases the value of $q$, and variation of $A$ and $k$ have no
important effect on $q$. The current observation of $q\approx-0.8$ indicated at current time $t\approx2$.\\
Plots of the Fig. 4 should compare with the Fig. 1 to find effect of
interaction on $G(t)$. In the Fig. 4 we vary $n$, $\alpha$, $b$ and
$\gamma$. We found that $G$ is increasing function of $t$ at the
early step, which is similar to the non interacting case. Then, at
the large stage, relating to the value of parameters, it may yield to a
constant or may diverges. For example, by choosing $A=1$,
$0.1\leq\alpha\leq0.8$, $b=0.04$, $n=2.5$, $k=0.5$ and $\gamma=0.05$
as well as $A=1.5$, $\alpha=2.5$, $b=0.01$, $n=1.5$, $k=1.5$ and
$\gamma=0.05$ we obtain constant $G$. It is clear from the Fig. 4
that increasing $\alpha$, $b$ and $n$ increase value of $G$, but
increasing $\gamma$ decreases value of $G$.
We can see that variation of $G$ with $\alpha$ is completely different with the case of non interacting components where,
as illustrated in the Fig. 1, $\alpha$ decreased value of $G$. At the current stage ($t\approx2$) the value of the $G$ is infinitesimal in agreement with current observations.\\\\
Plots of the Fig. 5 show behavior of $\omega_{tot}$ versus $t$ with
variation of $n$, $\alpha$, $b$ and $\gamma$ for the case of
interacting components. We found that $\omega_{tot}$ is totaly
negative and increasing after initial time and yields to negative
constant at the large $t$. It is clear from the Fig. 5 that
increasing of $b$ increases value of $\omega_{tot}$, but increasing of
$\alpha$, $n$ and $\gamma$ decrease the value of $\omega_{tot}$. We
can conclude that choosing $b=0.04$, $\gamma=0.05$, $A=k=n=1.5$ and
$\alpha=0.5$ give $\omega_{tot}\rightarrow-1$.\\
Plots of the Fig. 6 show behavior of the declaration parameter $q$
versus $t$ for the case of interacting components. We found that $q$
is totaly negative which confirm accelerating expansion of Universe
as well as non interacting case. We found that this parameter
decreases with time in the initial stage to reach a minimum, then
increases with time to reach a constant value at the late stage which
is near -1. It is clear from the Fig. 6 that increasing $b$, and
$\alpha$ increase value of $q$, but increasing $n$ and $\gamma$
decrease the value of $q$, which are different with the non
interacting case. It is illustrated that the $q\approx-0.8$ agree with current stage about $t=2$.\\
Now, we consider second set of quantities. Figures 7, 8, and 9 are
corresponding to the non interacting case. Plots of the Fig. 7 show that
the scalar field $\phi$ is increasing at initial stage and then yields to a constant suddenly. So it seems that the scalar field is constant at present.
In another word, the scalar field $\phi$ grows suddenly and reaches
to the stable phase at the present. It means that the Tachyon field is unstable at the
early stage. We found that $\phi$ increased by $A$ while decreased
by $n$, $k$ and $\alpha$.\\ These plots suggest $0<\alpha\leq2$ is
necessary to obtain non trivial scalar field.\\
Plots of the Fig. 8 show behavior of the Tachyon potential versus the cosmic time
which is increasing initially and decreasing function at the late
stage. This is clear from the last plot which obtained for $k=1.5$,
$\alpha=0.5$, $A=0.5$ and $0.5\leq n\leq7$.  The maximum value of
potential, which is obtained at initial stage, tells that Tachyon
field is unstable in early stages of evolution. This may be because
of some irreversible processes such as a particle creation and
annihilation. However, more deep analysis and studies are needed in
order to conclude with right physics. We also found that increasing
$A$ and $\alpha$ decrease value of potential.\\
Pressure of cosmological constant for the case of non interacting
component illustrated in the plots of the Fig. 9. As expected,
pressure of cosmological constant is negative and decreasing
function of time. The second and third plots of the Fig. 9 show that
$\alpha$ and $k$ decrease value of cosmological constant pressure,
respectively. It is illustrated that the pressure of cosmological
constant diverges at initial stage. This infinite negative pressure
is corresponding to suddenly expansion of the early Universe.\\
Plots of the Fig. 10 show that scalar field $\phi$ is totally
increasing function of time in the case of interacting components.
The first and second plots suggest that increasing $n$ and $\alpha$
decrease value of scalar field, which is similar to the non
interacting case, however there are no differences between them at
initial stage.\\
Two lost plots represent the effect of interaction
term. It is clear that $b$ increases while $\gamma$ decreases value
of scalar field.\\
Tachyon potential of the interaction case illustrated in the plots
of the Fig. 11, which is similar to the non interaction case. We see
transition from unstable to stable state at inial stage. We found
that $n$, $\alpha$ and $b$ decrease value of Tachyon potential but
$\gamma$ increases value of Tachyon potential. As we expected the Tachyon potential vanishes at the late time.\\
Plots of the Fig. 12 are corresponding to time evolution of
cosmological constant pressure which is negative during time. In
contradict with non interacting case we see that the cosmological
constant pressure increased at initial stage to a negative minimum
and then decreased to zero at the late stage. We can see that $n$
and $b$ decrease pressure while $\gamma$ increase it.\\
Then, cosmological constant density represented by plots of the Fig.
13 which show that it is increasing at initial tile and decreasing
until now, which is in agreement with presence accelerating
expansion of Universe. It is found that $n$, $\alpha$ and $b$,
decrease value of time-dependent density but $\gamma$ increases
one.\\
Density of pressureless matter in presence of interaction term
plotted in the Fig. 14 which is decreasing function of time and yields to infinitesimal value at the late time as expected. We
found that $n$ and $\alpha$ decrease vale of density but $\gamma$
increases one. Variation of density with $b$ is depend on its value.
So small values of $b$ increases density but larger value than
$b=0.02$ decreased one.\\
Results for statefinder parameters $(s-r)$ for our model illustrated
in the Fig. 15. We draw diagram for various values of $n$ and see
that the standard $\Lambda$CDM fixed point is $(s=0, r=0)$ for all
$n$. For $r\geq1$ we found that increasing $n$ decreases $s$ while
in the range of $r\leq1$ there is no difference between curves with
different $n$. Also, we see different behavior for $n>1$ and $0<n<1$
in the range of $r>1$.\\
Finally, we can obtain suitable condition to have stable model. Our
numerical analysis shows that the square of sound speed
$C_{s}^{2}=\dot{P}/\dot{\rho}$ is always a positive constant for
$\gamma=1$ and $b>\gamma$. Therefore, our model will be stable
during evolution of Universe. In the special case of interaction
parameters discussed in section of analytical analysis ($\gamma=b$)
also there are stable region which illustrated in the Fig. 16. We
can see that there are small region in the early Universe where
$C_{s}^{2}<0$ and our model is Unstable. These regions are
corresponding to maximum of potential in the Fig. 11 which discussed
above and interpreted as transition from unstable to stable state.\\
Therefore, we obtained effects of variable $G$ and $\Lambda$ on the model which suggested as a model of our Universe. In summary, we proposed interacting Tachyon dark energy with variable $G$ and $\Lambda$ as a toy model of our Universe which is closest to real world.

\section*{Acknowledgments}
Martiros Khurshudyan has been supported by EU fonds in the frame of the program FP7-Marie Curie Initial Training Network INDEX NO.289968.

\end{document}